\documentclass[12pt]{article}
\usepackage{amsmath}

\usepackage{amssymb}
\usepackage{bm}
\usepackage{braket}
\usepackage[dvips]{graphicx}
\usepackage{physics}
\usepackage{comment}

\newcommand{\D}{ \not \hspace{-4pt}D}

\begin{document}

\baselineskip=17pt

\begin{titlepage}
\rightline{\texttt{arXiv:2405.09243}}
\rightline{\tt KUNS-3004}
\begin{center}
\vskip 1.5cm
\baselineskip=22pt
{\Large \bf {Noether's theorem and Ward-Takahashi identities\\ from homotopy algebras}}
\end{center}
\begin{center}
\vskip 1.0cm
{\large Keisuke Konosu$^{1}$ and Jojiro Totsuka-Yoshinaka$^{2}$}
\vskip 1.0cm
{${}^{1}$\it {Graduate School of Arts and Sciences, The University of Tokyo}}\\
{\it {3-8-1 Komaba, Meguro-ku, Tokyo 153-8902, Japan}}\\
konosu-keisuke@g.ecc.u-tokyo.ac.jp,\\
{${}^{2}$\it {Department of Physics, Kyoto University,}}\\
{\it {Kyoto 606-8502, Japan}}\\
george.yoshinaka@gauge.scphys.kyoto-u.ac.jp,
\vskip 2.0cm

{\bf Abstract}
\end{center}

\noindent
We derive the new identity in homotopy algebras which directly corresponds to the Schwinger-Dyson equations in quantum field theory. As an application, we derive the Ward-Takahashi identities. We demonstrate that the Ward-Takahashi identities are reproduced in several examples. In general, our formula contains divergence. We mediate this problem by introducing stubs known in the context of string field theory. With the regularization, we can calculate the anomaly such as  axial U(1) anomaly in vector-like U(1) gauge theory.

\end{titlepage}

\tableofcontents
\section{Introduction}
\setcounter{equation}{0}
Homotopy algebras such as $A_\infty$ algebras~\cite{Stasheff:I,Stasheff:II,Getzler-Jones,Markl,Penkava:1994mu,Gaberdiel:1997ia} and $L_\infty$ algebras~\cite{Zwiebach:1992ie,Markl:1997bj} have contributed in the context of constructing the action of string field theory~\cite{Zwiebach:1992ie,Maccaferri:2022yzy,Maccaferri:2023gcg,Kajiura:2004xu,Kajiura:2005sn,Erler:2013xta,Erler:2016ybs,Kunitomo:2019glq,Kunitomo:2022qqp}. The homotopy structure ensures the gauge invariance of the action explicitly.
Recently, homotopy algebras have been used for various purposes beyond constructing the action. For instance, we can reproduce the Feynman diagram expansion~\cite{Kajiura:2003ax,Sen:2016qap,Erbin:2020eyc,Koyama:2020qfb,Arvanitakis:2020rrk,Arvanitakis:2021ecw,Bonezzi:2023xhn,Konopka:2015tta,Kunitomo:2020xrl}. We can also relate covariant and light-cone string field theories by homotopy algebras~\cite{Erler:2020beb}.
 Notably, they are applicable to not only  string field theory but also ordinary quantum field theories~\cite{Hohm:2017pnh,Jurco:2018sby,Nutzi:2018vkl,Arvanitakis:2019ald,Macrelli:2019afx,Jurco:2019yfd,Saemann:2020oyz}. 
 Particularly, recent proposals utilizing homotopy algebras to express correlation functions~\cite{Okawa:2022sjf,Konosu:2023pal,Konosu:2023rkm} are noteworthy.\footnote{In the context of the Batalin-Vilkovisky formalism~\cite{Batalin:1981jr,Batalin:1983ggl,Schwarz:1992nx}, see~\cite{Gwilliam:2012jg,Chiaffrino:2021pob}. See also~\cite{Masuda:2020tfa,Doubek:2017naz} for related works.}
One advantage of describing quantum field theory using homotopy algebras is that it allows descriptions independent of the details of the theory. 
For instance, string field theory and scalar field theory have different structures and the former is far more complex than the latter. Expressions based on homotopy algebras, however,  formally represent both of the theories in a similar form.
While solving string field theory poses significant challenges due to its complexity, various analytical techniques are available in quantum field theory.
By translating the properties and analytical methods of quantum field theory into the language of homotopy algebras, we expect that we can directly apply or extend them for applications in string field theory.

Symmetry plays a crucial role in the analysis of quantum field theory.
In particular, global symmetry is an important concept related to conserved charges and anomalies.
In the context of homotopy algebras, there are several discussions regarding global symmetry~\cite{Erler:2016rxg,Erbin:2020eyc}, but they remain at the classical level.
In this paper, we provide an expression of the Ward-Takahashi identity in terms of homotopy algebras, based on the expression of correlation functions given in~\cite{Okawa:2022sjf,Konosu:2023pal,Konosu:2023rkm}.\footnote{Though the approach is different from ours, there is an unpublished work deeply related to our work by Hiroaki Matsunaga~\cite{Matsunaga}.} 
To derive this, we use a new identity that we refer to as the \textit{algebraic Schwinger-Dyson equation}. 
Moreover, this equation improves the proof of the Schwinger-Dyson equation from the point of the formula for correlation functions in terms of homotopy algebras presented in~\cite{Okawa:2022sjf,Konosu:2023pal,Konosu:2023rkm}, which enable us to extend to general theories easier.
The Ward-Takahashi identity in terms of homotopy algebras we derived contains a term corresponding to an anomaly.
This term is generally ill-defined and requires regularization.
We applied regularization with stubs~\cite{Zwiebach:1992ie}, which is widely used in the context of string field theory.
In the context of homotopy algebras, research on adding stubs has seen significant development in recent years~\cite{Schnabl:2023dbv,Erbin:2023hcs,Schnabl:2024fdx,Maccaferri:2024puc}.
By adding stubs, we can perform the heat kernel regularization while preserving the structure of homotopy algebras, making it the optimal regularization for the discussions in this paper. The regularization method we introduce is related to the work~\cite{Chiaffrino:2021uyd}, in which the anomaly and the regularization with stubs are discussed in the framework of the Batalin-Vilkovisky formalism.\footnote{The approach is a bit different since we are directly evaluating  correlation functions.}  

The rest of the paper is organized as follows.
In section~2, we briefly review $A_\infty$ algebras and quantum $A_\infty$ algebras.
In section~3, we review the formula for correlation functions in terms of homotopy algebras for the scalar field theories~\cite{Okawa:2022sjf,Konosu:2023pal}.
Subsequently, we discuss correlation functions for the general theories and introduce the \textit{algebraic Schwinger-Dyson equation}, which are important in our discussion.
In section~4, we provide expressions for Noether's theorem and the Ward-Takahashi identity in terms of homotopy algebras.
We then illustrate these expressions with three examples, the U(1) symmetry in complex scalar field theories, the translation symmetry in scalar field theories, and the axial $U(1)$ symmetry in vector-like gauge theories.
In section~5, we explain the regularization with stubs and calculate the values of anomalies in the above examples.
As is well known, non-zero anomalies can be calculated only in the third example. 

Through the paper, we mostly follow the convention~\cite{srednicki2007quantum} when we describe quantum field theory.

\section{$A_{\infty}$ algebras and quantum $A_{\infty}$ algebras}\label{A-infty-review}
\setcounter{equation}{0}
In this section, we explain $A_\infty$ algebras and their extensions, commonly referred to as quantum $A_\infty$ algebras.
These algebras are defined within the context of graded vector space\footnote{Strictly speaking, this space is not the vector space, but the module when we consider Grassmann odd fields. In this paper, these differences do not contribute to our discussions.} denoted as $\mathcal{H}$.
In our setup, we consider the graded vector space spanned by the basis $\{e_a(x),\, \widetilde{e}^{\,a}(x)\}$.
Here the index $a$ represents discrete labels and $x$ represents continuous label.
In this paper, we associate the discrete label with the types of fields or the indices of Lorentz and gauge groups, and the continuous label is associated with the coordinates of a $d$-dimensional spacetime.
We denote the subspace generated by the basis $\{e_a(x)\}$ as $\mathcal{H}_1$, and the subspace generated by $\{\widetilde{e}^{\,a}(x)\}$ as $\mathcal{H}_2$.
The general element  of $\mathcal{H}$ can be written as 
\begin{align}
    F=\sum_a\int\dd[d]x\,\qty[\,f^a(x)\, e_a(x)+\widetilde{f}_{\,a}(x)\, 
 \widetilde{e}^{\,a}(x)\,]\,,
\end{align}
where coefficients $f^a$ and $\widetilde{f_a}$ are functions.\footnote{In this paper, we assume that spacetime is flat. On a curved spacetime, it is necessary to replace the integral measure with $\sqrt{g}\,\dd^dx$.}
 The identity map on $\mathcal{H}$ is denoted by $\mathbb{I}$. 
The degree is determined by the Grassmann parity and the degree of $\widetilde{e}^{\,a}(x)$ is defined to be opposite to that of $e_{\,a}(x)$. 
If the basis vector $e_a(x)$ has an even degree, $\widetilde{e}^{\,a}(x)$ has an odd degree, and vice versa.
We assume that the degree of coefficients and basis vectors do not depend on continuous labels $x$, and denote their degree as $\mathrm{deg}(f^a)$ and $\mathrm{deg}(e_a)$. 
Furthermore, when the element $F$ of $\mathcal{H}$ is homogeneous with respect to the degree, we denote it as $\mathrm{deg}(F)$.
For the sake of simplicity in notation, we write the sign factor $(-1)^{\mathrm{deg}(f^a)}$ as $(-1)^{f^a}$.

We introduce the three kinds of operator acting on this $\mathcal{H}$,  the degree-odd linear map $Q:\mathcal{H}\rightarrow\mathcal{H}$, the degree-odd multilinear map $m_n:\mathcal{H}^{\otimes n}\rightarrow\mathcal{H}$ for $n\geq 0$, and the degree-odd symplectic form $\omega:\mathcal{H}\otimes\mathcal{H}\rightarrow\mathbb{C}$.
These operators are graded commutative with the coefficients.
For instance, $Q$ acts as
\begin{align}
     Q\,\big(\,f^a(x)\,e_a(x)\,\big)&=(-1)^{f^a}\,f^a(x)\,Q\,e_a(x)\,.
\end{align}
The operator $Q$ is defined to satisfy the following conditions:
\begin{align}
    Q^2&=0\,\label{nil-Q}\\
    \omega\,(\,F_1\,,Q\,F_2\,)&=-(-1)^{F_1}\,\omega\,(\,Q\,F_1\,,F_2\,)\,.\label{cyclicity-Q}
\end{align}
The first condition means that the operator $Q$ is nilpotent, and the second condition is called \textit{cyclicity}.
The symplectic form $\omega$ is defined for the $e_a(x)$ and $\widetilde{e}^{\,a}(x)$ as follows:
\begin{equation}\label{normalization}
\begin{split}
    \omega\, (\,e_a(x)\,, \widetilde{e}^{\,b}(y)\,)&=(-1)^{e_a}\,\delta_a^b\, \delta(x-y)\,,\\
    \omega\, (\,\widetilde{e}^{\,a}(x)\, , e_b(y)\, )&=(-1)^{\widetilde{e}^{a}}\,\delta_b^a\, \delta(x-y)\,,\\
    \omega\,(\,e_a(x)\,,e_b(y)\,)&=\omega\,(\,\widetilde{e}^{\,a}(x),\,\widetilde{e}^{\,b}(y)\,)=0\,.
\end{split}
\end{equation}
When $x=y$, the above two equations are defined by taking a limit $x\rightarrow y$.
As it can be seen from the definition, the symplectic form $\omega$ satisfies the graded antisymmetric property:
\begin{align}
    \omega\,(\,F_1,\,F_2\,)=-(-1)^{F_1F_2}\,\omega\,(\,F_2\,,F_1\,)\,.
\end{align}
It is convenient to define the bra notation of the symplectic form $\bra{\omega}$ as
\begin{align}
    \bra{\omega}\,F_1\otimes F_2=\omega(\,F_1,\, F_2\,)\,.
\end{align}

In terms of homotopy algebras, the action is given by
\begin{align}
    S=-\frac{1}{2} \, \omega \, ( \, \Phi, Q \, \Phi \, )
-\sum_{n=0}^\infty \, \frac{1}{n+1} \,
\omega \, ( \, \Phi \,, m_n \, ( \underbrace{\, \Phi\,, \dots \,, \Phi \,}_n ) \, ) \,, \label{homotopy-action}
\end{align}
where we used the operators defined above, and $\Phi$ represents the degree-even element of $\mathcal{H}_1$:
\begin{align}
    \Phi=\sum_a\,\int\dd[d]x\, \phi^a(x)\,e_a(x)\,,
\end{align}
where $\{\phi^a(x)\}$ are fields.
The condition that the element $\Phi$ has even degree is required so that the first term on the right-hand side of~\eqref{homotopy-action} does not vanish by the cyclicity of $Q$.
Since the element $\Phi$ has even degree,
\begin{align}
    \mathrm{deg}(\phi^a)=\mathrm{deg}(e_a)\,.
\end{align}
For the simple descriptions, it is convenient to define $M_n$ by
\begin{align}
    M_1=Q+m_1\,,\quad
    M_n=m_n\,,
\end{align}
for $n\neq1$.
The action \eqref{homotopy-action} is invariant under
the gauge transformation of the element $\Phi$ given by
\begin{equation}
   \delta_\Lambda \Phi=\sum_{k=1}^{\infty} \sum_{l=0}^{k-1}M_{k}(\underbrace{\Phi\otimes\ldots\otimes\Phi}_{l}\otimes\Lambda\otimes\underbrace{\Phi\otimes\ldots\otimes\Phi}_{k-l-1})\,, \label{gauge}
\end{equation}
where $\Lambda$ is the degree-odd gauge parameter,
if the symplectic form $\omega$ and operators $M_{n}$ satisfy the following two kinds of conditions, which are extensions of the nilpotency and the cyclicity of $Q$.
The former one is called $A_\infty$ \textit{relations}.\footnote{More precisely, this relation is called \textit{weak} $A_\infty$ \textit{relations}, and, in particular, when $M_0=0$, it is called $A_\infty$ \textit{relations}.}
A few of them are
\begin{subequations}
\begin{align}
0=&M_1(M_0)\,,\\
0=&M_1 ( M_1 ( \Phi_1 ) ) +M_2(M_0\,,\Phi_1)+(-1)^{\Phi_{1}}M_2(\Phi_1\,,M_0)\,, \label{A_1}\\
0=& M_1 ( M_2 ( \Phi_1\,, \Phi_2 ) ) +M_2 ( M_1 ( \Phi_1 ) \,, \Phi_2 )
+(-1)^{\Phi_1} M_2 ( \Phi_1 \,, M_1 ( \Phi_2 ) ) \nonumber\\
&+M_3(M_0\,,\Phi_1\,,\Phi_2)+(-1)^{\Phi_1}M_3(\Phi_1\,,M_0\,,\Phi_2)+(-1)^{\Phi_1+\Phi_2}M_3(\Phi_1\,,\Phi_2\,,M_0)\,, \label{A_2}\\
0=& M_1 ( M_3 ( \Phi_1 \,, \Phi_2 \,, \Phi_3 ) )
+M_2 ( M_2 ( \Phi_1 \,, \Phi_2 ) \,, \Phi_3 )\nonumber \\
& +(-1)^{\Phi_1} M_2 ( \Phi_1 \,, M_2 ( \Phi_2 \,, \Phi_3 ) )
+M_3 ( M_1 ( \Phi_1 ) \,, \Phi_2 \,, \Phi_3 )\nonumber \\
&+(-1)^{\Phi_1} M_3 ( \Phi_1 \,, M_1 ( \Phi_2 ) \,, \Phi_3 )
+(-1)^{\Phi_1+\Phi_2}
M_3 ( \Phi_1 \,, \Phi_2 \,, M_1 ( \Phi_3 ) )\nonumber\\
&+M_4(M_0\,,\Phi_1\,,\Phi_2\,,\Phi_3)+(-1)^{\Phi_1}M_4(\Phi_1\,,M_0\,,\Phi_2,\Phi_3)\nonumber\\
&+(-1)^{\Phi_1+\Phi_2}M_4(\Phi_1\,,\Phi_2\,,M_0\,,\Phi_3)+(-1)^{\Phi_1+\Phi_2+\Phi_3}M_4(\Phi_1\,,\Phi_2\,,\Phi_3\,,M_0)\,. \label{A_3}
\end{align}
\end{subequations}
Another one is cyclicity of $M_n$:
\begin{align}
    \omega \, ( \, \Phi_1 \,, M_n \, ( \, \Phi_2 \,, \ldots \,, \Phi_{n+1} \, ) \, )
= {}-(-1)^{\Phi_1} \,
\omega \, ( \, M_n ( \, \Phi_1 \,, \ldots \,, \Phi_n \, ) \,, \Phi_{n+1} \, ) \,.
\label{cyclic}
\end{align}
When an algebra satisfies $A_\infty$ relations, it is called an \textit{$A_\infty$ algebra}. Additionally, if it possesses cyclicity, it is referred to as a \textit{cyclic $A_\infty$ algebra}.

In order to manipulate  $A_{\infty}$ algebras, it is convenient to introduce the \textit{coalgebra representation}.\footnote{
The coalgebra representation of $A_\infty$ algebras
is explained in detail in many papers. See, for example, appendix~A of~\cite{Erler:2015uba} and~\cite{Koyama:2020qfb}.
We mostly follow the conventions used in these papers and \cite{Konosu:2023pal}.} In the context of coalgebra representation, we extend the operators on $\mathcal{H}^{\otimes n}$ to those on the tensor algebra:
\begin{align}
    T\mathcal{H}=\mathcal{H}^{\otimes 0}\oplus\mathcal{H}\oplus\mathcal{H}^{\otimes 2}\oplus\mathcal{H}^{\otimes 3}\oplus\cdots\,.
\end{align}
Here $\mathcal{H}^{\otimes 0}$ is isomorphic to $\mathbb{C}$, and we denote the single basis vector $\vb{1}$, which satisfies
\begin{equation}
\begin{split}
  &\vb{1}\otimes\vb{1}=\vb{1}\,,\\
    &\vb{1}\otimes F=F\otimes \vb{1}=F\,,
    \end{split}
\end{equation}
for $F\in\mathcal{H}$. We denote the projection from $T\mathcal{H}$ to $\mathcal{H}^{\otimes n}$ by $\pi_n$.
The tensor algebra equips a product $\triangledown:T\mathcal{H}\otimes'T\mathcal{H}\rightarrow T\mathcal{H}$ defined by
\begin{align}
    \bigtriangledown\, (\,A\otimes'B\,)=A\otimes B\,,
\end{align}
for $A,B \in T\mathcal{H}$,
and a coproduct $\triangle:T\mathcal{H}\rightarrow T\mathcal{H}\otimes'T\mathcal{H}$ defined by
\begin{equation}
\begin{split}
\triangle\, \vb{1}&=\vb{1}\otimes'\vb{1}\,,\\
    \triangle\, (\,F_1\otimes F_2\otimes\cdots\otimes F_n\,)&=\sum_{k=1}^{n-1}\,(\,F_1\otimes\cdots\otimes F_k)\otimes'(F_{k+1}\otimes\cdots\otimes F_n\,)\\
    &+\,\vb{1}\otimes'(\,F_1\otimes F_2\otimes\cdots\otimes F_n\,)+(\,F_1\otimes F_2\otimes\cdots\otimes F_n\,)\otimes'\vb{1}\,,
\end{split}
\end{equation}
for $F_i\in\mathcal{H}$. 
Here $\otimes'$ is the tensor product for the tensor algebra.
The following relationship holds between the projection $\pi_n$, the product $\bigtriangledown$, and the coproduct $\triangle$:
\begin{align}
    \pi_n=\bigtriangledown\,(\,\pi_k\otimes'\pi_l\,)\,\triangle\,,
\end{align}
for $k+l=n$.
The operator that plays a central role in the coalgebra representation is called the \textit{coderivation}, and it is a linear map $\vb{c}:T\mathcal{H}\rightarrow T\mathcal{H}$ satisfying
\begin{align}
    \triangle\, \vb{c}=\big(\,\vb{I}\otimes'\vb{c}+\vb{c}\otimes'\vb{I}\,\big)\,\triangle\,,\label{def of coderivation}
\end{align}
where $\vb{I}$ is the identity map on the tensor algebra defined by
\begin{align}
    \vb{I}=\vb{1}+\mathbb{I}+\mathbb{I}^{\otimes 2}+\mathbb{I}^{\otimes 3}+\cdots\,.
\end{align}
An important property of the coderivation is that the graded commutator of two coderivations  also becomes a coderivation. 
We can easily confirm that  the graded commutator for two coderivation $\vb{c},\vb{c}'$
\begin{align}
    [\,\vb{c},\,\vb{c}'\,]=\vb{c}\,\vb{c}'-(-1)^{\vb{c}\vb{c}'}\,\vb{c}'\,\vb{c}\,,
\end{align} 
satisfies the condition~\eqref{def of coderivation}.
The degree-odd coderivation $\vb{M}_{n}$ can be constructed from the multilinear map $M_n$ as follows:
\begin{equation}
\begin{split}
    &\vb{M}_{n}\,(\,F_1\otimes\cdots\otimes F_m\,)=0\,, \quad \text{for} \quad m < n \,,\\
    &\vb{M}_{n}\,(\,F_1\otimes\cdots\otimes F_n\,)=M_n(\,F_1,\dots,F_n\,)\,,\\
    &\vb{M}_{n}\,(\,F_1\otimes\cdots\otimes F_m\,)=\sum_{k=0}^{m-n}\,(-1)^{F_1+\dots+F_k}\,F_1\otimes\dots F_k\otimes M_n\,(\,F_{k+1},\dots, F_{k+n}\,)\\
    &\qquad\qquad\qquad\qquad\qquad\qquad\otimes F_{k+n+1}\otimes\cdots \otimes F_m\, \quad \text{for} \quad m > n\,,
    \end{split}
\end{equation}
where $F_1,\dots,F_n \in \mathcal{H}$. 
We denote the sum of $\vb{M}_{n}$ as $\vb{M}$:
\begin{align}
    \vb{M}=\sum_{n=1}^\infty\, \vb{M}_n\,.
\end{align}
The multilinear map $M_n$ can be written as
\begin{align}
    M_n=\pi_1\,\vb{M}_{n}=\pi_1\,\vb{M}\,\pi_n\,.
\end{align}
It is clear from this construction that, in general, the coderivation $\vb{c}$ is completely determined by $\pi_1$-projection $\pi_1\,\vb{c}$.
Under the above notation, the $A_\infty$ relation can be written as the following simple equation:
\begin{align}
    \pi_1\,\vb{M}^2=0\,.\label{pi1-A-infty}
\end{align}
This equation implies 
\begin{align}
    \vb{M}^2=0\,,\label{A-infty}
\end{align}
i.e., $\vb{M}$ is nilpotent. This is because $\vb{M}^2=\frac{1}{2}\,[\,\vb{M}\,,\vb{M}\,]$ is a coderivation and is completely determined by the $\pi_1$-projection.
Furthermore, the cyclicity~\eqref{cyclic} can be written as
\begin{align}
    \bra{\omega}\,\pi_2\,\vb{M}_{n}=0\,.
\end{align}
In some cases, it is convenient to consider $Q$ and $m_{1}$ separately instead of considering $M_{1}$. We introduce degree-odd coderivations $\vb{Q}$ and $\vb{m}_{n}$ associated to $Q$ and $m_{n}$, respectively. We also define
\begin{align}
    \vb{m} = \sum_{n=0}^{\infty}\,\vb{m}_{n}\,.
\end{align}
Then, the relation~\eqref{nil-Q} is promoted to
\begin{align}
    \vb{Q}^2=0\,,
\end{align}
and the $A_{\infty}$ relation~\eqref{A-infty} is rewritten as
\begin{align}
    (\,\vb{Q}+\vb{m}\,)^{2}=0\,.
\end{align}
We define the group-like element $\frac{1}{1-\Phi}$ on $T\mathcal{H}$ corresponding to the degree even element of $\Phi$ of $\mathcal{H}_1$ that constitutes the action as
\begin{align}
    \frac{1}{1-\Phi}=\vb{1}+\Phi+\Phi\otimes\Phi+\dots\,,
\end{align}
which satisfies
\begin{align}
    \triangle\, \frac{1}{1-\Phi}=\frac{1}{1-\Phi}\otimes'\frac{1}{1-\Phi}\,.\label{propertyofgroup-likeelement}
\end{align}
Moreover, as it is clear from the definition, this group element is invariant under the permutation of elements in $\mathcal{H}$.
In other words, for the permutation map defined by $T^{(\sigma)}:T\mathcal{H}\rightarrow T\mathcal{H}$
\begin{align}
  T^{(\sigma)}(F_1\otimes\cdots\otimes F_n)=(-1)^{\epsilon(\sigma)}\,F_{\sigma(1)}\otimes\dots\otimes F_{\sigma(n)}  \,,
\end{align}
where $\sigma\in S_n$ and the sign factor $(-1)^{\epsilon(\sigma)}$ represents the appropriate sign that arises from the permutation, then
\begin{align}
    T^{(\sigma)}\,\frac{1}{1-\Phi}=\frac{1}{1-\Phi}\,.
\end{align}
The action~\eqref{homotopy-action} can be written as
\begin{align}
    S=-\sum_{n=0}^\infty \frac{1}{n+1}\,\bra{\omega}\,\bigtriangledown\,(\,\pi_1\otimes'\pi_1 \vb{M}_{n}\,)\,\triangle\, \frac{1}{1-\Phi}
\end{align}
with the group-like element.
When there is no confusion, $\bigtriangledown$ and $\triangle$ are often omitted and the action is simply written as
\begin{align}
    S=-\sum_{n=0}^\infty\, \frac{1}{n+1}\,\bra{\omega}\,(\,\mathbb{I}\otimes\pi_1 \vb{M}_{n}\,)\,\frac{1}{1-\Phi}\,.
\end{align}

We have presented all the ingredients to describe the quantum field theory at the tree level. The action with cyclic $A_\infty$ algebra is a solution of the classical master equation in the Batalin-Vilkovisky formalism.
At the quantum level, the action must rather be a solution of the quantum master equation. 
To satisfy the quantum master equation, we need to extend the algebra to the \textit{quantum $A_{\infty}$ algebra}, which is defined by the relation
\begin{align}
    \qty(\,\vb{M}+i\hbar\,\vb{U}\,)^{2}=0\,, \label{qA-infty}
\end{align}
or
\begin{align}
    \qty(\,\vb{Q}+\vb{m}+i\hbar\,\vb{U}\,)^{2}=0\,, \label{qA-infty2}
\end{align}
where $\hbar$ is the parameter and physically the Dirac constant. 
Here, the multilinear map $m_n$, which is $\pi_1$-projection of the coderivation $\vb{m}_{n}$, generally differs from the classical case. In this case, the multilinear map $m_n$ can be written as 
\begin{align}
    m_n=\sum_{g=0}^\infty\, m_{g,n}\,,
\end{align} 
in the power series expansion of $\hbar$, and the zeroth order  $m_{0,n}$ coincides with the classical case. 
Here and in what follows, we use the same notation to denote the operators before and after the $\hbar$ correction.
Next, let us explain the operator $\vb{U}$. The operator $\vb{U}$ is the degree-odd second-order coderivation\footnote{For the detailed definition of higher-order coderivation, see~\cite{Markl:1997bj}.} promoted by the map $U$ from $\mathcal{H}^{\otimes 0}$ to $\mathcal{H}^{\otimes 2}$ given by
\begin{align}
    U=\sum_{a}\int \dd[d] x \, \big(\,e_a(x)\, \otimes \widetilde{e}^{\,a}(x)+\,\widetilde{e}^{\,a}(x)\, \otimes e_a(x)\,\big).
\end{align}
The second-order coderivation $\vb{U}$ is constructed by
\begin{equation}
\begin{split}
    \vb{U}\, \vb{1}&=U=\sum_{a}\int \dd[d] x \, \big(\,e_a(x)\, \otimes \widetilde{e}^{\,a}(x)+\,\widetilde{e}^{\,a}(x)\, \otimes e_a(x)\,\big)\\
    \vb{U}\,(F_1\otimes\cdots\otimes F_n)&=\sum_{\substack{i,j,k\geq0\\i+j+k=n}}\sum_a\int \dd[d] x\, (-1)^{F_{1}+\cdots+F_{i}+\widetilde{e}^{\,a}(F_{i+1}+\cdots+F_{i+j})}\\
    &\qquad\qquad\qquad\qquad\big(\, F_1\otimes\cdots\otimes F_i\, \otimes\, e_a(x)\,\otimes\, F_{i+1}\otimes\cdots\otimes F_{i+j}\\
    &\qquad\qquad\qquad\qquad\otimes\,  \widetilde{e}^{\,a}(x)\,\otimes F_{i+j+1}\otimes\cdots\otimes F_{i+j+k}\,\big)\, +\, (e_a\, \leftrightarrow \,\widetilde{e}^{\,a})\,,
    \end{split}
\end{equation}
for $F_1,\dots,F_n\in\mathcal{H}$, 
where $(e_a\, \leftrightarrow \,\widetilde{e}^{\,a})$ represents the term that the operators $e_a$ and $\widetilde{e}^{\,a}$ are exchanged in the first term.
Unlike usual coderivations (also called the first-order coderivation), $\vb{U}$ is not determined by $\pi_1$-projection : $\pi_1\,\vb{U}=0$, but is determined by $\pi_2$-projection : $\pi_2\,\vb{U}=U$.
If we use the coderivations $\vb{e}_a(x)$ and $\widetilde{\vb{e}}^{\,a}(x)$ defined by
\begin{equation}
\begin{split}
        \vb{e}_a(x)\,\vb{1}&=e_a(x)\\
        \vb{e}_a(x)\,(\,F_1\otimes\cdots\otimes F_n\,)&= 
    \sum_{k=0}^{n}\,(-1)^{e_a(F_1+\cdots+F_k)} F_1\otimes\cdots\otimes F_k \otimes e_a(x) \otimes F_{k+1}\otimes\cdots\otimes F_n
\end{split}
\end{equation}
for $F_1,\dots,F_n\in\mathcal{H}$,
and similarly for $\widetilde{\vb{e}}^{\,a}(x)$,
then the second-order coderivation $\vb{U}$ can be rewritten as
\begin{align}
    \vb{U}=\sum_a\,\int \dd[d] x\, \vb{e}_a(x)\, \vb{\widetilde{e}}^a(x)\,.\label{defU}
\end{align} 
An important property of $U$ is that it is normalized by
\begin{equation}
\begin{split}
    \mathbb{I}&=-(\,\mathbb{I}\,\otimes\,\bra{\omega}\,)(\,U\,\otimes\,\mathbb{I}\,)\\
            &=\sum_a\int\dd[d] x\,\big(\,(-1)^{\widetilde{e}^{\,a}}\,e_a(x)\,\otimes\,\bra{\omega}\,(\,\widetilde{e}_a(x)\,\otimes\,\mathbb{I}\,)+(-1)^{e_a}\,\widetilde{e}^{\,a}(x)\,\otimes\,\bra{\omega}\,(\,e_a(x)\,\otimes\,\mathbb{I}\,)\big)\,.\label{nomalization of U}
\end{split}
\end{equation}
By using this property, the cyclicity~\eqref{cyclic} can also be expressed as
\begin{equation}
\begin{split}
    \sum_a\int\dd[d]x\,&\big[\,M_n\,(e_a(x)\otimes \mathbb{I}^{\otimes n-1})\,\otimes\,\widetilde{e}^{\,a}(x)
    +(-1)^{e_a}\,\,e_a(x)\,\otimes\,M_{n}\,(\mathbb{I}^{\otimes n-1}\otimes \widetilde{e}^{\,a}(x))\,\big]\\
    &+\, (e_a(x)\, \leftrightarrow \,\widetilde{e}^{\,a}(x))
    =0\,,\label{cyclic another}
\end{split}  
\end{equation}
The equivalence of two expression \eqref{cyclic} and \eqref{cyclic another} is proved in appendix \ref{app: cyclic}.
The relation \eqref{qA-infty} contains two types of relations.
One is called the \textit{main identity}:
\begin{align}
    \pi_1\,(\, \vb{M}+i\hbar\,\vb{U}\,)^2=0\ .
\end{align}
This is an extension of the $A_\infty$ relations~\eqref{pi1-A-infty}.
In particular, the contribution from $i\hbar\,\pi_1\,\vb{M\,U}$ can often be canceled by the appropriate renormalization, in which case those satisfying $A_\infty$ relations~\eqref{pi1-A-infty} satisfy this main identity as well.
Another one is
\begin{equation}
\begin{split}
    \sum_a\int\dd[d]x\,&\big[\,M_n\,\pi_{n}\,\vb{e}_a(x)\,\otimes\,\widetilde{e}^{\,a}(x)
    +(-1)^{e_a}\,\,e_a(x)\,\otimes\,M_{n}\,\pi_n\,\vb{\widetilde{e}}^{\,a}(x)\,\big]\\
    &+\, (e_a(x)\, \leftrightarrow \,\widetilde{e}^{\,a}(x))
    =0\,,
\end{split}
\end{equation}
which is derived from $\pi_2(\, \vb{M}+i\hbar\vb{U}\,)^2=0$.
When the multilinear map $M_n$ is commutative, as in the case of $C_\infty$ algebras or $L_\infty$ algebras, this relation is equivalent to the cyclicity~\eqref{cyclic another}.
In the case of general $A_\infty$ algebras, this relation does not imply the cyclicity.
However, in this paper, we assume the cyclicity of $M_n$. 

\section{Correlation functions from quantum $A_{\infty}$ algebras} \label{correlation function}
\setcounter{equation}{0}
In this section, we describe correlation functions in terms of quantum $A_{\infty}$ algebras. In \S \ref{scalar}, we first review the case for scalar field theories following \cite{Okawa:2022sjf,Konosu:2023pal,Konosu:2023rkm}.\footnote{As mentioned in the introduction, there are several approaches to describe correlation functions. In this paper, we take the approach first described in \cite{Okawa:2022sjf} We mostly follow the convention in \cite{Konosu:2023pal,Konosu:2023rkm}.}. In \S \ref{cor-trivial}, we show an important identity, the \textit{algebraic Schwinger-Dyson equation}, which contains the Schwinger-Dyson equations. This identity plays a key role in proving the algebraic version of the Ward-Takahashi identities. In \S\ref{equiv-SD}, we confirm that we can derive the Schwinger-Dyson equation from the algebraic Schwinger-Dyson equation. 
\setcounter{equation}{0}
\subsection{Correlation functions 
for scalar field theories}\label{scalar}
In this subsection, we give a brief review on correlation functions for scalar field theories in terms of quantum $A_{\infty}$ algebras following \cite{Konosu:2023pal,Konosu:2023rkm}. As in \S\ref{A-infty-review}, we consider the graded vector space
\begin{equation}
    \mathcal{H}=\mathcal{H}_{1}\oplus\mathcal{H}_{2}\,.
\end{equation}
When we consider theories without gauge symmetry, we only need basis vectors corresponding to the spacetime ghost number 0 sector and the counterparts. In scalar field theory, we denote the degree-even basis vector of $\mathcal{H}_{1}$ by $c(x)$ and the degree-odd basis vector of $\mathcal{H}_{2}$ by $d(x)$. 

In terms of homotopy algebra, we denote the action by
\begin{equation}
    S = {}-\frac{1}{2} \, \omega \, ( \, \Phi, Q \, \Phi \, )
        -\sum_{n=0}^\infty \, \frac{1}{n+1} \,
        \omega \, ( \, \Phi \,, m_n \, ( \, \Phi,\,\ldots\,, \Phi \, ) \, ) \,,
\end{equation}
where $\Phi$ is the degree-even element of $\mathcal{H}_{1}$. The element $\Phi$ is expanded as
\begin{equation}
    \Phi = \int \dd[d] x \, \varphi (x) \, c(x)\,,
\end{equation}
where  $\varphi(x)$ is the real scalar field.  We define the symplectic form $\omega$ by
\begin{equation}
\biggl(
\begin{array}{cc}
\omega \, ( \, c(x_1) \,, c(x_2) \, ) & \omega \, ( \, c(x_1) \,, d(x_2) \, ) \\
\omega \, ( \, d(x_1) \,, c(x_2) \, ) & \omega \, ( \, d(x_1) \,, d(x_2) \, )
\end{array}
\biggr)
= \biggl(
\begin{array}{cc}
0 & \delta^d ( x_1-x_2 ) \\
{}-\delta^d ( x_1-x_2 ) & 0
\end{array}
\biggr) \,.
\end{equation}
We define the degree-odd operator $Q$ by
\begin{equation}
Q \, c(x) = ( {}-\partial^2 +m^2 \, ) \, d(x) \,, \qquad
Q \, d(x) = 0 \,, \label{def-s-q}
\end{equation}
where $m$ is the mass of the scalar field. From these definitions, the kinetic term is reproduced in terms of homotopy algebras:
\begin{equation}
    -\frac{1}{2} \, \omega \, ( \, \Phi, Q \, \Phi \, )={}-\frac{1}{2} \int \dd[d] x \, \bigl[ \, \partial_\mu \varphi (x) \, \partial^\mu \varphi (x)
  +m^2 \, \varphi (x)^2 \, \bigr]\,.
\end{equation}
Note that $Q$ is niloptent:
\begin{equation}
    Q^2=0\,.
\end{equation}
Operators $m_{n}$ are degree-odd map from $\mathcal{H}^{\otimes n}$ to $\mathcal{H}$. For example, the cubic interaction in $\varphi^{3}$ theory is described by
\begin{equation}
    m_{2} \, ( \, c \, (x_{1}),\, c \, (x_{2}) \, )
= {}-\frac{g}{2} \, \delta^d (x_{1}-x_{2}) \, d \, (x_{1})\,,
\end{equation}
where $g$ is the coupling constant. We define $m_{2} \, ( \, c \, (x_{1}) ,\, d \, (x_{2}) \, ),\,m_{2} \, ( \, d \, (x_{1}) ,\, c \, (x_{2}) \, )$ and\\ $m_{2} \, ( \, d \, (x_{1}) ,\, d \, (x_{2}) \, )$ to be zero. In theories without gauge symmetry, operators $m_{n}$ are defined to be nonzero only when $m_{n}$ map from $\mathcal{H}_{1}^{\otimes n}$ to $\mathcal{H}_{2}$. 

Associated to the symplectic form $\omega$, we can define the operator ${\vb U}$ by
\begin{equation}
{\vb U} = \int \dd[d] x \, {\vb c} (x) \, {\vb d} (x) \,,
\end{equation}
where ${\vb c} (x)$ and ${\vb d}(x)$ are coderivations given by
\begin{equation}
    \begin{split}
        &\pi_1 \, {\vb c} (x) \, {\bf 1} = c(x) \,, \quad
        \pi_1 \, {\vb c}(x) \, \pi_n = 0 \,, \quad\\
        &\pi_1 \, {\vb d} (x) \, {\bf 1} = d(x) \,, \quad
        \pi_1 \, {\vb d} (x) \, \pi_n = 0
    \end{split}
\end{equation}
for $n > 0 \,$. Then, the main identity
\begin{equation}
    \pi_{1}\,\qty(\vb{Q}+\sum_{n}\vb{m}_{n}+i\hbar\vb{U})^2=0
\end{equation}
holds trivially, where $\vb{Q}$ and $\vb{m}_{n}$ are coderivations associates with $Q$ and $m_{n}$, respectively.
Since we can easily confirm that the operators $\vb{Q}$ and  $\vb{m}_{n}$ meet the cyclicity~\eqref{cyclicity-Q} and~\eqref{cyclic}, these operators satisfy quantum $A_{\infty}$ relations.

Let us consider the description of correlation functions. To extract the physical information from homotopy algebras, the projection operator $P$ onto the subspace of $\mathcal{H}$ plays the key role. As in \cite{Okawa:2022sjf}, we take $P$ to be
\begin{equation}
    P=0 \label{projection-cor}
\end{equation}
to calculate correlation functions. Intuitively, the operator $\mathbb{I}-P$ corresponds to the region where we carry out the path integral and the definition~\eqref{projection-cor} means that we carry out the full path integration. 
In the coalgebra representation, the operator $\vb{P}$ associated with $P$ is given by 
\begin{equation}
    \vb{P}=\pi_{0}\,. \label{zero-projection}
\end{equation}
Associated to the projector $P$, we define the \textit{contracting homotopy} $h$ by
\begin{equation}
Q \, h +h \, Q = \mathbb{I} -P \,, \qquad
h \, P = 0 \,, \qquad
P \, h = 0 \,, \qquad
h^2 = 0 \,. \label{Hodge-Kodaira}
\end{equation}
In addition, we impose
\begin{align}
    \omega\,(\,F_1,h\,F_2\,)=(-1)^{F_1}\omega\,(\,h\,F_1\,,F_2\,)\,.\label{anticyclic-h}
\end{align}
In the current case, the second and third definitions in~\eqref{Hodge-Kodaira} are trivially  satisfied due to~\eqref{projection-cor}. Then, we choose
\begin{equation}
h \, c(x) = 0 \,, \qquad
h \, d(x) = \int \dd[d] y \, \Delta (x-y) \, c (y) \,,
\label{scalar-h}
\end{equation}
where $\Delta(x-y)$ is the Feynman propagator of scalar field theories given by
\begin{equation}
\Delta (x-y)
= \int \frac{\dd[d] k}{(2 \pi)^d} \,
\frac{e^{ik \, (x-y)}}{k^2+m^2-i \epsilon}\,.
\end{equation}
Associated with the contracting homotopy $h$, we define the operator $\vb{h}$ in the coalgebra representation  by\footnote{In general, the projector $P$ is involved with the action of $\vb{h}$ on $T\mathcal{H}$. Since the projection operator is defined by~\eqref{zero-projection}, terms that contain $P$ are trivially dropped.}
\begin{equation}
    \vb{h} = h \, \pi_1
    +\sum_{n=2}^\infty ( \, \mathbb{I}^{\otimes (n-1)} \otimes h \, ) \, \pi_n \,.\label{coalgebra-h}
\end{equation}
The definitions~\eqref{Hodge-Kodaira} can be extended in the coalgebra representations by
\begin{equation}
\vb{Q} \, \vb{h} +\vb{h} \, \vb{Q} = \vb{I}-\vb{P} \,, \quad
\vb{h} \, \vb{P} = 0 \,, \quad
\vb{P} \, \vb{h} = 0 \,, \quad
\vb{h}^2 = 0 \,. \label{Hodge-Kodaira-coalgebra}
\end{equation}
Then, we can describe the formula for correlation functions in terms of quantum $A_{\infty}$ algebras by operators defined above. The formula is given by 
\begin{equation}
\langle \, \Phi^{\otimes n} \, \rangle = \pi_n \, \vb{f} \, {\bf 1}\,,
\label{scalar-correlation-functions}
\end{equation}
where
\begin{equation}
\Phi^{\otimes n} = \underbrace{\, \Phi \otimes \Phi \otimes \ldots \otimes \Phi \,}_n
\end{equation}
and
\begin{equation}
    \vb{f} = \frac{1}{\vb{I} +\vb{h} \, \vb{m} +i \hbar \, \vb{h} \, \vb{U}} \,.
\end{equation}
The inverse of $\vb{I} +\vb{h} \, \vb{m} +i \hbar \, \vb{h} \, \vb{U}$ is defined by the formal power series:
\begin{equation}
\frac{1}{\vb{I} +\vb{h} \, \vb{m} +i \hbar \, \vb{h} \, \vb{U}}
= {\bf I} +\sum_{n=1}^\infty \, (-1)^n \,
( \, \vb{h} \, {\bm m} +i \hbar \, \vb{h} \, \vb{U} \, )^n \,.
\end{equation}
Intuitively, operators $\pi_{n}$, $\vb{h}$, $\vb{m}$, and $\vb{U}$ correspond to the number of external lines, propagators, vertices, and loops, respectively in Feynman diagrams. In the formula~\eqref{scalar-correlation-functions}, the usual correlation functions in  quantum field theory appear as the coefficients of basis vectors:
\begin{equation}
\begin{split}
\langle \, \Phi^{\otimes n} \, \rangle
& = \langle \, \underbrace{\, \Phi \otimes \Phi \otimes \ldots \otimes \Phi \,}_n \, \rangle \\
& = \int \dd[d] x_1 \dd[d] x_2 \ldots \dd[d] x_n \,
\langle \, \varphi (x_1) \, \varphi (x_2) \, \ldots \, \varphi (x_n) \, \rangle \,
c (x_1) \otimes c(x_2) \otimes \ldots \otimes c(x_n) \,.
\end{split}
\end{equation}
We define\footnote{Notice that we use slightly different notations from the papers~\cite{Okawa:2022sjf},~\cite{Konosu:2023pal} and~\cite{Konosu:2023rkm}. These two notations are essentially the same.}
\begin{equation}
    \bra{\,\omega_{n}(x_{1},\,\ldots,x_{n})}=(-1)^{n}\bra{\omega}(\,d(x_{1})\otimes\mathbb{I}\,)\cdots\bra{\omega}(\,d(x_{n})\otimes\mathbb{I}\,)\,. \label{scalar-correlation-omega-refined}
\end{equation}
Then, we can extract ordinary correlation functions in quantum field theory by
\begin{equation}
    \langle\,\varphi(x_{1})\,\cdots\,\varphi(x_{n})\,\rangle = \bra{\,\omega_{n}(x_{1},\,\ldots,x_{n})}\pi_{n}\,\vb{f}\,\vb{1}\,. \label{scalar-correlation-function}
\end{equation}
This is verified by
\begin{equation}
    \varphi(x_{1})\,\ldots\,\varphi(x_{n})=\bra{\,\omega_{n}(x_{1},\,\ldots,x_{n})}\frac{1}{1-\Phi}\,,
\end{equation}
and
\begin{equation}
    \Phi^{\otimes n}=\pi_{n}\,\frac{1}{1-\Phi}\,.
\end{equation}
Another important property of the formula~\eqref{scalar-correlation-functions} is that the quantity $\pi_{n}\,\vb{f}\,\vb{1}$ is symmetric under the permutations of tensor products. This is confirmed indirectly through~\eqref{scalar-correlation-functions}. 

As an example, we calculate the two-point function of the free scalar theory. From the formula~\eqref{scalar-correlation-function}, the two-point function for the free scalar theory is given by
\begin{equation}
    \langle \, \varphi (x_1) \, \varphi (x_2) \rangle
        = \bra{\omega_{2}(x_{1},x_{2})}\pi_2 \, \vb{f} \, \vb{1} \,, \label{scalar-2pt}
\end{equation}
where
\begin{equation}
    \vb{f} = \frac{1}{\vb{I} +i \hbar \, \vb{h} \, \vb{U}} \,.
\end{equation}
In fact, $\, \pi_{2} \, \vb{f} \, \vb{1} \,$ for the free theory can be simplified as
\begin{equation}
\pi_{2} \, \vb{f} \, {\bf 1} \,
= {}-i \hbar \, \pi_{2} \, \vb{h} \, \vb{U} \, \vb{1} \,.
\end{equation}
Since $\vb{U} \, \vb{1} $ is given by
\begin{equation}
    \vb{U} \, \vb{1} = \int d^d x \, ( \, c(x) \otimes d(x) +d(x) \otimes c(x) \, )\,, 
\end{equation}
and the contracting homotopy $\vb{h}$ acts on $\mathcal{H}^{\otimes 2}$ as
\begin{equation}
    \vb{h} \, \pi_2 = ( \, \mathbb{I} \otimes h \, ) \, \pi_2 \,,
\end{equation}
we obtain
\begin{equation}
    \vb{h} \, \vb{U} \, \vb{1} = \int \dd[d] x \,\left[\, c(x) \otimes h \, d(x)+ d(x) \otimes h \, c(x)\,\right]\,.
\end{equation}
Since the contacting homotopy $h$ act on the basis vector $c(x)$ as $h\,c(x)=0$,
we obtain
\begin{equation}
    \vb{h} \, \vb{U} \, \vb{1} = \int \dd[d] x \,\left[\, c(x) \otimes h \, d(x)\,\right]\,.
\end{equation}
Therefore, we obtain
\begin{equation}
\begin{split}
    \pi_{2} \, \vb{f} \, {\bf 1}
    & = {}-i \hbar \, \pi_{2} \, \vb{h} \, \vb{U} \, \vb{1} \\
    & = {}-i \hbar \int \dd[d] x \int \dd[d] y \, [ \, c(x) \otimes \Delta(x-y) \, c(y) \, ] \,,
\end{split}
\end{equation}
When we substitute the above results into~\eqref{scalar-2pt}, we reproduce the result of ordinary quantum field theory:
\begin{equation}
    \begin{split}
    \bra{\omega_2}   \pi_2 \, \vb{f} \, \vb{1} 
        &=\frac{\hbar}{i}\,\Delta\,(x_1\,-\,x_2\,)\\
         &= \langle \, \varphi (x_1) \, \varphi (x_2) \rangle\,.
       \end{split}
\end{equation}

In general, the formula~\eqref{scalar-correlation-functions} is verified by showing that correlation functions from~\eqref{scalar-correlation-functions} satisfy the Schwinger-Dyson equations. In~\cite{Okawa:2022sjf} and the following papers, it follows from the trivial identity
\begin{equation}
    \vb{f}^{-1}\,\vb{f}\,\vb{1}=\vb{1}\,.\label{okawa-SD}
\end{equation}
We present this proof in the Appendix~\ref{old-SD}.
In this paper, we prove this fact using the \textit{algebraic Schwinger-Dyson equations} in the next subsection.

\subsection{Correlation functions and the algebraic Schwinger-Dyson equation}\label{cor-trivial}

Let us extend the above discussions of correlation functions for scalar field theory to general gauge-fixed theories. 
For a gauge-fixed theory, the kinetic term of the action~\eqref{homotopy-action} is nondegenerate, so it is possible to construct $h$ such that it makes $P$ in~\eqref{Hodge-Kodaira} equal to zero using the propagator.
At this time, the vector space $\mathcal{H}$ can be decomposed into two subspaces projected by $h\,Q$ and $Q\,h$, with the former being $\mathcal{H}_1$ and the latter being $\mathcal{H}_2$:
\begin{equation}
\begin{split}
    &h\,Q\,e_a(x)=e_a(x)\,,\quad Q\,h\,e_a(x)=0\,,\\
    &h\,Q\,\widetilde{e}^{\,a}(x)=0\,,\quad Q\,h\,\widetilde{e}^{\,a}(x)=\widetilde{e}^{\,a}(x)\,.
    \end{split}
\end{equation}
Notice that we need to add ghost fields in the gauge-fixed theory. In the homotopy algebraic point of view, we need to add an additional basis associated with the ghost. Then, the basis $\{e_{a}(x)\}$ contains both degree-even and degree-odd basis vectors in general.
Since the element $\Phi$ in the action is in $\mathcal{H}_1$, this implies  
\begin{align}
    h\,\Phi=0\,.
\end{align}
This corresponds to choosing the gauge fixing condition~\cite{Kajiura:2001ng,Kajiura:2003ax,Erbin:2020eyc}.

As in the scalar field example above, consider the quantity 
\begin{align}
    \vb{f}=\frac{1}{\vb{I}+\vb{h\,m}+i\hbar\,\vb{h\,U}}\,,
\end{align}
under $P=0$.
Here, the second-order coderivation $\vb{U}$ is defined in~\eqref{defU}.
As in the example above, we claim that correlation functions in terms of homotopy algebras for the gauge-fixed field theory are given by
\begin{align}
   \expval{\,\phi^{a_1}(x_1)\phi^{a_2}(x_2)\dots\phi^{a_n}(x_n)\,}= \bra{\omega_n^{a_1,\dots,a_n}(x_1,\dots x_n)}\,\pi_n\,\vb{f\,1}\,,\label{homotopyexpval}
\end{align}
where the operator $\bra{\omega_n^{a_1,\dots,a_n}(x_1,\dots x_n)}$ is defined by
\begin{align}
    \bra{\omega_n^{a_1,\dots,a_n}(x_1,\dots x_n)}=(-1)^n\bra{\omega}(\widetilde{e}^{\,a_1}(x_1)\otimes\pi_1)\otimes\cdots\otimes\bra{\omega}(\widetilde{e}^{\,a_n}(x_n)\otimes\pi_1)\,.\label{def omega_n}
\end{align}
In the same way as scalar field theory, the formula can also be written as
\begin{align}
    \expval{\Phi^{\otimes n}}=\pi_n\,\vb{f\,1}\,.\label{pif1}
\end{align}
The quantity $\vb{f\,1}$ satisfies the following important relation:
\begin{align}
    (\,\vb{M}+i\hbar\,\vb{U}\,)\,\vb{f\,1}=(\,\vb{Q}+\vb{m}+i\hbar\,\vb{U}\,)\,\vb{f\,1}=0\ .\label{alg schwinger}
\end{align}
We call this identity the \textit{algebraic Schwinger-Dyson equation} because we can derive the well-known Schwinger-Dyson equations from this identity. The proof is given in the next subsection. In~\cite{Okawa:2022sjf} and the following papers, the Schwinger-Dyson equations are obtained from~\eqref{okawa-SD}. This derivation works well, but as it can be seen in the proof described in Appendix~\ref{old-SD}, we need to act the concrete kinetic operators on~\eqref{okawa-SD} according to the model. In our case, we do not need such an operation and it is written by using only operators in homotopy algebras.
Thus, it is more suitable to carry out the proof in general theories. Let us prove~\eqref{alg schwinger}:
\begin{equation}
    \begin{split}
    \vb{Q\,f\,1}&=[\,\vb{Q},\,\vb{f}\,]\,\vb{1}\\
    &=\left[\,\vb{Q},\,\frac{1}{\vb{I}+\vb{h}\,(\vb{m}+i\hbar\,\vb{U})}\,\right]\,\vb{1}\\
    &=-\frac{1}{\vb{I}+\vb{h}\,(\vb{m}+i\hbar\,\vb{U})}\,\left[\,\vb{Q},\vb{h}\,(\vb{m}+i\hbar\,\vb{U})\,\right]\,\frac{1}{\vb{I}+\vb{h}\,(\vb{m}+i\hbar\,\vb{U})}\,\vb{1}\\
    &=-\vb{f}\,(\,\vb{m}+i\hbar\vb{U}\,)\,\vb{f}\,\vb{1}
    +\vb{f}\,\vb{h}\,\left[\,\vb{Q},(\vb{m}+i\hbar\,\vb{U})\,\right]\,\vb{f\,1}\\
    &=-\vb{f}\,(\,\vb{m}+i\hbar\vb{U}\,)\,\vb{f}\,\vb{1}-\vb{f}\,\vb{h}\,(\,\vb{m}+i\hbar\,\vb{U}\,)^2\,\vb{f}\,\vb{1}\\
    &=-\vb{f}\big[\,\vb{I}+\vb{h}\,(\,\vb{m}+i\hbar\,\vb{U}\,)\,\big]\,(\,\vb{m}+i\hbar\vb{U}\,)\,\vb{f\,1}\\
    &=-(\vb{m}+i\hbar\vb{U})\,\vb{f\,1}\ .
    \end{split}
\end{equation}
We used the quantum $A_\infty$ relation~\eqref{qA-infty2} at the fifth equality.

As it can be inferred from \eqref{pif1}, the quantity $\vb{f\,1}$ is an element of $T\mathcal{H}$ that has properties similar to the group-like element $\frac{1}{1-\Phi}$.
In fact, it vanishes when $\vb{h}$ acts on it:
\begin{align}
    \vb{h\,f\,1}=0\,.
\end{align}
In other words, each output is an element of $\mathcal{H}_1$.
Furthermore, it is invariant under the permutation of elements in $\mathcal{H}$:
\begin{align}
    T^{(\sigma)}\,\vb{f\,1}=\vb{f\,1}\,.\label{perm-f}
\end{align}
Previously, this property was not proven by the direct computation of $\vb{f\,1}$, however, this can be verified at each order by using the cyclicity~\eqref{cyclic another}  of $m_n$ and the formula
\begin{align}
    \sum_a\int\dd^dx\,e_a(x)\otimes h\,\widetilde{e}^{\,a}(x)=\sum_a\int\dd^dx\,(-1)^{e_a}\,h\,\widetilde{e}^{\,a}(x)\otimes e_a(x)\,,
\end{align}
which is derived as
\begin{equation}
    \begin{split}
        \sum_a\,\int\dd^dx\,e_a(x)\otimes h\,\widetilde{e}^{\,a}(x)&=\sum_{a,b}\,\int\dd^dx\,\dd^dy\,(-1)^{\widetilde{e}^{\,b}}\,e_a(x)\otimes e_b(y)\otimes \omega\big(\,\widetilde{e}^{\,b}(y),h\,\widetilde{e}^{\,a}(x)\,\big)\\
        &=-\sum_{a,b}\,\int\dd^dx\,\dd^dy\,e_a(x)\otimes e_b(y)\,\omega\big(\,\widetilde{e}^{\,a}(x),h\,\widetilde{e}^{\,b}(y)\,\big)\\
        &=\sum_b\,\int\dd^dy\,(-1)^{e_b}\,h\,\widetilde{e}^{\,b}(y)\otimes e_b(y)\,,
    \end{split}
\end{equation}
where we used the normalization~\eqref{nomalization of U} and the property~\eqref{anticyclic-h} of $h$.
For instance, the $\hbar^2$ order of $\pi_2\,\vb{f\,1}$ is
\begin{equation}
    \begin{split}
\hbar^2\sum_{a,b}\int\dd^dx\,\dd^dy\, \big[&(-1)^{e_a+e_b}\, e_b(y)\otimes\,hm_3\big(\,e_a(x),\,h\,\widetilde{e}^{\,a}(x),\,h\,\widetilde{e}^{\,b}(y)\,\big)\\
+&(-1)^{e_a+e_b+e_ae_b}\, e_a(x)\otimes\, hm_3\big(\,e_b(y),h\,\widetilde{e}^a(x),h\,\widetilde{e}^{\,b}(y)\,\big)\\
+&(-1)^{e_a+e_b}\, e_a(x)\otimes\,hm_3\big(\,h\,\widetilde{e}^a(x),e_b(y),h\,\widetilde{e}^b(y)\,\big)\big]\,.
   \end{split}
\end{equation}
Due to the cyclicity of $m_n$ and the property~\eqref{anticyclic-h} of $h$, we get
\begin{equation}
    \begin{split}
    \hbar^2\,\sum_{a,b}\,\int\dd^dx\,\dd^dy\, \big[&\,(-1)^{e_b}\, hm_3\,\big(\,h\,\widetilde{e}^{\,a}(x),\,e_b(y),\,h\,\widetilde{e}^{\,b}(y)\big)\otimes\,e_a(x)\\
    +&(-1)^{e_b+e_ae_b}\, hm_3\big(e_b(y),\,h\,\widetilde{e}^{\,a}(x),h\widetilde{e}^{\,b}(y)\big)\otimes\, e_a(x)\\
    +&(-1)^{e_a}\, hm_3\,\big(\,e_a(x),\,h\,\widetilde{e}^{\,a}(x),\,h\widetilde{e}^{\,b}(y)\,\big)\otimes e_b(y)
\big]\,.
    \end{split}
\end{equation}
This is just a form where the output has been permuted. We can verify~\eqref{perm-f} in each order as well in principle.
On the other hand, the action of the coproduct $\triangle$ is different:
\begin{align}
    \triangle\,\vb{f\,1}\neq\vb{f\,1}\otimes'\vb{f\,1}\,.
\end{align}
This corresponds to the fact that the correlation function is generally not factorized.

\subsection{The algebraic derivation of the Schwinger-Dyson equations}\label{equiv-SD}
Let us show that the Schwinger-Dyson equations can be derived from the algebraic Schwinger-Dyson equation~\eqref{alg schwinger}.
In the path integral formalism, correlation functions are defined by
\begin{equation}
    \,\langle\,\phi^{a_{1}}(x_{1})\phi^{a_{2}}(x_{2})\ldots\phi^{a_{n-1}}(x_{n-1})\,\rangle=\frac{1}{Z}\int\mathcal{D}\phi\,\phi^{a_{1}}(x_{1})\phi^{a_{2}}(x_{2})\ldots\phi^{a_{n-1}}(x_{n-1})e^{\frac{i}{\hbar}S}\,,
\end{equation}
where
\begin{equation}
    Z=\int\mathcal{D}\phi\,e^{\frac{i}{\hbar}S}\,.
\end{equation}
Since
\begin{equation}
    \frac{1}{Z}\int\mathcal{D}\phi\,\frac{\delta}{\phi^{a}(x_{n})}\qty[\,\phi^{a_{1}}(x_{1})\phi^{a_{2}}(x_{2})\ldots\phi^{a_{n-1}}(x_{n-1})e^{\frac{i}{\hbar}S}\,]=0\,,
\end{equation}
the Schwinger-Dyson equations are obtained:
\begin{equation}
    \begin{split}
        &\sum_{i=1}^{n-1}\,(-1)^{\phi^{a}\sum_{j=1}^{i}\phi^{a_{j-1}}}\langle\,\phi^{a_{1}}(x_{1})\ldots\phi^{a_{i-1}}(x_{i-1})\delta_a^{a_{i}}\delta^{d}(x_{i}-x_{n})\phi^{a_{i+1}}(x_{i+1})\ldots\phi^{a_{n}}(x_{n})\,\rangle\\
        &+\frac{i}{\hbar}\,(-1)^{\phi^{a}\sum_{j=1}^{n-1}\phi^{a_{j}}}\left\langle\,\phi^{a_{1}}(x_{1})\ldots\phi^{a_{n-1}}(x_{n-1})\frac{\delta S}{\delta \phi^{a}(x_{n})}\,\right\rangle\,=0\,,\label{Schwinger-Dyson}
    \end{split}
\end{equation}
where we introduced auxiliary parameter $\phi^{a_{0}}=0$.

To show the equivalence between~\eqref{alg schwinger} and~\eqref{Schwinger-Dyson}, let us introduce the operator
\begin{equation}
    \pi_{n}(\,\mathbb{I}^{\otimes(n-1)}\otimes\,P_{\,\widetilde{e}^{a}}\,)\,,
\end{equation}
where $P_{\,\widetilde{e}^{\,a}}$ is the projection onto the subspace of $\mathcal{H}_{2}$ spanned by $\widetilde{e}^{\,a}$ for fixed label $a$. For example, in scalar field theory described in \S\ref{scalar}, $P_{\,d}$ project onto the space spanned by $d(x)$.
Acting the above operator on the algebraic Schwinger-Dyson equation~\eqref{alg schwinger}, we obtain
\begin{equation}
    \pi_{n}(\,\mathbb{I}^{\otimes(n-1)}\otimes\,P_{\,\widetilde{e}^{a}}\,)\,\vb{M}\,\vb{f\,1}+i\hbar\,\pi_{n}(\,\mathbb{I}^{\otimes(n-1)}\otimes\,P_{\,\widetilde{e}^{a}}\,)\,\vb{U}\,\pi_{n-2}\,\vb{f\,1}=0\,.
\end{equation}
Since
\begin{equation}
    \pi_{n}\,\vb{f}\,{\vb 1}=\langle\,\Phi^{\otimes n}\,\rangle\,,
\end{equation}
we can derive
\begin{equation}
    \sum_{k=0}^{\infty}\,\langle\,\Phi^{\otimes(n-1)}\otimes\,P_{\,\widetilde{e}^{a}}\,M_{k}(\Phi\otimes\ldots\otimes\Phi)\,\rangle+i\hbar\,\pi_{n}(\,\mathbb{I}^{\otimes(n-1)}\otimes\,P_{\,\widetilde{e}^{a}}\,)\,\vb{U}\,\langle\Phi^{\otimes(n-2)}\,\rangle=0\,.\label{alg-schwinger2}
\end{equation}
Let us consider the first term on the left-hand side of~\eqref{alg-schwinger2}.
From the relation~\eqref{nomalization of U}, we obtain
\begin{equation}
    M(\Phi\otimes\ldots\otimes\Phi)={}-\sum_a\int \dd[d]x\, \widetilde{e}^{\,a}(x)\frac{\delta S}{\delta \phi^{\,a}(x)}\,.
\end{equation}
Then, the first term of~\eqref{alg-schwinger2} is written as
\begin{equation}
    \sum_{k=0}^{\infty}\,\langle\,\Phi^{\otimes(n-1)}\otimes\,P_{\,\widetilde{e}^{\,a}}\,M_{k}(\Phi\otimes\ldots\otimes\Phi)\,\rangle={}-\left\langle\,\Phi^{\otimes(n-1)}\otimes\int\dd[d]y\,\widetilde{e}^{\,a}(y)\,\frac{\delta S}{\delta \phi^{\,a}(y)}\,\right\rangle\,.
\end{equation}
Notice that the superscript $a$ is not summed over all the fields, but corresponds to the field labeled by $a$.
Let us next consider the second term of~\eqref{alg-schwinger2}. We obtain 
\begin{equation}
    \begin{split}
          &i\hbar\,\pi_{n}(\,\mathbb{I}^{\otimes(n-1)}\otimes\,P_{\,\widetilde{e}}\,)\,\vb{U}\,\langle\Phi^{\otimes(n-2)}\,\rangle\\
          &\quad=i\hbar\,\sum_{i=1}^{n-1}\,\int\dd[d]y\,(\,\mathbb{I}^{\otimes (i-1)}\otimes e_{a}(y)\otimes\mathbb{I}^{\otimes (n-i-1)}\otimes\widetilde{e}^{\,a}(y)\,)\,\pi_{n-2}\,\vb{f}\,\vb{1}\\
          &\quad=i\hbar\,\sum_{i=1}^{n-1}\,\int\dd[d]y\,\langle\,\Phi^{\otimes (i-1)}\otimes e_{a}(y)\otimes\Phi^{\otimes (n-i-1)}\otimes\widetilde{e}^{\,a}(y)\,\rangle\,. \label{gen-sd-temp}
    \end{split}
\end{equation}
Let us comment on the last line.  
Since
\begin{equation}
    \langle\,\Phi^{\otimes (n-2)}\,\rangle=(-1)^{\sigma}\int\dd[d]y_{1}\ldots\dd[d]y_{n-2}\langle\,\phi^{a_{1}}(y_{1})\,\ldots\,\phi^{a_{n-2}}(y_{n-2})\,\rangle (\,e_{a_{1}}(y_{1})\otimes\ldots\otimes e_{a_{n-2}}(y_{n-2})\,)\,,
\end{equation}
where $(-1)^{\sigma}$ is the appropriate sign factor, the term
\begin{equation}
    (\,\mathbb{I}^{\otimes (i-1)}\otimes e_{a}(y)\otimes\mathbb{I}^{\otimes (n-i-1)}\otimes\widetilde{e}^{\,a}(y)\,)\,\pi_{n-2}\,\vb{f}\,\vb{1}=(\,\mathbb{I}^{\otimes (i-1)}\otimes e_{a}(y)\otimes\mathbb{I}^{\otimes (n-i-1)}\otimes\widetilde{e}^{\,a}(y)\,)\,\langle\,\Phi^{\otimes (n-2)}\,\rangle
\end{equation}
should be understood as
\begin{equation}
    \begin{split}
        &\int\dd[d]y_{1}\ldots\dd[d]y_{n-2}(-1)^{\sigma+ {e}_{a}( e_{a_i}+\ldots e_{a_{n-2}})}\langle\,\phi^{a_{1}}(y_{1})\,\ldots\,\phi^{a_{n-2}}(y_{n-2})\,\rangle\\
        &\qquad\qquad\qquad\times(\,e_{a_{1}}(y_{1})\otimes\ldots\otimes e_{a_{i-1}}(y_{i-1})\otimes e_{a}(y)\otimes e_{a_{i}}(y_{i})\otimes\ldots\otimes\widetilde{e}^{\,a}(y)\,)\,.\label{comp-sd-temp1}
    \end{split}
\end{equation}
In the last line of~\eqref{gen-sd-temp}, we carry out the formal calculation:
\begin{equation}
    \begin{split}
        &(\,\mathbb{I}^{\otimes (i-1)}\otimes e_{a}(y)\otimes\mathbb{I}^{\otimes (n-i-1)}\otimes\widetilde{e}^{\,a}(y)\,)\,\langle\,\Phi^{\otimes (n-2)}\,\rangle\\
        &\qquad=\langle\,(\,\mathbb{I}^{\otimes (i-1)}\otimes e_{a}(y)\otimes\mathbb{I}^{\otimes (n-i-1)}\otimes\widetilde{e}^{\,a}(y)\,)\,\Phi^{\otimes (n-2)}\,\rangle\\
        &\qquad=\langle\,(\,\Phi^{\otimes (i-1)}\otimes e_{a}(y)\otimes\Phi^{\otimes (n-i-1)}\otimes\widetilde{e}^{\,a}(y)\,)\,\rangle\,.\label{comp-sd-temp2}
    \end{split}
\end{equation}
In fact, we find that the calculation in~\eqref{comp-sd-temp1} coincides with that of~\eqref{comp-sd-temp2}.

Then, the equation~\eqref{alg-schwinger2} becomes
\begin{equation}
    -\left\langle\,\Phi^{\otimes(n-1)}\otimes\int\dd[d]y\,\widetilde{e}^{\,a}(y)\,\frac{\delta S}{\delta \phi^{\,a}}\,\right\rangle+i\hbar\,\sum_{i=1}^{n-1}\,\int\dd[d]y\,\langle\,\Phi^{\otimes (i-1)}\otimes e_{a}(y)\otimes\Phi^{\otimes (n-i-1)}\otimes\widetilde{e}^{\,a}(y)\,\rangle=0\,.\label{alg-schwinger3}
\end{equation}
To extract ordinary correlation functions, we use
\begin{equation}
    \bra{\omega_n^{\,a_1,\dots,a_n}(x_1,\dots x_n)}=(-1)^n\bra{\omega}(\widetilde{e}^{\,a_1}(x_1)\otimes\pi_1)\otimes\cdots\otimes\bra{\omega}(\widetilde{e}^{\,a_n}(x_n)\otimes\pi_1)\,.
\end{equation}
defined in~\eqref{def omega_n}.
Let us act the operator
\begin{equation}
    \bra{\omega_n^{\,a_1,\dots,a_{n-1}}(x_1,\dots x_{n-1})}\otimes\bra{\omega}(\,e_{a}(x_{n})\otimes\mathbb{I}\,)\label{alg-schwinger-op}
\end{equation}
from the left on the both sides in~\eqref{alg-schwinger3}.
After the action of the operator~\eqref{alg-schwinger-op} on the first term of the left-hand side, we obtain
\begin{equation}
    \begin{split}
        &\bra{\omega_n^{\,a_1,\dots,a_{n-1}}(x_1,\dots x_{n-1})}\otimes\bra{\omega}(\,e_{a}(x_{n})\otimes\mathbb{I}\,)\,\left[-\left\langle\,\Phi^{\otimes(n-1)}\otimes\int\dd[d]y\,\widetilde{e}^{\,a}(y)\,\frac{\delta S}{\delta \phi^{\,a}(y)}\,\right\rangle\right]\\
        &={}-\left\langle\,\phi^{a_{1}}(x_{1})\ldots\phi^{a_{n-1}}(x_{n-1})\int\dd[d]y\bra{\omega}\big(\,e_{a}(x_{n})\otimes\widetilde{e}^{\,a}(y)\,\frac{\delta S}{\delta \phi^{\,a}(y)}\big)\,\right\rangle\\
        &=(-1)^{e^{a}+1}\left\langle\,\phi^{a_{1}}(x_{1})\ldots\phi^{a_{n-1}}(x_{n-1})\frac{\delta S}{\delta \phi^{\,a}(x_{n})}\,\right\rangle\,.
    \end{split}
\end{equation}
Let us next consider the second term of the left-hand side of~\eqref{alg-schwinger3} after the action of the operator~\eqref{alg-schwinger-op}. We obtain
\begin{equation}
    \begin{split}
        &\bra{\omega_n^{\,a_1,\dots,a_{n-1}}(x_1,\dots x_{n-1})}\otimes\bra{\omega}(e_{a}(x_{n})\otimes\mathbb{I})\\
        &\qquad\left[i\hbar\sum_{i=1}^{n-1}\int\dd[d]y\,\langle\,\Phi^{\otimes (i-1)}\otimes e_{a}(y)\otimes\Phi^{\otimes (n-i-1)}\otimes\widetilde{e}^{\,a}(y)\rangle\right]\\
        &=i\hbar\,\sum_{i=1}^{n-1}\int\dd[d]y\,\bra{\omega_n^{\,a_1,\dots,a_{n-1}}(x_1,\dots x_{n-1})}\langle\,\Phi^{\otimes (i-1)}\otimes e_{a}(y)\otimes\Phi^{\otimes(n-i-1)}\otimes\langle\,\omega|(\,e_{a}(x_{n})\otimes \widetilde{e}^{\,a}(y)\,)\rangle\\
        &=(-1)^{e^{a}}\,i\hbar\,\sum_{i=1}^{n-1}\bra{\omega_n^{\,a_1,\dots,a_{n-1}}(x_1,\dots x_{n-1})}\langle\,\Phi^{\otimes (i-1)}\otimes e_{a}(x_{n})\otimes\Phi^{\otimes(n-i-1)}\,\rangle\\
        &=(-1)^{e^{a}}\,i\hbar\,\sum_{i=1}^{n-1}(-1)^{e^{a}\sum_{j=i+1}^{n-1}e^{a_{j}}}\\
        &\quad\times\langle\,[-\bra{\omega}(\widetilde{e}^{\,a_{1}}(x_{1})\otimes\Phi)]\ldots[-\bra{\omega}(\widetilde{e}^{\,a_{i}}(x_{i})\otimes e^{\,a}(x_{n})]\ldots[-\bra{\omega}(\widetilde{e}^{\,a_{n-1}}(x_{n-1})\otimes\Phi)]\,\rangle
    \end{split}
\end{equation}
In the third line, no sign factor appears since $\bra{\omega}(\,e_{a}(x_{n})\otimes\mathbb{I}\,)$ is degree $e^{a}+1$ and $e^{a}(x)$ has degree $e^{a}(x)$, that is, either $\bra{\omega}(\,e_{a}(x_{n})$ or $e^{a}(x)$ is degree even.
In the fifth line, the sign factor arise due to the exchange of $e^{a}$ and $\bra{\omega}(\widetilde{e}^{\,a_{j-1}}(x_{n-1})\otimes\mathbb{I})$ for $j>i$. Then, the normalization~\eqref{normalization} yields 
\begin{equation}
    \begin{split}
        \,i\hbar\,\sum_{i=1}^{n-1}(-1)^{e^{a}\sum_{j=i+1}^{n-1}e^{a_{j}}}\langle\,\phi^{a_{1}}(x_{1})\ldots\delta_a^{a_{i}}\delta^{d}(x_{i}-x_{n})\ldots\phi^{a_{n-1}}(x_{n-1})\,\rangle\,.
    \end{split}
\end{equation}
Therefore, the equation~\eqref{alg-schwinger3} is reduced to
\begin{equation}
    \begin{split}
    &(-1)^{e^{a}+1}\left\langle\,\phi^{a_{1}}(x_{1})\ldots\phi^{a_{n-1}}(x_{n-1})\frac{\delta S}{\delta \phi^{\,a}(x_{n})}\,\right\rangle\\
    &\quad+i\hbar\,\sum_{i=1}^{n-1}(-1)^{e^{a}\sum_{j=i+1}^{n-1}e^{a_{j}}}\langle\,\phi^{a_{1}}(x_{1})\ldots\delta_a^{a_{i}}\delta^{d}(x_{i}-x)\ldots\phi^{a_{n-1}}(x_{n-1})\,\rangle=0\,.
    \end{split}
\end{equation}
By multiplying $(-1)^{e^{a}\sum_{j=1}^{n-1}{e^{a_j}}}$ on the above equation, we obtain
\begin{equation}
    \begin{split}
    &-\frac{i}{\hbar}(-1)^{e^{a}\sum_{j=1}^{n-1}e^{a_{j}}}\left\langle\,\phi^{a_{1}}(x_{1})\ldots\phi^{a_{n-1}}(x_{n-1})\frac{\delta S}{\delta \phi^{\,a}(x_{n})}\,\right\rangle\\
    &\quad-\sum_{i=1}^{n-1}(-1)^{e^{a}\sum_{j=1}^{i}e^{a_{j-1}}}\langle\,\phi^{a_{1}}(x_{1})\ldots\delta_a^{a_{i}}\delta^{d}(x_{i}-x)\ldots\phi^{a_{n-1}}(x_{n-1})\,\rangle=0\,,
    \end{split}
\end{equation}
where we used the relation $(-1)^{e^{a}}=(-1)^{e^{a}e^{a_{i}}}$ in the second term of the left-hand side. Since the degree of fields and their associated basis vectors of $\mathcal{H}_{1}$ coincides, this is exactly the Schwinger-Dyson equations in quantum field theory.

\section{Noether's theorem and Ward-Takahashi identity}\label{sec:noether}
\setcounter{equation}{0}
In quantum field theory, one of the most important formulas relating to symmetries is the Ward-Takahashi identity, which is given by
\begin{align}
     \expval{\,\phi^{a_1}(x_1)\dots\phi^{a_n}(x_n)\,\partial_\mu j^\mu(x)\,}={}-i\hbar\,\sum_{i=1}^n\,\delta^d(x-x_i)\,\expval{\,\phi^{a_1}(x_1)\dots\delta\phi^{a_i}(x_i)\dots\phi^{a_n}(x_n)\,}\,.\label{ward-takahashi}
\end{align}
The above Ward-Takahashi identity holds only when no anomaly exists.
In the remainder of this paper, we will investigate how this formula is described in terms of homotopy algebras.

\subsection{Infinitesimal symmetry in homotopy algebras}
 In this subsection, we introduce a degree-even multilinear operator $A_n:\mathcal{H}^{\otimes n}\rightarrow \mathcal{H}$, which corresponds to an infinitesimal transformation:
 \begin{align}
     \delta\Phi=\sum_a\,\int\dd^dx\, \delta\phi^a(x)\,e_a(x)=A_0(\vb{1})+A_1(\Phi)+A_2(\Phi,\Phi)+\cdots\,.\label{sym transf}
 \end{align}
 We assume that the operator $A_n$ is cyclic:
 \begin{align}
     \omega \, ( \, \Phi_1 \,, A_n \, ( \, \Phi_2 , \ldots , \Phi_{n+1} \, ) \, )
= {}- \,
\omega \, ( \, A_n ( \, \Phi_1 , \ldots , \Phi_n \, ) \,, \Phi_{n+1} \, ) \,,\label{sym-cyclic}
 \end{align}
We denote the coderivations promoted from $A_n$ by $\vb{A}_n$. We also introduce the operator 
 \begin{align}
     \vb{A}=\sum_{n=0}^\infty\,\vb{A}_n\,.
 \end{align}
Consequently, the transformation~\eqref{sym transf} can be compactly expressed by
 \begin{align}
     \delta\Phi=\pi_1\,\vb{A}\,\frac{1}{1-\Phi}\,.\label{transformation}
 \end{align}
In the same way, the cyclicity~\eqref{sym-cyclic} is written as
\begin{align}
    \bra{\omega}\,\pi_2\,\vb{A}=0\,.
\end{align}
The variation of the action~\eqref{homotopy-action} due to this transformation is given by
\begin{align}
    \delta S=-\sum_{n=0}^\infty\,\frac{1}{n+1}\,\omega\qty(\,\Phi,\pi_1[\vb{M},\vb{A}]\pi_n\frac{1}{1-\Phi}\,)\,.
\end{align}
From this expression, we find that it is sufficient for $\vb{A}$ to satisfy 
\begin{align}
     [\,\vb{M},\vb{A}\,]=0\,.\label{sym-com}
\end{align}
for the infinitesimal transformation generated by  $\vb{A}$ to be the symmetry transformation~\cite{Erler:2016rxg,Erbin:2020eyc}.
Here and in what follows, we only consider a symmetry transformation by $\vb{A}$ which satisfies~\eqref{sym-com}.
Furthermore, we assume that if all of the inputs of $A_n$ are elements of $\mathcal{H}_1$, $A_n$ maps to $\mathcal{H}_1$.
In other words, it satisfies 
\begin{align}
    h\,A_n(\,\Phi_{1}\otimes\Phi_{2}\otimes\ldots\otimes\Phi_{n}\,)=0\,,
\end{align}
where $\Phi_{1},\,\Phi_{2}$ and $\Phi_{n}$ are elements of $\mathcal{H}_{1}$. This assumption implies that this transformation is compatible with the gauge fixing.
To define $A_n$ cyclically, it is important to note that non-trivial values should generally be assigned even when only one of the inputs is an element of $\mathcal{H}_2$. 

\subsection{Noether's theorem in homotopy algebra}
Before discussing the homotopy description of the Ward-Takahashi identity, let us consider the homotopy description of Noether's theorem. 
As it is well known, when the action remains invariant under an infinitesimal transformation, a conserved current $j^\mu(x)$ associated with the transformation exists. The conserved current satisfies the conservation law:
\begin{equation}
    \partial_\mu \,j^\mu(x)\approx0\,,
\end{equation}
where we used $\approx$ to represent that the equality holds under the equation of motion.
In the ordinary procedure in quantum field theory, one way to obtain the conserved current is to replace the infinitesimal parameter $\epsilon$ to $\epsilon(x)$. In this process, the infinitesimal transformation can be modified as follows:
For a symmetry transformation $\delta\phi(x)$, let us define $\delta_{\epsilon}\phi(x)$ by
\begin{align}
    \delta_{\epsilon}\phi(x)=\epsilon(x)\,\delta\phi(x)\,,
\end{align}
where $\epsilon(x)$ is arbitrary function.
The conserved current is obtained from the variation of the action under this transformation:
\begin{align}
    \delta_\epsilon\, S= -\int \dd^dx\,\epsilon(x)\,\partial_\mu\, j^\mu(x)\,.\label{variant current}
\end{align}

Let us translate the above procedure in terms of homotopy algebras.
As seen in the previous subsection, the symmetry transformation can be described by a degree-even cyclic operator $A_n$ satisfying the condition~\eqref{sym-com}.
We define a degree-even cyclic operator $A^\epsilon_n$ with respect to the transformation $\delta_\epsilon\phi(x)$ by
\begin{align}
    \delta_\epsilon\Phi=\sum_a\int \dd^dx\, \epsilon(x)\,\delta\phi^a(x)\,e_a(x)=A^\epsilon_0(\vb{1})+A^\epsilon_1(\Phi)+A^\epsilon_2(\Phi,\Phi)+\cdots\,,
\end{align}
The coderivation promoted by $A^\epsilon_n$ is denoted as $\vb{A}^\epsilon_n$ and we define $\vb{A}^\epsilon$ as
\begin{align}
    \vb{A}^\epsilon=\sum_{n=0}^\infty \vb{A}^\epsilon_n\,.
\end{align}
If we define the operator $\mathcal{O}_\epsilon$ as 
\begin{align}
    \mathcal{O}_\epsilon\,e_a(x)=\epsilon(x)\,e_a(x),\quad \mathcal{O}_\epsilon\, \widetilde{e}^{\,a}(x)=\epsilon(x)\,\widetilde{e}^{\,a}(x)\,,\label{def of epsilon}
\end{align}
then $\vb{A}^\epsilon_n$ can be written as
\begin{align}
     A^\epsilon_n\,(\,\Phi_1,\dots,\Phi_n\,)&=\mathcal{O}_\epsilon\,A_n\,(\,\Phi_1,\dots,\Phi_n\,)\,,
\end{align}
on $\mathcal{H}_1^{\otimes n}$.
In addition, for this $A^\epsilon_n$ to be cyclic it should be defined as 
\begin{align}
    A^\epsilon_n\,(\,\Phi_1,\dots,\widetilde{\Phi},\dots,\Phi_{n-1}\,)&=A_n\,(\,\Phi_1,\dots,\mathcal{O}_\epsilon\,\widetilde{\Phi},\dots,\Phi_{n-1}\,)\,,
\end{align}
for $\Phi_1,\dots,\Phi_{n-1}\in\mathcal{H}_1$ and $\widetilde{\Phi}\in \mathcal{H}_2$. 
The operator $\vb{A^\epsilon}$ no longer satisfies $[\,\vb{M},\vb{A^\epsilon}\,]=0$.
The variation of the action for the transformation is written as
\begin{equation}
    \begin{split}
    \delta_\epsilon S &=-\sum_{n=0}^\infty\,\frac{1}{n+1}\,\omega\,\qty(\,\Phi,\pi_1\,[\,\vb{M},\,\vb{A^\epsilon}\,]\,\pi_n\,\frac{1}{1-\Phi}\,)\\
                      &=-\sum_{n=0}^\infty \frac{1}{n+1}\bra{\omega}\,(\,\mathbb{I}\,\otimes\,\pi_1[\,\vb{M},\,\vb{A^\epsilon}\,]\pi_n\,)\,\frac{1}{1-\Phi}\,.\label{variant action homotopy}
    \end{split}
\end{equation}
Since this is the counterpart of~\eqref{variant current} in terms of homotopy algebras, we find
\begin{align}
    \int\dd^dx\,\epsilon(x)\,\partial_\mu\, j^\mu(x)=\sum_{n=0}^\infty\, \frac{1}{n+1}\bra{\omega}\,(\,\mathbb{I}\,\otimes\,\pi_1\,[\,\vb{M},\,\vb{A^\epsilon}\,]\,\pi_n\,)\,\frac{1}{1-\Phi}\,.\label{conservation low homotopy}
\end{align}
The operator $\mathcal{O}_\epsilon$ can also be written as
\begin{align}
    \mathcal{O}_\epsilon=\sum_a\int\dd^dx\,\epsilon(x)\,\big[(-1)^{\,\widetilde{e}^{\,a}}e_a(x)\otimes\bra{\omega}(\,\widetilde{e}^{\,a}(x)\otimes\mathbb{I}\,)+(-1)^{e_a}\,\widetilde{e}^{\,a}(x)\otimes\bra{\omega}\,(\,e_a(x)\otimes\mathbb{I}\,)\big]\,,\label{O-U}
\end{align}
where we used the normaliztion of $U$ described in~\eqref{nomalization of U}. 
We can rewrite the right-hand side of the equation~\eqref{conservation low homotopy} utilizing~\eqref{O-U}.
The cyclicity of operators $\vb{M}$ and $\vb{A}$ allows us to express~\eqref{conservation low homotopy} as
\begin{align}
    \int\dd^dx\,\epsilon(x)\,\partial_\mu\, j^\mu(x)=\bra{\omega}(\,\pi_1\vb{A^\epsilon}\otimes\pi_1\vb{M}\,)\,\frac{1}{1-\Phi}\,.\label{current1}
\end{align}
Since the action on $\mathcal{H}^{\otimes n}_{1}$ of $\pi_1\vb{A^\epsilon}$ can be written as 
\begin{align}
  \pi_1\vb{A^\epsilon}=\mathcal{O}_\epsilon\,\pi_1\,\vb{A}=\sum_a\,\int\dd^dx\,\epsilon(x)\,(-1)^{\widetilde{e}^{\,a}}\,e_a(x)\otimes\bra{\omega}(\widetilde{e}^{\,a}(x)\otimes \pi_1\vb{A})\,,
\end{align}
substituting this into \eqref{current1} yields  
\begin{align}
   \int\dd^dx\,\epsilon(x)\,\partial_\mu\, j^\mu(x)={}-\sum_a\int\dd^dx\,\epsilon(x)\big(\bra{\omega}(e_a(x)\otimes \pi_1\vb{M})\otimes \bra{\omega}(\widetilde{e}^{\,a}(x)\otimes \pi_1\vb{A})\big)\frac{1}{1-\Phi}\,.
\end{align}
Here, we omitted the product and the coproduct. To be more precise, we can write
\begin{align}
    -\sum_a\int\dd^dx\,\epsilon(x)\,\bigtriangledown\,\big(\bra{\omega}(e_a(x)\otimes \pi_1\vb{M})\otimes' \bra{\omega}(\,\widetilde{e}^{\,a}(x)\otimes \pi_1\vb{A}\,)\big)\,\triangle\,\frac{1}{1-\Phi}\,.\label{homotopy-Noether}
\end{align}
From this expression, we can see that $\partial_\mu j^{\mu}(x)$ can be written in terms of homotopy algebras by
\begin{align}
    \partial_\mu\, j^\mu(x)=-\sum_a\big(\,\bra{\omega}\,(\,e_a(x)\otimes \pi_1\vb{M}\,)\otimes\bra{\omega}\,(\,\widetilde{e}^{\,a}(x)\otimes \pi_1\vb{A})\,\big)\frac{1}{1-\Phi}\,,\label{conservation law}
\end{align}
From the property of group-like element~\eqref{propertyofgroup-likeelement}, it is understood that the right-hand side of~\eqref{homotopy-Noether} is proportional to the equation of motion
\begin{equation}
    \pi_1\,\vb{M}\,\frac{1}{1-\Phi}=0\,.
\end{equation}
Then, Noether's theorem is rederived in terms of homotopy algebras:
\begin{equation}
    \partial_\mu\, j^\mu(x)=0\,.
\end{equation}
Furthermore, integrating the right-hand side and using $\eqref{nomalization of U}$, we obtain 
\begin{align}
   \bra{\omega}(\,\pi_1\vb{A}\otimes\pi_1\vb{M}\,)\frac{1}{1-\Phi}\,.
\end{align}
This can also be written as
\begin{align}
    \sum_{n=0}^\infty\, \frac{1}{n+1}\,\bra{\omega}(\,\mathbb{I}\otimes\pi_1\,[\,\vb{M},\vb{A}\,]\,\pi_n\,)\,\frac{1}{1-\Phi}\,.
\end{align}
Now, since $\vb{A}$ is a symmetry transformation, it commutes with $\vb{M}$, so this is zero. 
Therefore, this is consistent with the fact that the right-hand side of~\eqref{conservation law} is a total derivative.

\subsection{Ward-Takahashi identity}\label{general-WT} 
Now that we have determined the expression corresponding to the derivative of the conserved current $\partial_\mu \,j^\mu(x)$, we can proceed to consider the expectation value $\expval{\,\partial_\mu\, j^\mu(x)\,}$.
Recall that the equation
\begin{equation}
    \,\left\langle\,\frac{1}{1-\Phi}\,\right\rangle=\,\vb{f}\,\vb{1}\label{exp-temp}
\end{equation}
holds, it is understood that replacing the group-like element $\frac{1}{1-\Phi}$  with $\vb{f\,1}$ corresponds to taking the expectation value. 
Hence, the expectation value of $\partial_\mu\, j^\mu(x)$ is expressed as
\begin{align}
    \expval{\,\partial_\mu\, j^\mu(x)\,}=-\sum_a\big(\,\bra{\omega}\,(\,e_a(x)\otimes \pi_1\vb{M}\,)\otimes\bra{\omega}\,(\,\widetilde{e}^{\,a}(x)\otimes \pi_1\vb{A})\,\big)\,\vb{f\,1} \,.\label{exp conservation low}
\end{align}
This quantity should vanish in the absence of anomalies. When an anomaly exists, it should be non-zero.
Let us examine this from the perspective of homotopy algebras. 
Since the operator $\vb{M}$ is the coderivation, the right-hand side of~\eqref{exp conservation low} can be expressed as
\begin{equation}
    \begin{split}
        \expval{\,\partial_\mu\, j^\mu(x)\,}=&{}-\sum_a(-1)^{e_a}\big(\,\bra{\omega}\,(\,e_a(x)\otimes \pi_1\,)\otimes\bra{\omega}\,(\,\widetilde{e}^{\,a}(x)\otimes \pi_1\vb{A})\,\big)\,\vb{M}\,\vb{f\,1}\\
        &\quad+\sum_a(-1)^{e_a}\big(\,\bra{\omega}\,(\,e_a(x)\otimes \pi_1\,)\otimes\bra{\omega}\,(\,\widetilde{e}^{\,a}(x)\otimes \pi_1\vb{A}\vb{M}\,)\,\big)\,\vb{f\,1}\,.
    \end{split}
\end{equation}
Notice that the second term vanishes due to the definition of the symplectic form.
Here, using the algebraic Schwinger-Dyson equation~\eqref{alg schwinger}, the first term  can be expressed as
\begin{align}
    i\hbar\,\sum_{a,b}\int\dd[d]y\, (-1)^{e_a}\big(\,\bra{\omega}\,(\,e_{\,a}(x)\otimes \pi_1\,)\otimes\bra{\omega}\,(\,\widetilde{e}^{\,a}(x)\otimes \pi_1\vb{A})\,\big)\,\vb{e}_b(y)\,\vb{\widetilde{e}}_{\,b}(y)\,\vb{f\,1}\,.
\end{align}
Since
\begin{equation}
    \bra{\omega}\,(\,e_a(x)\otimes e_b(y)\,)=0\,,
\end{equation}
we obtain
\begin{align}
     i\hbar\,\sum_{a,b}\,\int\dd[d]y\, (-1)^{e_a+\widetilde{e}^{\,b}\, e_a}\big(\,\bra{\omega}\,(\,e_a(x)\otimes \widetilde{e}_b(y)\,)\otimes\bra{\omega}\,(\,\widetilde{e}^{\,a}(x)\otimes \pi_1\,\vb{A}\,\vb{e}_b(y))\,\big)\,\vb{f\,1}\,.
\end{align}
By the normalization~\eqref{normalization}, we ultimately obtain 
\begin{align}
     \expval{\,\partial_\mu\, j^\mu(x)\,}=-i\hbar\,\sum_a\,\bra{\omega}(\pi_1\,\vb{A\,e}_a(x)\otimes \widetilde{e}^{\,a}(x))\,\vb{f\,1}\,.\label{single-WT}
\end{align}
In general, the right-hand side of this equation includes divergences.
For instance, if a symmetry transformation $\vb{A}$ is simply linear and scales by constant, this quantity is proportional to $\delta^d(0)$.
Hence, some appropriate regularization is necessary.
We will discuss regularization in detail later.
From these facts, this quantity is expected to represent anomalies.\footnote{In addition to the contribution of anomalies, these divergences are related to the renormalization of composite operators. We will discuss this point in \S\ref{discussion}.}
Indeed, as we will see in section~\ref{sec:reg}, it can be confirmed to coincide with known anomalies in specific examples.

As an extension of \eqref{single-WT}, the expectation value between the fields and $\partial_\mu \,j^\mu(x)$ can be written as 
\begin{equation}
\begin{split}
    &\expval{\phi^{a_1}(x_1)\dots\phi^{a_n}(x_n)\,\partial_\mu\, j^\mu(x)}\\
    &\quad={}-\sum_a\big(\bra{\omega_n^{a_1,\dots,a_n}(x_1,\dots x_n)}\pi_n\, \otimes \bra{\omega}(\,e_a(x)\otimes \pi_1\vb{M}\,)\otimes\bra{\omega}(\,\widetilde{e}^{\,a}(x)\otimes \pi_1\vb{A}\,)\big)\,\vb{f\,1}\,.
    \end{split}
\end{equation}
From this expression, we can derive the Ward-Takahashi identity as follows.
As before, the coderivation $\vb{M}$ can be move to just in front of $\vb{f\,1}$ and using the algebraic Schwinger-Dyson equation~\eqref{alg schwinger}, we obtain
\begin{equation}
    \begin{split}
         &i\hbar\sum_{a,b}\int\dd[d]y(-1)^{e_a}\big(\bra{\omega_n^{a_1,\dots,a_n}(x_1,\dots x_n)}\pi_n\,\\
        &\qquad\qquad\qquad\qquad\otimes \bra{\omega}(\,e_a(x)\otimes \pi_1\,)\otimes\bra{\omega}(\,\widetilde{e}^{\,a}(x)\otimes \pi_1\vb{A}\,)\big)\,\vb{e}_{\,b}(y)\,\vb{\widetilde{e}}^{\,b}(y)\,\vb{f\,1}\,.
    \end{split}
\end{equation}
Due to the normalization~\eqref{normalization}, 
\begin{equation}
    \bra{\,\omega}\,(e_{a}(x)\otimes\pi_{1})
\end{equation}
does not vanish only when the input is the element of $\mathcal{H}_{2}$. Then, we obtain
\begin{equation}
    \begin{split}
        &\expval{\,\phi^{a_1}(x_1)\dots\phi^{a_n}(x_n)\,\partial_\mu\, j^\mu(x)\,}\\
        &\quad={}-i\hbar\sum_a\big(\bra{\omega_n^{a_1,\dots,a_n}(x_1,\dots x_n)}\pi_n\otimes\bra{\omega}(\pi_1\vb{A}\otimes\widetilde{e}^{\,a}(x) )\big)\vb{e}_a(x)\,\vb{f\,1}\\
        &\quad={}-i\hbar\sum_a\,(-1)^{e_a}\big(\bra{\omega_n^{a_1,\dots,a_n}(x_1,\dots x_n)}\pi_n\,\vb{e}_a(x)\otimes\bra{\omega}(\pi_1\vb{A}\otimes\widetilde{e}^{\,a}(x) )\big)\,\vb{f\,1}\\
        &\qquad\qquad-i\hbar\sum_a\big(\bra{\omega_n^{a_1,\dots,a_n}(x_1,\dots x_n)}\pi_n\otimes\bra{\omega}(\pi_1\,\vb{A}\,\vb{e}_a(x)\otimes\widetilde{e}^{\,a}(x) )\big)\,\vb{f\,1}\,.
    \end{split}
\end{equation}
The second term on the right-hand side corresponds to the anomaly.
Let us rewrite the first term to make its meaning clear. 
According to the definition of $\bra{\omega_n^{a_1,\dots,a_n}(x_1,\dots x_n)}$, this term can be expressed as the sum of cases where $e_a(x)$ enters one of the $n$ slots,
\begin{equation}
    \begin{split}
    &-i\hbar\sum_a\sum_{i=1}^n(-1)^{n+e_a+e_a(e_{a_{i+1}}+\cdots+e_{a_n})}
    \Big(\bra{\omega}(\widetilde{e}^{\,a_1}(x_1)\otimes \pi_1)\otimes\cdots\otimes\bra{\omega}(\widetilde{e}^{\,a_i}(x_i)\otimes e_a(x))\otimes\cdots\\
    &\qquad\qquad\qquad\qquad\qquad\qquad\qquad\qquad\qquad\otimes\bra{\omega}(\widetilde{e}^{\,a_n}\otimes \pi_1)\otimes\bra{\omega}(\pi_1\vb{A}\otimes\widetilde{e}^{\,a}(x))\Big)\,\vb{f\,1}\,.
    \end{split}
\end{equation} 
With the careful treatment of sign factors, we obtain
\begin{equation}
    \begin{split}
    -i\hbar \sum_{i=1}^n (-1)^n\,\delta^d(x-x_i)\Big(\,\bra{\omega}(\widetilde{e}^{\,a_1}(x_1)\otimes\pi_1)\otimes\cdots&\otimes\bra{\omega}(\widetilde{e}^{\,a_i}(x_i)\otimes \pi_1\vb{A})\otimes\cdots\\
    &\otimes\bra{\omega}(\widetilde{e}^{\,a_n}(x_n)\otimes\pi_1)\Big)\,\vb{f\,1}\,,       
    \end{split}
\end{equation}
where we used the normalization~\eqref{normalization}. 
Again, from the definition of $\bra{\,\omega_n^{a_1,\dots,a_n}(x_1,\dots x_n)}$, it can be expressed as
\begin{align}
     -i\hbar \sum_{i=1}^n \delta^d(x-x_i)\bra{\omega_n^{a_1,\dots,a_n}(x_1,\dots x_n)}(\pi_{i-1}\otimes\pi_1\vb{A}\otimes\pi_{n-i})\,\vb{f\,1}\,.
\end{align}
Since the operator $\vb{A}$ generates the infinitesimal transformation~\eqref{transformation} and the replacement of $\frac{1}{1-\Phi}$ by $\vb{f\,1}$ corresponds to taking the expectation value, we find that this term is a contact term,
\begin{equation}
    \begin{split}
        &-i\hbar\sum_a\,(-1)^{e_a}\big(\bra{\omega_n^{a_1,\dots,a_n}(x_1,\dots x_n)}\pi_n\,\vb{e}_a(x)\otimes\bra{\omega}(\pi_1\vb{A}\otimes\widetilde{e}^{\,a}(x) )\big)\,\vb{f\,1}\\
        &\quad=-i\hbar\sum_{i}\int\dd^dx\,\epsilon(x)\,\delta^d(x-x_i)\expval{\phi^{a_1}(x_1)\dots\delta\phi^{a_i}(x_i)\dots\phi^{a_n}(x_n)}\,.
    \end{split}
\end{equation}
In summary, the Ward-Takahashi identity can be viewed in terms of homotopy algebras as
\begin{equation}\label{algebraic WT}
    \begin{split}
    &-\sum_a\big(\bra{\omega_n^{a_1,\dots,a_n}(x_1,\dots, x_n)}\pi_n\, \otimes \bra{\omega}(e_{\,a}(x)\otimes \pi_1\vb{M})\otimes\bra{\omega}(\widetilde{e}^{\,a}(x)\otimes \pi_1\vb{A})\big)\,\vb{f\,1}\\
    &=-i\hbar\, \sum_{i=1}^n \delta^d(x-x_i)\bra{\omega_n^{a_1,\dots,a_n}(\,x_1,\dots, x_n\,)}(\pi_{i-1}\otimes\pi_1\vb{A}\otimes\pi_{n-i})\,\vb{f\,1}\\
    &\quad-i\hbar\,\sum_a\big(\bra{\omega_n^{a_1,\dots,a_n}(x_1,\dots x_n)}\pi_n\otimes\bra{\omega}(\,\pi_1\,\vb{A}\,\vb{e}_a(x)\otimes\widetilde{e}^{\,a}(x)\,)\big)\,\vb{f\,1}\,.
    \end{split}
\end{equation}
The left-hand side represents the expectation value of fields and the derivative of the current $\partial_\mu\, j^\mu(x)$, while the first term on the right-hand side corresponds to the contact term, and the second term represents the anomaly. 

\subsection{Examples}\label{example}
In this subsection, we give several examples of the main formula~\eqref{algebraic WT}.

\subsubsection{The U(1) symmetry in complex scalar field theories}
Let us consider the U(1) symmetry in  complex scalar field theory. We consider the action given by
 \begin{equation}
 	S \,= {}-\int\,\dd[d]x\,\qty[\varphi^{\dagger}(x)(\,-\partial^{2}+m^{2})\,\varphi(x)+\sum_{n=1}^{\infty}\frac{1}{2n}\,g_{n}(\varphi^{\dagger}(x)\varphi(x))^{n}]\,.\label{action-complex-scalar}
 \end{equation}
  The action is invariant under the $U(1)$ transformation given by
  \begin{equation}
      \begin{split}
 	\varphi(x)\,\rightarrow\,&\varphi'(x)\,=\,e^{-i\alpha}\,\varphi(x)\,,\\
 	\varphi^{\dagger}(x)\,\rightarrow\,&\varphi'^{\dagger}(x)\,=\,e^{i\alpha}\,\varphi^{\dagger}(x)\,.
      \end{split}
  \end{equation}
 The infinitesimal transformations $\varphi(x)\rightarrow\varphi(x)+\delta\varphi(x)$ and $\varphi^{\dagger}(x)\rightarrow\varphi^{\dagger}(x)+\delta\varphi^{\dagger}(x)$ are given by
 \begin{equation}\label{U1-phi}
 \begin{split}
 	&\delta\varphi(x)\,\rightarrow\,-\,i\,\alpha\,\varphi(x)\,, \\
 	&\delta\varphi^{\dagger}(x)\,\rightarrow\,\,i\,\alpha\,\varphi^{\dagger}(x)\,, 
 \end{split}
 \end{equation}
  In terms of quantum $A_{\infty}$ algebras, we consider an action of the form
  \begin{equation}
        S = {}-\frac{1}{2} \, \omega \, ( \, \Phi, Q \, \Phi \, )
        -\sum_{n=1}^\infty \, \frac{1}{2n} \,
        \omega \, ( \, \Phi \,, m_{2n-1} \, ( \, \Phi , \ldots , \Phi \, ) \, ) \,,
\end{equation}
where
 \begin{equation}
 	\Phi=\int \dd[d]x\,[\,\varphi(x) \,c^{\dagger}(x)\,+\,\varphi^{\dagger}(x) \,c(x)\,]\,,
 \end{equation}
 is the degree-even element of $\mathcal{H}_{1}$ and $c(x)$ is the basis vector of $\mathcal{H}_{1}$ and $c^{\dagger}(x)$ is its hermitian conjugate. Associated with the basis vector $c(x)$, we introduce a degree-odd basis vector $d(x)$ of $\mathcal{H}_{2}$ and its hermitian conjugate $d^{\,\dagger}(x)$. Then, the non-vanishing symplectic form $\omega$ is defined by
 \begin{equation}\label{def-omega-cpx}
 \begin{split}
	\omega \, ( \, c (x_1) \,, d^{\,\dagger} (x_2) \, )
	& =\, \delta^d ( x_1-x_2 ) \,, \\
	\omega \, ( \, c^{\dagger} (x_1) \,, d (x_2) \, )
	& =\, \delta^d ( x_1-x_2 ) \,, \\
	\omega \, ( \,d^{\,\dagger} (x_2)  \,, c (x_1) \, )
	& =\, -\delta^d ( x_1-x_2 ) \,, \\
	\omega \, ( \, d (x_2) \,, c^{\dagger} (x_1) \, )
	& =\, -\delta^d ( x_1-x_2 ) \,.
\end{split}
 \end{equation}
We define the operator $Q$ by
\begin{equation}\label{def-Q-cpx}
\begin{split}
& Q \, c (x)
= {}(-\,\partial^{2}\,+\,m^{2})\,d(x) \,, \quad
Q \, d (x) = 0 \,, \\
& Q \, c^{\dagger} (x)
= {}(-\,\partial^{2}\,+\,m^{2})\,d^{\,\dagger}(x)\,, \quad
Q \,d^{\dagger}(x) = 0 \,,
\end{split}
\end{equation}
to obtain
\begin{equation}
    -\frac{1}{2} \, \omega \, ( \, \Phi, Q \, \Phi \, )=-\int\,\dd[d]x\,\varphi^{\dagger}(x)(\,-\partial^{2}+m^{2})\,\varphi(x)\,.
\end{equation}
To define the operators $m_{2n-1}$, which describe interaction terms, let $e_1(x_1),\dots, e_{2n-1}(x_{2n-1})$ be the $2n-1$ inputs of $m_{2n-1}$.
If the operator $m_{2n-1}$ contains $n$ pieces of $c$ and $n-1$ pieces of  $c^\dag$, we define
\begin{align}
    m_{2n-1}(\,e_1(x_1),\ldots, e_{2n-1}(x_{2n-1})\,)=
        \frac{n!\,(n-1)!\,}{2\,(2n-1)!\,}\,g_n\,\delta^d(x_1-x_2)\cdots \delta^d(x_1-x_{2n-1})\,d(x_1)\,,
\end{align}
and if the operator $m_{2n-1}$ contains $n-1$ pieces of $c$ and $n$ pieces of  $c^\dag$, we define
\begin{align}
    m_{2n-1}(\,e_1(x_1),\dots, e_{2n-1}(x_{2n-1})\,)=
        \frac{n!\,(n-1)!\,}{2\,(2n-1)!\,}\,g_n\,\delta^d(x_1-x_2)\cdots \delta^d(x_1-x_{2n-1})\,d^{\,\dag}(x_1)\,.
\end{align}
Otherwise, we define
\begin{align}
    m_{2n-1}(\,e_1(x_1),\dots,e_{2n-1}(x_{2n-1})\,)=0\,.
\end{align}
Associated with the operator $Q$, we introduce the contracting homotopy $h$ by
\begin{equation}
    \begin{split}
        & h \, c (x) = 0 \,, \qquad
        h \, d (x) = {}\int d^d y \, \Delta(x-y)\, c(y) \,, \\
        & h \, c^{\dagger} (x) = 0 \,, \quad\,\,\,\,
        h \, d^{\dagger} (x)
        = \int d^d y \, \Delta(x-y)\,c^{\dagger}(y) \,,
        \end{split}
    \end{equation}
where $\Delta(x-y)$ is the scalar propagator given by
\begin{equation}
\Delta(x-y)=\int\,\frac{d^{d}k}{(2\pi)^{d}}\,\frac{e^{ik(x-y)}}{k^{2}+m^{2}-i\epsilon}\,.
\end{equation}
The correlation functions of this theory in terms of homotopy algebras are given by
the formula~\eqref{homotopyexpval} with the operators defined above.

 Next, let us consider the symmetry transformations in terms of homotopy algebras. We introduce the degree-even operator $A$ by
\begin{equation}\label{def-A-cpx}
\begin{split}
& A\, c (x) = i\,\alpha\,c(x) \,, \qquad\,\,\,
A\, d (x) = i\,\alpha\,d(x) \,, \\
&A\, c^{\dagger} (x) = -i\,\alpha\,c^{\dagger}(x) \,, \quad
A\, d^{\,\dagger} (x) = -i\,\alpha\,d^{\,\dagger}(x)\,.
\end{split}
\end{equation}
This operator correctly reproduces the symmetry transformation~\eqref{U1-phi}.
\begin{equation}
    \begin{split}
        A\Phi&=\int \dd[d]x\,[\,\varphi(x) \,A\,c^{\dagger}(x)\,+\,\varphi^{\dagger}(x) \,A\,c(x)]\\
             &=\int \dd[d]x\,\qty{\,\varphi(x) \,\qty[-i\,\alpha\, c^{\dagger}(x)\,]\,+\,\varphi^{\dagger}(x) \,\qty[\,i\,\alpha\, c(x)\,]}\\
             &=\int \dd[d]x\,\qty{\,\qty[\,-i\,\alpha\,\varphi(x)\,] \, c^{\dagger}(x)\,+\,\qty[\,i\,\alpha\,\varphi^{\dagger}(x)\,] \, c(x)}\\
             &=\int \dd[d]x\,[\,\delta\varphi(x) \,c^{\dagger}(x)\,+\,\delta\varphi^{\dagger}(x) \,c(x)]\\
             &=\delta\Phi\,. \label{complex-U1}
    \end{split}
\end{equation}
It can be easily verified that $[\,\vb{M}\,,\vb{A}\,]=0$ holds.

The Ward-Takahashi identity in terms of homotopy algebras~\eqref{algebraic WT} are described by
\begin{equation}\label{WT-complex}
    \begin{split}
        &-\sum_{a\in\{c,c^\dag\}}\big(\bra{\,\omega_n^{a_1,\dots,a_n}(x_1,\dots x_n)}\pi_n\, \otimes \bra{\,\omega}(\,e_a(x)\otimes \pi_1\vb{M}\,)\otimes\bra{\,\omega}(\,\widetilde{e}^{\,a}(x)\otimes \pi_1\vb{A}\,)\big)\,\vb{f\,1}\\
        &={}-i\hbar \sum_{i=1}^n \delta^d(x-x_i)\bra{\,\omega_n^{a_1,\dots,a_n}(x_1,\dots x_n)}(\,\pi_{i-1}\otimes\pi_1\vb{A}\otimes\pi_{n-i}\,)\,\vb{f\,1}\\
        &\quad\quad-i\hbar\sum_{a\in\{c,c^\dag\}}\big(\bra{\,\omega_n^{a_1,\dots,a_n}(x_1,\dots x_n)}\pi_n\otimes\bra{\,\omega}(\,\pi_1\,\vb{A}\,\vb{e}_a(x)\otimes\widetilde{e}^{\,a}(x)\,)\big)\,\vb{f\,1}\,,
    \end{split}
\end{equation}
where
\begin{equation}
    e_{a}(x) = \left\{
    \begin{array}{cc}
        c(x) & (a=c)\\
        \,c^{\dagger}(x) & \,\,(a=c^{\dagger})
    \end{array}
    \right.
\end{equation}
and
\begin{equation}
    \widetilde{e}^{\,a}(x) = \left\{
    \begin{array}{cc}
        d^{\,\dagger}(x) & (a=c)\\
        \,d(x) & \,\,(a=c^{\dagger})
    \end{array}
    \right.\,.
\end{equation}
Let us first consider the case for free theory. In this case, the identities~\eqref{WT-complex} can be written as
\begin{equation}\label{WT-cpx-free}
    \begin{split}
    &-\sum_{a\in\{c,c^\dag\}}\big(\bra{\,\omega_n^{a_1,\dots,a_n}(x_1,\dots x_n)}\pi_n\, \otimes \bra{\,\omega}(e_{a}(x)\otimes \pi_1\vb{Q})\otimes\bra{\omega}(\widetilde{e}^{\,a}(x)\otimes \pi_1\vb{A})\big)\,\vb{f\,1}\\
    &={}-i\hbar \sum_{i=1}^n \delta^d(x-x_i)\bra{\,\omega_n^{a_1,\dots,a_n}(x_1,\dots x_n)}(\pi_{i-1}\otimes\pi_1\vb{A}\otimes\pi_{n-i})\,\vb{f\,1}\\
    &\quad\quad-i\hbar\sum_{a\in\{c,c^\dag\}}\big(\bra{\,\omega_n^{a_1,\dots,a_n}(x_1,\dots x_n)}\pi_n\otimes\bra{\,\omega}(\pi_1\,\vb{A}\,\vb{e}_{\,a}(x)\otimes\widetilde{e}^{\,a}(x) )\big)\,\vb{f\,1}\,.
    \end{split}
\end{equation} 
For simplicity,
we take $n=2k\,,a_{1}=a_{2}=\ldots=a_{k}=c^{\dagger}$ and $a_{k+1}=a_{k+2}=\ldots=a_{2k}=c$. Then, the operator $\bra{\,\omega_n^{a_1,\dots,a_n}(x_1,\dots x_n)}$ is
\begin{equation}
    \begin{split}
    \bra{\,\omega_{2k}^{a_1,\dots,a_{2k}}(x_1,\dots x_{2k})}=&\bra{\,\omega}(d(x_1)\otimes\pi_1)\otimes\cdots\otimes\bra{\,\omega}(d(x_{k})\otimes\pi_1)\\
    &\otimes\bra{\,\omega}(d^{\,\dag}(x_{k+1})\otimes\pi_1)\otimes\cdots\otimes\bra{\,\omega}(d^{\,\dag}(x_{2k})\otimes\pi_1)\,.
    \end{split}
\end{equation} 
Since
\begin{equation}
    \pi_n \, \vb{f} \, {\bf 1}  =  \langle \, \Phi^{\otimes n} \, \rangle=\expval{\Big[\int d^{d}x\,\big(\,\varphi(x) \,c^{\dagger}(x)\,+\,\varphi^{\dagger}(x) \,c(x)\big)\Big]^{\otimes n} }\,, 
\end{equation}
the left-hand side of \eqref{WT-cpx-free} is given by
\begin{equation}
    \begin{split}
        -\int\dd[d]y\,\dd[d]z\, \big\langle\,&\varphi(x_1)\dots\varphi(x_k)\,\varphi^\dag(x_{k+1})\dots\varphi^\dag(x_{2k})\\
        &\big[\,\varphi(y)\,\varphi^\dag(z)\bra{\omega}(\,c(x)\otimes Q\,c^\dag(y)\,)\otimes \bra{\omega}(d^{\,\dag}(x)\otimes A\, c(z))\\
        &+\varphi^\dag(y)\,\varphi(z)\,\bra{\omega}(\,c^\dag(x)\otimes Q\,c(y)\,)\otimes \bra{\omega}(\,d(x)\otimes A \,c^\dag(z)\,)\,\big]\,\big\rangle\,.\label{cpx-WT-lhs}
    \end{split}
\end{equation}
The quantity inside the square brackets can be calculated as follows.
From the definition \eqref{def-Q-cpx} and \eqref{def-A-cpx}, we obtain
\begin{equation}
    \begin{split}
    &-\int\dd[d]y\,\dd[d]z\, \varphi(y)\,\varphi^\dag(z)\,\bra{\omega}(\,c(x)\otimes Q\,c^\dag(y)\,)\otimes \bra{\omega}(\,d^{\,\dag}(x)\otimes A\, c(z))\\
    &=-\int\dd[d]y\,\dd[d]z\, \varphi(y)\,\varphi^\dag(z)\bra{\omega}(\,c(x)\otimes (-\partial^2_{y}+m^2)\,d^\dag(y)\,)\otimes \bra{\omega}(\,d^{\,\dag}(x)\otimes i\alpha\, c(z))\\
    &=-i\alpha\,\int\dd[d]y\,\dd[d]z\, (\,-\partial^2_{y}+m^2\,)\,\varphi(y)\,\varphi^\dag(z)\,\bra{\omega}(\,c(x)\otimes d^{\,\dag}(y)\,)\otimes \bra{\omega}(\,d^{\,\dag}(x)\otimes  c(z)\,)\,,
    \end{split}
\end{equation}
and 
\begin{equation}
    \begin{split}
    &-\int\dd[d]y\,\dd[d]z\, \varphi^\dag(y)\,\varphi(z)\,\bra{\omega}(\,c^\dag(x)\otimes Q\,c(y)\,)\otimes \bra{\omega}(\,d(x)\otimes A\, c^\dag(z)\,)\\
    &=-\int\dd[d]y\,\dd[d]z\, \varphi^\dag(y)\,\varphi(z)\,\bra{\omega}(\,c^\dag(x)\otimes (-\partial^2_{y}+m^2\,)\,d(y))\otimes \bra{\omega}\,(\,d(x)\otimes (-i\alpha)\, c^\dag(z)\,)\\
    &=i\alpha\,\int\dd[d]y\,\dd[d]z\, (\,-\partial^2_{y}+m^2\,)\,\varphi^\dag(y)\,\varphi(z)\,\bra{\omega}(\,c^\dag(x)\otimes d(y)\,)\otimes \bra{\omega}\,(\,d(x)\otimes  c^\dag(z)\,)\,.
    \end{split}
\end{equation} 
Then, the quantity inside the square brackets in~\eqref{cpx-WT-lhs} can be expressed as
\begin{align}
   i\alpha\qty(\,\varphi(x)\,\partial^2\varphi^\dag(x)-\varphi^\dag(x)\,\partial^2\varphi(x)\,)\,.
\end{align}
We find that this equals to $\partial_\mu\, j^\mu(x)$, where $j^{\mu}(x)$ denotes the Noether current  associated with the symmetry~\eqref{U1-phi}.
Hence, the left-hand side of \eqref{WT-cpx-free} is equal to
\begin{align}
    \langle\,\varphi(x_{1})\,\varphi(x_{2})\ldots\varphi(x_{k})\,\varphi^{\dagger}(x_{k+1})\,\varphi^{\dagger}(x_{k+2})\ldots\varphi^{\dagger}(x_{2k})\,\partial_{\mu}\,j^{\mu}(x)\,\rangle\,.
\end{align} 
Next, let us consider the first term of the right-hand side of~\eqref{WT-cpx-free}.
The relation~\eqref{complex-U1} yields
\begin{equation}
        (\pi_{i-1}\otimes\pi_1\vb{A}\otimes\pi_{2k-i})\,\vb{f}\,\vb{1}
            =\langle\, \Phi^{\otimes i-1}\otimes\delta\Phi\otimes\Phi^{\otimes(2k-i)}\,\rangle\,,
\end{equation}
and we find that the first term of the right-hand side of~\eqref{WT-cpx-free} is given by 
\begin{equation}
    \begin{split}
        &-i\hbar \sum_{i=1}^{2k} \delta^d(x-x_i)\bra{\,\omega_{2k}^{a_1,\dots,a_{2k}}(x_1,\dots x_{2k})}(\pi_{i-1}\otimes\pi_1\vb{A}\otimes\pi_{2k-i})\,\vb{f\,1}\\
        &=-i\hbar \sum_{i=1}^{k}\delta^d(x-x_i)\langle\,\varphi(x_{1})\ldots\varphi(x_{i-1})\,\delta\varphi(x_{i})\,\varphi(x_{i+1})\ldots\varphi(x_{k})\,\varphi^{\dagger}(x_{k+1})\ldots\varphi^{\dagger}(x_{2k})\,\rangle\\
        &\quad-i\hbar \sum_{i=1}^{k}\delta^d(x-x_i)\langle\,\varphi(x_{1})\ldots\varphi(x_{k})\,\varphi^{\dagger}(x_{k+1})\ldots\varphi^{\dagger}(x_{k+i-1})\,\delta\varphi^{\dagger}(x_{k+i})\,\varphi^{\dagger}(x_{k+i+1})\ldots\varphi^{\dagger}(x_{2k})\,\rangle\,.
    \end{split}
\end{equation}
The second term of the right-hand side of~\eqref{WT-cpx-free} is given by
\begin{equation}
    \begin{split}
        &-i\hbar\,\sum_{a\in\{c,c^\dag\}}\big(\bra{\omega_{2k}^{a_1,\dots,a_n}(x_1,\dots x_{2k})}\pi_{2k}\otimes\bra{\omega}(\pi_1\,\vb{A}\,\vb{e}_{\,a}(x)\otimes\widetilde{e}^{\,a}(x) )\big)\,\vb{f\,1}\\
        &=-i\hbar\expval{\varphi(x_1)\dots\varphi(x_k)\,\varphi^\dag(x_{k+1})\dots\varphi^\dag(x_{2k})
    \big[\bra{\omega}(Ac(x)\otimes d^{\,\dag}(x))+\bra{\omega}(Ac^\dag(x)\otimes d(x))\big]}\,.
    \end{split}
\end{equation}
We obtain
\begin{align}
    \bra{\omega}(Ac(x)\otimes d^{\,\dag}(x))+\bra{\omega}(Ac^\dag(x)\otimes d(x))=i\alpha\,\delta^d(0)-i\alpha\,\delta^d(0)\,.
\end{align}
This expression is ill-defined and requires regularization, however, for the convenience, we ignore the divergence and then, this term vanishes. Details regarding the regularization are discussed in \S\ref{sec:reg}. 
Then, the identities~\eqref{WT-cpx-free} can be expressed as
\begin{equation}\label{WT-free-cpx-result}
    \begin{split}
         &\langle\,\varphi(x_{1})\,\varphi(x_{2})\ldots\varphi(x_{k})\,\varphi^{\dagger}(x_{k+1})\varphi^{\dagger}(x_{k+2})\ldots\varphi^{\dagger}(x_{2k})\,\partial_{\mu}\,j^{\mu}(x)\,\rangle\\
         &=-i\hbar \sum_{i=1}^{k}\delta^d(x-x_i)\langle\,\varphi(x_{1})\ldots\varphi(x_{i-1})\,\delta\varphi(x_{i})\,\varphi(x_{i+1})\ldots\varphi(x_{k})\,\varphi^{\dagger}(x_{k+1})\ldots\varphi^{\dagger}(x_{2k})\,\rangle\\
          &\quad-i\hbar \sum_{i=1}^{k}\delta^d(x-x_i)\langle\,\varphi(x_{1})\ldots\varphi(x_{k})\,\varphi^{\dagger}(x_{k+1})\ldots\varphi^{\dagger}(x_{k+i-1})\,\delta\varphi^{\dagger}(x_{k+i})\,\varphi^{\dagger}(x_{k+i+1})\ldots\varphi^{\dagger}(x_{2k})\,\rangle\,.
    \end{split}
\end{equation}
This is exactly the Ward-Takahashi identity for the free complex scalar theory. 
Actually, for the action~\eqref{action-complex-scalar}, the interaction term does not affect the identity~\eqref{WT-complex}. In the interaction theory, the term that contains 
\begin{equation}
    \sum_{n=1}^\infty\sum_{a\in\{c,c^\dag\}}\bra{\omega}(e_{a}(x)\otimes\pi_1\vb{m}_{2n-1})\otimes\bra{\omega}(\widetilde{e}^{\,a}(x)\otimes \pi_1\vb{A})
\end{equation}
is added to the identities for the free theory.
We can verify that this term vanishes.
It is sufficient to evaluate its action on 
\begin{align}
    \Phi^{\otimes 2n}=\Big[\int d^{d}x\,\big(\,\varphi(x) \,c^{\dagger}(x)\,+\,\varphi^{\dagger}(x) \,c(x)\big)\Big]^{\otimes 2n}\,.
\end{align}
Terms in which the number of $c$'s and $c^\dag$'s is not equal vanish trivially due to the definition of $m_{2n-1}$. 
There are two types of non-trivial terms.
One type is the term where $\vb{m}_{2n-1}$ takes $n$ pieces of $c$ and $n-1$ pieces of $c^\dag$ as inputs, and $\vb{A}$ takes a $c$ as a input.
There are $\binom{2n}{n}$ such terms.
These terms are evaluated as
\begin{align}
    -i\alpha\,\frac{n!\,(n-1)!}{2(2n-1)!}\,g_n\,(\,\varphi(x)\varphi^\dag(x)\,)^n\times \binom{2n}{n}=-\frac{i\alpha g_n}{2}(\varphi(x)\varphi^\dag(x))^n\,.
\end{align}
Another type is the term where  $\vb{m}_{2n-1}$ takes $n-1$ pieces of $c$ and $n$ pieces of $c^\dag$ as input, and $\vb{A}$ takes a $c^\dag$ as input.
There are also $\binom{2n}{n}$ such terms.
These terms are evaluated as
\begin{align}
    i\alpha\,\frac{n!\,(n-1)!}{2(2n-1)!}\,g_n\,(\,\varphi(x)\varphi^\dag(x)\,)^n\times \binom{2n}{n}=\frac{i\alpha g_n}{2}(\varphi(x)\varphi^\dag(x))^n\,,
\end{align}
and cancel out with the former.
Hence, the interaction term does not affect the current and obtain the same form of the Ward-Takahashi identities as~\eqref{WT-free-cpx-result}.

\subsubsection{The energy-momentum tensor in scalar field theories} 
In this subsection, we consider the energy-momentum tensor in scalar field theories. We use the setup described in \S\ref{scalar}. We consider a scalar field theory whose action is given by
\begin{equation}
    S=\int\dd[d]{x}\qty(-\frac{1}{2}\,\partial^{\mu}\varphi(x)\,\partial_{\mu}\varphi(x)-\sum_{n=0}^{\infty}\,\frac{1}{n+1}\,g_{n+1}\,\varphi(x)^{n+1})\,,
\end{equation}
This theory has the spacetime transformation symmetry:
\begin{equation}
    \varphi(x)\rightarrow\varphi(x-a)\,,\label{spacetime-trsf}
\end{equation}
where $a$ is the constant spacetime transformation. The infinitesimal transformation $\varphi(x)\rightarrow\varphi(x)+\delta\phi(x)$  of~\eqref{spacetime-trsf}  is given by
\begin{equation}
    \delta\varphi(x)=-a^{\nu}\,\partial_{\nu}\,\varphi(x)\,.\label{inf-spacetime-trsf}
\end{equation}
From Noether's theorem, the conserved current associated with the transformarion~\eqref{spacetime-trsf} is given by
\begin{equation}
    j^{\mu}(x)=a_{\nu}\,T^{\mu\nu}\,,
\end{equation}
where $T^{\mu\nu}$ is the energy-momentum tensor. The energy-momentum tensor is given by
\begin{equation}
    T^{\mu\nu}=-\frac{\partial\mathcal{L}(x)}{\partial(\partial_{\mu}\,\varphi_{a}(x))}+\eta^{\mu\nu}\mathcal{L}(x)\,,
\end{equation}
where $\mathcal{L}(x)$ is the lagrangian density and $\eta^{\mu\nu}$ is a metric defined by
\begin{equation}
    \eta^{\mu\nu}=\mathrm{diag}(\,-1,1,1,1\,)\,.
\end{equation}

In terms of homotopy algebras, we can rewrite the above action to
\begin{equation}
S = {}-\frac{1}{2} \, \omega \, ( \, \Phi, Q \, \Phi \, )
-\sum_{n=0}^\infty \, \frac{1}{n+1} \,
\omega \, ( \, \Phi \,, m_n \, ( \, \Phi , \ldots , \Phi \, ) \, ) \,,
\end{equation}
where
\begin{equation}
Q \, c(x) = ( {}-\partial^2 +m^2 \, ) \, d(x) \,, \qquad
Q \, d(x) = 0 \,,
\end{equation}
and
\begin{equation}
    m_{n}(\,c(x_{1}),\ldots, c(x_{n})\,)=g_{n+1}\,\delta^{d}(x_{1}-x_{2})\ldots\delta^{d}(x_{1}-x_{n})\,d(x_{1})\,,
\end{equation}
for $n>0$ and
\begin{equation}
    m_{0}\,\vb{1}=\int \dd[d]{x}\,g_{1}\,d(x)\,.
\end{equation} 
To consider the symmetry tansformation~\eqref{inf-spacetime-trsf}, we introduce degree-even operators
\begin{equation}\label{trsf em tensor}
\begin{split}
    &A\,c(x)=a_{\nu}\,\partial^{\nu}c(x)\,,\\
    &A\,d(x)=a_{\nu}\,\partial^{\nu}d(x)\,.
\end{split}
\end{equation}
Since
\begin{equation}
    \begin{split}
        A\Phi&=\int \dd[d]x\,[\,\varphi(x) \,A\,c(x)\,]\\
             &=\int \dd[d]x\,a_{\nu}\varphi(x)\,\partial^{\nu}c(x)\\
             &=\int \dd[d]x\,\qty[-a_{\nu}\,\partial^{\nu}\,\varphi(x)]\,c(x)\\
             &=\int \dd[d]x\,\delta\varphi(x)\,c(x)\\
             &=\delta\Phi\,, 
    \end{split}
\end{equation}
this correctly reproduces the symmetry transformation~\eqref{inf-spacetime-trsf}. 
It can be easily verified that $[\,\vb{M}\,,\vb{A}\,]=0$ holds.

The Ward-Takahashi identities in terms of homotopy algebras~\eqref{algebraic WT} are described by
\begin{equation}\label{WT-scalar}
    \begin{split}
        &-\big(\bra{\,\omega_n(x_1,\dots x_n)}\pi_n\, \otimes \bra{\omega}(\,c(x)\otimes \pi_1\vb{M}\,)\otimes\bra{\,\omega}(\,d(x)\otimes \pi_1\,\vb{A})\,\big)\,\vb{f\,1}\\
        &=-i\hbar \sum_{i=1}^n\, \delta^d(x-x_i)\bra{\,\omega_n(x_1,\dots x_n)}(\pi_{i-1}\otimes\pi_1\,\vb{A}\otimes\pi_{n-i})\,\vb{f\,1}\\
        &\qquad-i\hbar\big(\bra{\,\omega_n(x_1,\dots x_n)}\,\pi_n\otimes\bra{\omega}(\,\pi_1\,\vb{A}\,\vb{c}(x)\otimes d(x) \,)\big)\,\vb{f\,1}\,.
    \end{split}
\end{equation} 
Since
\begin{equation}
    \pi_{n}\,\vb{f}\,\vb{1}=\langle\,\Phi^{\otimes n}\,\rangle=\expval{\left[\int\dd[d]x\, \varphi(x)c(x)\right]^{\otimes n}}\,,
\end{equation}
the left-hand side of \eqref{WT-scalar} is given by
\begin{equation}
    \begin{split}
    &-\big(\bra{\omega_n(x_1,\dots x_n)}\pi_n\, \otimes \bra{\,\omega}(\,c(x)\otimes \pi_1\vb{M}\,)\otimes\bra{\omega}(\,d(x)\otimes \pi_1\vb{A})\,\big)\,\vb{f\,1}\\
    &=-\expval{\,\varphi(x_1)\dots\varphi(x_n)\bra{\omega}\left(c(x)\otimes \pi_1\,\vb{M}\,\frac{1}{1-\Phi}\right)\otimes\bra{\omega}(\,d(x)\otimes A\,\Phi\,)\,}\,.
    \end{split}
\end{equation}
Let us calculate 
\begin{equation}
    \begin{split}
    &-\bra{\,\omega}\left(\,c(x)\otimes \pi_1\vb{M}\,\frac{1}{1-\Phi}\,\right)\otimes\bra{\,\omega}(\,d(x)\otimes A\,\Phi\,)\\
    &=-\bra{\omega}(c(x)\otimes Q\,\Phi)\otimes\bra{\omega}(\,d(x)\otimes A\,\Phi\,)\\
    &\qquad-\sum_{n=0}^\infty\bra{\omega}(c(x)\otimes m_n(\Phi,\ldots,\Phi))\otimes\bra{\omega}(\,d(x)\otimes A\,\Phi\,)\,.
    \end{split}
\end{equation}

Let us first consider the free part. For simplicity, we split $Q$ into two parts: 
\begin{equation}
    Q=Q^{\mathrm{kin}}+Q^{\mathrm{mass}}\,,
\end{equation}
\begin{align}
    &Q^{\mathrm{kin}}\,c(x)=-\partial^{2}\,d(x)\,,
    \quad Q^{\mathrm{mass}}\,c(x)=m^2\,d(x)\,.
\end{align}
The kinetic part can be calculated as follows, 
\begin{equation}
    \begin{split}
        &-\bra{\omega}(c(x)\otimes Q^{\mathrm{kin}}\,\Phi)\otimes\bra{\omega}(\,d(x)\otimes A\,\Phi\,)\\
        &=-\int\dd[d]y\,\dd[d]z\bra{\omega}\big(c(x)\otimes \varphi(y)\,(-\partial^2_{y}\,d(y))\big)\otimes\bra{\omega}\big(d(x)\otimes \varphi(z)\,a_\nu\, \partial^\nu_{z} \,c(z)\big)\\
        &=-a_\nu\int\dd[d]y\,\dd[d]z\, \partial^2_{y}\,\varphi(y)\,\partial^\nu_{z}\, \varphi(z)\,\bra{\omega}(c(x)\otimes d(y))\otimes\bra{\omega}(d(x)\otimes c(z))\\
        &=a_\nu\,\partial^2\varphi(x)\,\partial^\nu \varphi(x)\\
        &=a_\nu\,\partial_\mu\big(\,\partial^\mu\varphi(x)\,\partial^\nu\varphi(x)-\frac{1}{2}\,\eta^{\mu\nu}\,\partial^\rho\varphi(x)\,\partial_\rho\varphi(x)\,\big)\\
        &=a_\nu\,\partial_\mu\, T_{\mathrm{kin}}^{\mu\nu}(x)\,,
    \end{split}
\end{equation}
where $T^{\mu\nu}_{\mathrm{kin}}$ is the energy-momentum tensor which is the contribution from the kinetic term.
Since the contribution of the operator $Q^{\mathrm{mass}}$ can be calculated in the same way as that of the operator $m_{1}$,  we omit the calculation. 

Let us next consider the interaction part:
\begin{align}
    -\sum_{n=0}^\infty\bra{\omega}(c(x)\otimes m_n(\,\Phi,\ldots,\,\Phi\,))\otimes\bra{\omega}(d(x)\otimes A\,\Phi)\,.
\end{align}
This can be calculated as follows:
\begin{equation}
    \begin{split}
        &-\bra{\omega}\big(c(x)\otimes m_n(\,\Phi,\ldots,\Phi\,)\big)\otimes\bra{\omega}(\,d(x)\otimes A\,\Phi\,)\\
        &=-\int\dd[d]y\,\dd[d]z\,g_{n+1}\,\varphi(y)^n\,\varphi(z)\,\bra{\,\omega}(c(x)\otimes d(y))\otimes\bra{\omega}(d(x)\otimes a_\nu\,\partial^\nu c(z))\\
        &=-a_\nu\, g_{n+1}\,\varphi(x)^n\,\partial^\nu\, \varphi(x)\\
        &=-a_\nu \frac{g_{n+1}}{n+1}\partial_\mu(\,\eta^{\mu\nu}\varphi(x)^{n+1}\,)\\
        &=a_\nu\,\partial_\mu\, T^{\mu\nu}_{n+1}\,,
    \end{split}
\end{equation}
where $T^{\mu\nu}_{n+1}$ is the energy-momentum tensor which is the contribution from the $(n+1)$-point interactions.
Combining all the contributions, the left-hand side of~\eqref{WT-scalar} can be expressed as 
\begin{equation}    
    \begin{split}
    &-\big(\bra{\omega_n(x_1,\dots x_n)}\pi_n\, \otimes \bra{\omega}(c(x)\otimes \pi_1\vb{M})\otimes\bra{\omega}(d(x)\otimes \pi_1\vb{A})\big)\,\vb{f\,1}\\
    &=a_\nu\,\expval{\,\varphi(x_1)\dots\varphi(x_n)\,\partial_\mu\, T^{\mu\nu}(x)\,}\,.
    \end{split}
\end{equation}
The right-hand side of~\eqref{WT-scalar} can be evaluated in the same as the example of a complex scalar field and the first term can be written as
\begin{equation}
    \begin{split}
    &-i\hbar \sum_{i=1}^n \delta^d(x-x_i)\bra{\omega_n(x_1,\dots x_n)}(\pi_{i-1}\otimes\pi_1\vb{A}\otimes\pi_{n-i})\,\vb{f\,1}\\
    &=-i\hbar\sum_{i=1}^n\delta^d(x-x_i)\expval{\,\varphi(x_1)\dots\varphi(x_{i-1})\,\delta\varphi(x_i)\,\varphi(x_{i+1})\dots\varphi(x_n)\,}\,.
    \end{split}
\end{equation}
The second term includes an ill-defined factor
\begin{align}
    \bra{\omega}\big(A\,c(x)\otimes d(x)\big)=\bra{\,\omega}\big(a_\nu\,\partial^\nu\, c(x)\otimes d(x)\big)=\lim_{y\rightarrow x}a_\nu\,\partial^\nu_x\,\delta^d(x-y)\,.
\end{align}
Thus, it requires regularization, which will be described in \S\ref{sec:reg}. Here, we ignore this term. 
We finally find that the equation~\eqref{WT-scalar} is equivalent to 
\begin{equation}
    \begin{split}
        &a_\nu\langle\,\varphi(x_{1})\ldots\varphi(x_{n})\,\partial_{\mu}\,T^{\mu\nu}(x)\,\rangle\\
        &\qquad=-i\hbar\sum_{i=1}^{n}\,\delta^{d}(x-x_{i})\,\langle\,\varphi(x_{1})\ldots\varphi(x_{i-1})\,\delta\varphi(x_{i})\,\varphi(x_{i+1})\ldots\varphi(x_{n})\,\rangle\,.
     \end{split}
\end{equation}
This is exactly the Ward-Takahashi identity for the energy-momentum tensor of scalar field theory.

\subsubsection{The axial $U(1)$ symmetry in vector-like gauge theories } 
Finally, let us consider the axial $U(1)$ symmetry in a vector-like gauge theory.
Consider the action of a massless Dirac fermion given by
\begin{align}
    S=\int\dd[4]x\,\overline{\Psi}_s(x)\,i\D_{st}(x)\,\Psi_t(x)\,,
\end{align}
where $s$ and $t$ are spinor indices\footnote{We take Einstein summation convention. We also omit the spinor indices  if there is no confusion.} running from 1 to 4, $\D$ is defined by
\begin{align}
  \D(x) =\gamma^\mu(\,\partial_\mu-iA_\mu(x)\,)\,,
\end{align}
and $\gamma^\mu$ is a $\gamma$ matrix satisfying
\begin{align}
    \{\gamma^\mu,\gamma^\nu\}=-2\,\eta^{\mu\nu}\,.
\end{align}
Furthermore, we treat $A_\mu(x)$ as the background field.
This theory has a axial $U(1)$ symmetry:
\begin{equation}\label{chilal sym}
\begin{split}
    \Psi(x)\rightarrow &\ e^{-i\alpha \gamma_5}\ \Psi(x)\,,\\
    \overline{\Psi}(x)\rightarrow &\ \overline{\Psi}(x)\ e^{-i\alpha \gamma_5}\,,
\end{split}
\end{equation}
where $\gamma_5$ is defined by
\begin{align}
    \gamma_5=i\gamma^0\gamma^1\gamma^2\gamma^3\,.
\end{align} 
The infinitesimal transformations are given by
\begin{equation}\label{chiral sym infinitesimal}
\begin{split}
    \delta \Psi(x)&={}-i\alpha\,\gamma_5\, \Psi(x)\,,\\
    \delta \Bar{\Psi}(x)&={}-i\alpha\, \overline{\Psi}(x)\,\gamma_5\,.
\end{split}
\end{equation}

In terms of homotopy algebras, the above action can be rewritten as
\begin{align}
    S=-\frac{1}{2}\,\omega(\,\Phi,Q\,\Phi\,)\,,
\end{align}
where
\begin{align}
    \Phi=\int\dd[4]x\ \qty[\Psi_s(x)\,\overline{\theta}_s(x)+\overline{\Psi}_s(x)\,\theta_s(x)]\,,
\end{align} 
is the degree-even element of $\mathcal{H}_1$ and $\{\theta_s(x),\,\overline{\theta}_s(x)\}$ are the degree-odd basis vectors of $\mathcal{H}_1$.\footnote{In~\cite{Konosu:2023pal,Konosu:2023rkm}, the basis vector $\overline{\theta}_s(x)$ is the Dirac adjoint of $\theta_s(x)$ to be consistent with the reality condition. Our notation is slightly different from the previous papers and the ``adjoint'' of the basis vector should be  minus the Dirac adjoint.}
We introduce the degree-even basis vectors $\{\lambda_s(x)\,,\overline{\lambda}_s(x)\}$ of $\mathcal{H}_2$ satisfying the normalization:
\begin{equation}
\begin{split}
    \omega\big(\,\theta_s(x)\,,\overline{\lambda}_t(y)\,\big)&={}-\delta_{st}\,\delta^4(x-y)\,,\\
    \omega\big(\,\overline{\theta}_s(x)\,,\lambda_t(y)\,\big)&={}-\delta_{st}\,\delta^4(x-y)\,,\\
     \omega\big(\,\overline{\lambda}_s(x)\,,\theta_t(y)\,\big)&=\delta_{st}\,\delta^4(x-y)\,,\\
     \omega\big(\,\lambda_s(x)\,,\overline{\theta}_t(y)\,\big)&=\delta_{st}\,\delta^4(x-y)\,.
\end{split}
\end{equation}
We do not need to introduce a basis vector for the background field $A_\mu$. 
We define $Q$ by
\begin{equation}
\begin{split}
    Q\,\theta_s(x)=-i\D_{st}(x)\,\lambda_t(x)\,,&\quad Q\,\lambda_s(x)=0\,,\\
    Q\,\overline{\theta}_s(x)=i\overline{\lambda}_t(x)\overleftarrow{\D}_{ts}(x)\,,&\quad Q\,\overline{\lambda}_s(x)=0\,,
\end{split}
\end{equation}
where $\overleftarrow{\D}$ is defined by
\begin{align}
    \overleftarrow{\D}(x)=\gamma^\mu(-\overleftarrow{\partial}_\mu-iA_\mu(x))\,,
\end{align}
and $\overleftarrow{\partial}_\mu$ denotes the differential acting from the right. 
Then, we reproduce the action in terms of homotopy algebras:
\begin{equation}
    -\frac{1}{2}\,\omega(\,\Phi,Q\,\Phi\,)=\int\dd[4]x\,\overline{\Psi}(x)\,i\D(x)\,\Psi(x)\,.
\end{equation} 
Associated with the operator $Q$, we introduce the contracting homotopy $h$ by
\begin{equation}
    \begin{split}
    h\,\theta_s(x)=0\,,&\quad h\,\lambda_s(x)=i\,\left[\,\D(x)\int_0^{\infty}\dd\tau\,e^{-\tau \not D^2(x)}\,\right]_{st}\theta_t(x)\,,\\
    h\,\overline{\theta}_s(x)=0\,,&\quad h\,\overline{\lambda}_s(x)={}-i\, \overline{\theta}_t(x)\left[\,\overleftarrow{\D}(x)\int_0^{\infty}\dd\tau\,e^{-\tau \overleftarrow{\not D}^2(x)}\,\right]_{ts}\,.
    \end{split}
\end{equation} 
The degree-even operator $A$ corresponding to the symmetry transformation~\eqref{chiral sym infinitesimal} is defined as
\begin{equation}\label{chiral sym A}
\begin{split}
    A\,\theta_s(x)=-i\alpha\,(\gamma_5)_{st}\,\theta_t(x)\,,&\quad A\,\lambda_s(x)=i\alpha\,(\gamma_5)_{st}\,\lambda_t(x)\,,\\
    A\,\overline{\theta}_s(x)=-i\alpha\,\overline{\theta}_t(x)\,(\gamma_5)_{ts}\,,&\quad A\,\overline{\lambda}_s(x)=i\alpha\,\overline{\lambda}_t(x)\,(\gamma_5)_{ts}\,.
\end{split}
\end{equation}
It reproduces the  symmetry transformation as
\begin{equation}
    \begin{split}
    A\,\Phi&=\int\dd[4]x\,\qty[\,\Psi_s(x)\,A\,\overline{\theta}_s(x)+\overline{\Psi}_s(x)\,A\,\theta_s(x)\,]\\
    &=-i\alpha\int\dd[4]x\,\qty[\,\Psi_s(x)\,(\gamma_5)_{ts}\,\overline{\theta}_t(x)+\overline{\Psi}_s(x)\, (\gamma_5)_{st}\,\theta_t(x)\,]\\
    &=\int\dd[4]x\,\qty[\,\delta\Psi_{t}(x)\,\overline{\theta}_t(x)+\delta\overline{\Psi}_{t}(x)\,\theta_t(x)\,]\\
    &=\delta\Phi\,.
    \end{split}
\end{equation}
Moreover, we can confirm that the operator $A$ describes the symmetry since
\begin{equation}
    \begin{split}
       [\,Q,A\,]\,\theta_s(x)&={}-i\alpha\,(\gamma_5)_{st}\,Q\,\theta_t(x)+i\D_{st}(x)\,A\,\lambda_t(x)\\
        &={}-\alpha\,(\gamma_5)_{st}\,\D_{tu}(x)\,\lambda_u(x)-\alpha\,\D_{st}(x)\,(\gamma_5)_{tu}\,\lambda_u(x)\\
        &={}-\alpha\,(\gamma_5)_{st}\,\D_{tu}(x)\,\lambda_u(x)+\alpha\,(\gamma_5)_{st}\,\D_{tu}(x)\,\lambda_u(x)\\
        &=0\,,
    \end{split}
\end{equation}
where we used $\{\gamma_5\,,\gamma^\mu\}=0$. 
In the same way, we obtain
\begin{equation}
    [\,Q,A\,]\,\overline{\theta}_s(x)=0\,.
\end{equation}

The Ward-Takahashi identity in terms of homotopy algebras~\eqref{algebraic WT} is described by
\begin{equation}\label{WT chiral}
    \begin{split}
        &-\sum_{a\in\{\theta_s,\overline{\theta}_s\}}\big(\bra{\,\omega_n^{a_1,\dots,a_n}(x_1,\dots x_n)}\pi_n\, \otimes \bra{\,\omega}(e_a(x)\otimes \pi_1\vb{Q})\otimes\bra{\omega}(\widetilde{e}^{\,a}(x)\otimes \pi_1\vb{A})\big)\,\vb{f\,1}\\
        &=-i\hbar\, \sum_{i=1}^n\, \delta^4(x-x_i)\bra{\,\omega_n^{a_1,\dots,a_n}(x_1,\dots x_n)}(\pi_{i-1}\otimes\pi_1\vb{A}\otimes\pi_{n-i})\,\vb{f\,1}\\
        &\qquad-i\hbar\sum_{a\in\{\theta_s,\Bar{\theta}_s\}}\big(\bra{\,\omega_n^{a_1,\dots,a_n}(x_1,\dots x_n)}\pi_n\otimes\bra{\omega}(\pi_1\,\vb{A}\,\vb{e}_a(x)\otimes\widetilde{e}^{\,a}(x) )\big)\,\vb{f\,1}\,,
    \end{split}
\end{equation}
where
\begin{equation}
    e_{a}(x) = \left\{
    \begin{array}{cc}
        \theta_s(x) & (a=\theta_s)\\
        \,\overline{\theta}_s(x) & \,\,(a=\Bar{\theta}_s)
    \end{array}
    \right.
\end{equation}
and
\begin{equation}
    \widetilde{e}^{\,a}(x) = \left\{
    \begin{array}{cc}
        \overline{\lambda}_s(x) & (a=\theta_s)\\
        \,\lambda^s(x) & \,\,(a=\Bar{\theta}_s)
    \end{array}
    \right.\,.
\end{equation}  
For simplicity, we take $n=2k$, $a_i=\Bar{\theta}_{s_i}$ for $i=1,\dots,k$ and $a_i=\theta_{s_i}$ for $i=k+1,\dots,2k$.
Then, the operator $\bra{\omega_n^{a_1,\dots,a_n}(x_1,\dots x_n)}$ is
\begin{equation}
    \begin{split}
    \bra{\omega_{2k}^{a_1,\dots,a_{2k}}(x_1,\dots x_{2k})}=&\bra{\omega}(\lambda_{s_1}(x_1)\otimes\pi_1)\otimes\cdots\otimes\bra{\omega}(\lambda_{s_k}(x_{k})\otimes\pi_1)\\
    &\otimes\bra{\omega}(\overline{\lambda}_{s_{k+1}}(x_{k+1})\otimes\pi_1)\otimes\cdots\otimes\bra{\omega}(\overline{\lambda}_{s_{2k}}(x_{2k})\otimes\pi_1)\,.
    \end{split}
\end{equation} 
Since
\begin{align}
    \pi_n\,\vb{f\,1}=\expval{\,\Phi^{\otimes n}\,}=\expval{\,\left[\int\dd[4]x\ \big(\Psi_s(x)\,\overline{\theta}_s(x)+\overline{\Psi}_s(x)\,\theta_s(x)\big)\right]^{\otimes n}\,}\,,
\end{align}
the left-hand side of~\eqref{WT chiral} is given by
\begin{equation}
    \begin{split}
    -\int\dd[4]y\,\dd[4]z\,\langle&\,\Psi_{s_1}(x_1)\dots\Psi_{s_k}(x_k)\,\overline{\Psi}_{s_{k+1}}(x_{k+1})\dots\overline{\Psi}_{s_{2k}}(x_{2k})\\
    &\big[\,\Psi_t(y)\,\overline{\Psi}_u(z)\bra{\omega}\big(\theta_s(x)\otimes Q\, \overline{\theta}_t(y)\big)\otimes\bra{\omega}\big(\overline{\lambda}_s(x)\otimes A\,\theta_u(z)\big)\\
    &+\overline{\Psi}_t(y)\,\Psi_u(z)\bra{\omega}\big(\overline{\theta}_s(x)\otimes Q\,\theta_t(y)\big)\otimes\bra{\omega}\big(\lambda_s(x)\otimes A\,\overline{\theta}_u(z)\big)\big]\,\rangle\,.
    \end{split}
\end{equation}
Since 
\begin{equation}
    \begin{split}
     &-\int\dd[4]y\,\dd[4]z\,\Psi_t(y)\,\overline{\Psi}_u(z)\,\bra{\omega}\big(\theta_s(x)\otimes Q\, \overline{\theta}_t(y)\big)\otimes\bra{\omega}\big(\overline{\lambda}_s(x)\otimes A\,\theta_u(z)\big)\\
     &=\alpha\,\overline{\Psi}(x)\,\D(x)\,\gamma_5\,\Psi(x)\,,
     \end{split}
\end{equation}
and 
\begin{equation}
    \begin{split}
    &-\int\dd[4]y\,\dd[4]z\,\overline{\Psi}_t(y)\,\Psi_u(z)\,\bra{\omega}\big(\overline{\theta}_s(x)\otimes Q\, \theta_t(y)\big)\otimes\bra{\omega}\big(\lambda_s(x)\otimes A\,\overline{\theta}_u(z)\big)\\
    &=-\alpha\,\overline{\Psi}(x)\,\overleftarrow{\D}(x)\,\gamma_5\,\Psi(x)\,,
    \end{split}
\end{equation} 
we obtain
\begin{equation}
    \begin{split}
    &-\int\dd[4]y\,\dd[4]z\,\Big[\,\Psi_t(y)\,\overline{\Psi}_u(z)\,\bra{\omega}\big(\theta_s(x)\otimes Q\, \overline{\theta}_t(y)\big)\otimes\bra{\omega}\big(\overline{\lambda}_s(x)\otimes A\,\theta_u(z)\big)   \\
&\quad+\,\overline{\Psi}_t(y)\,\Psi_u(z)\,\bra{\omega}\big(\overline{\theta}_s(x)\otimes Q\, \theta_t(y)\big)\otimes\bra{\omega}\big(\lambda_s(x)\otimes A\,\overline{\theta}_u(z)\big)\,\Big]\\
    &=\partial_\mu\qty[\,\alpha\,\overline{\Psi}_s(x)\,\gamma^\mu\, \gamma_5\,\Psi(x)\,]\\
    &=\partial_\mu\, j^\mu_5(x)\,,
    \end{split}
\end{equation}
where $j_5^\mu(x)$ denotes the Noether current associated with the symmetry~\eqref{chiral sym infinitesimal}.
Hence, the left-hand side of~\eqref{WT chiral} is equal to
\begin{align}
    \expval{\,\Psi_{s_1}(x_1)\dots\Psi_{s_{k}}(x_k)\,\overline{\Psi}_{s_{k+1}}(x_{k+1})\dots\overline{\Psi}_{s_{2k}}(x_{2k})\,\partial_\mu\, j^\mu_5\,}\,.
\end{align}
The first term on the right-hand side of~\eqref{WT chiral} can be expressed similarly to the previous examples as
\begin{equation}
    \begin{split}
    &-i\hbar\, \sum_{i=1}^n\, \delta^4(x-x_i)\,\bra{\,\omega_n^{a_1,\dots,a_n}(x_1,\dots x_n)}(\,\pi_{i-1}\otimes\pi_1\vb{A}\otimes\pi_{n-i}\,)\,\vb{f\,1}\\
    &=-i\hbar \sum_{i=1}^k \delta^4(x-x_i)\langle\,\Psi_{s_1}(x_1)\dots\Psi_{s_{i-1}}(x_{s-1})\,\delta\Psi_{s_i}(x_i)\,\Psi_{s_{i+1}}(x_{i+1})\dots\Psi_{s_k}(x_k)\\
    &\qquad\qquad\qquad\qquad\qquad\overline{\Psi}_{s_{k+1}}(x_{k+1})\dots\overline{\Psi}_{s_{2k}}(x_{2k})\,\rangle\\
    &\quad
    -i\hbar \sum_{i=k+1}^{2k} \delta^4(x-x_i)\langle\, \Psi_{s_1}(x_1)\dots\Psi_{s_k}(x_k)\\
    &\qquad\qquad\qquad\qquad\qquad\quad\overline{\Psi}_{s_{k+1}}(x_{k+1})\dots\overline{\Psi}_{s_{i-1}}(x_{i-1})\,\delta\overline{\Psi}_{s_i}(x_i)\,\overline{\Psi}_{s_{i+1}}(x_{i+1})\dots\overline{\Psi}_{s_{2k}}(x_{2k})\,\rangle\,.
    \end{split}
\end{equation}
The second term still contains an ill-defined factor 
\begin{equation}
    \begin{split}
        &\bra{\omega}\big(A\,\theta_s(x)\otimes \overline{\lambda}_s(x)\big)+\bra{\omega}(A\,\overline{\theta}_s(x)\otimes \lambda_s(x)\big)\\
        &=2\,i\,\alpha\lim_{y\rightarrow x} \mathrm{tr}\,\gamma_5\,\delta^4(x-y)\,.
    \end{split}
\end{equation}
As is well known, the axial symmetry is anomalous, and this factor does not vanish after the regularization.
We will explore how to regularize this term from the homotopy algebraic point of view in the next section. 

\section{Regularization with stubs}\label{sec:reg}
\setcounter{equation}{0}
As we mentioned in the previous section, we need to regularize the term
\begin{align}
     -i\hbar\,\sum_a\,\bra{\omega}(\pi_1\,\vb{A\,e}_{a}(x)\otimes \widetilde{e}^{\,a}(x))\,\vb{f\,1}\,,
\end{align}
to mediate the divergence.
To preserve the homotopy algebraic structure, we introduce the  regularization with \textit{stubs}.
The concept of stubs~\cite{Zwiebach:1992ie} is widely used in the context of string field theory, and they can be used for the regularization.
Stubs in quantum field theory are discussed in~\cite{Chiaffrino:2021uyd}.
In this regularization, the operators $\vb{M}$ and $\vb{U}$ are modified by the heat kernel in the way to preserve the  homotopy algebraic structure.
By repeating the previous arguments using these modified operators, we obtain the correct anomalous term. 
Let us comment on  technical facts about the heat kernel. In the Lorentzian signature, the heat kernel does not regularize path integrals. 
The solution is that we formally calculate and when we carry out the integral involving the heat kernel, we introduce the Wick rotation. 

Let us explain the detailed treatment of stubs in homotopy algebras. 
In~\cite{Schnabl:2023dbv,Erbin:2023hcs,Schnabl:2024fdx,Maccaferri:2024puc,Chiaffrino:2021uyd}, adding stubs in terms of homotopy algebras is described by using the homological perturbation lemma (HPL)~\cite{crainic2004perturbation}. 
Before considering stubs, we explain the statement of the HPL.
Consider two degree-even maps $\pi:\mathcal{H}\rightarrow\mathcal{H}'$ and $\iota:\mathcal{H}'\rightarrow\mathcal{H}$, where $\mathcal{H}'$ denotes image of $\pi$, and degree-odd map $\eta:\mathcal{H}\rightarrow\mathcal{H}$ satisfying
\begin{align}
    \pi\,\iota&=\mathbb{I}\,,\label{piiota1}\\
    Q\,\eta+\eta \,Q&=\mathbb{I}-\iota\,\pi\,,\label{hodge-kodaire eta}\\
    \eta^2=\eta\,\iota&=\pi\,\eta=0\,.\label{anihilation}
\end{align}
From the condition~\eqref{piiota1}, it follows that $\iota\,\pi$ becomes projection:
\begin{align}
    (\,\iota\,\pi\,)\,(\,\iota\,\pi\,)=\iota\,(\,\pi\,\iota\,)\,\pi=\iota\,\pi\,.
\end{align}
These maps can be promoted to operators on the tensor algebra as follows:
\begin{equation}
\begin{split}
    \bm{\pi}&=\sum_{n=1}^\infty\,\pi^{\otimes n}\,,\\
    \bm{\iota}&=\sum_{n=1}^\infty\,\iota^{\otimes n}\,,\\
    \bm{\eta}&=\sum_{n=1}^\infty\sum_{k=0}^{n-1}\,\mathbb{I}^{\otimes k}\otimes \eta\otimes (\iota\pi)^{\otimes n-k-1}\,.
\end{split}
\end{equation}
These operators also satisfy
\begin{align}
     \bm{\pi\,\iota}&=\vb{I}\,,\\
    \vb{Q}\,\bm{\eta}+\bm{\eta}\, \vb{Q}&=\vb{I}-\bm{\iota\,\pi}\,,\\
    \bm{\eta}^2=\bm{\eta\,\iota}&=\bm{\pi\,\eta}=0\,.
\end{align}
According to the HPL, if the operator $\vb{M}$ satisfies the quantum $A_\infty$ relations
\begin{equation}
    (\,\vb{M}+i\hbar\, \vb{U}\,)^2=0\,,
\end{equation}
the operators 
\begin{equation}
\begin{split}
    \bm{\pi}'&=\bm{\pi}\,\frac{1}{\vb{I}+(\vb{m}+i\hbar\, \vb{U})\,\bm{\eta}}\,,\\
    \bm{\iota}'&=\frac{1}{\vb{I}+\bm{\eta}\,(\vb{m}+i\hbar\, \vb{U})}\,\bm{\iota}\,,\\
    \bm{\eta}'&=\frac{1}{\vb{I}+\bm{\eta}\,(\vb{m}+i\hbar\, \vb{U})}\,\bm{\eta}\,,
\end{split}
\end{equation}
correspond to the perturbed operators of $\bm{\pi},\bm{\iota}$ and $\bm{\eta}$, respectively, and satisfy
\begin{align}
    \bm{\pi}'\,\bm{\iota}'&=\vb{I}\,,\\
    (\,\vb{M}+i\hbar\, \vb{U}\,)\,\bm{\eta}'+\bm{\eta}' \,(\,\vb{M}+i\hbar\, \vb{U}\,)&=\vb{I}-\bm{\iota}'\,\bm{\pi}'\,,\label{M commute eta}\\
    (\bm{\eta}')^2=\bm{\eta}'\,\bm{\iota}'&=\bm{\pi}'\,\bm{\eta}'=0\,.
\end{align} 
When we can define a coderivation $\vb{M}'$ and  a second-order coderivation $\vb{U}'$ as
\begin{equation}
    \begin{split}
    \vb{M}'&=\bm{\pi}\,\vb{M}\,\bm{\iota}'\,,\\
    \vb{U}'&=\bm{\pi}\,\vb{U}\,\bm{\iota}\,,
    \end{split}
\end{equation}
these operators satisfy
\begin{align}
    (\,\vb{M}'+i\hbar\,\vb{U}'\,)^2=0\,,
\end{align}
that is, operator $\vb{M}'$ satisfies the quantum $A_\infty$ relations on $\mathcal{H}'$.
Furthermore, operators $\bm{\pi}'$ and $\bm{\iota}'$ meet the chain map conditions:
\begin{equation}\label{chain map}
    \begin{split}
    (\,\vb{M}'+i\hbar\vb{U}'\,)\,\bm{\pi}'&={\bm{\pi}}'\,(\vb{M}+i\hbar\vb{U})\,,\\
    {\bm{\iota}}'\,(\,\vb{M}'+i\hbar\vb{U}'\,)&=(\,\vb{M}+i\hbar\,\vb{U}\,)\,\bm{\iota}'\,.
    \end{split}
\end{equation}
To apply the HPL to adding stub, we define two degree-even maps $\pi,\iota$ by
\begin{align}
    \pi\,e_a(x)=\iota\,e_a(x)=K_{\lambda/2}\,e_a(x)\,,\quad
    \pi\,\widetilde{e}^{\,a}(x)=\iota\,\widetilde{e}^{\,a}(x)=K_{\lambda/2}\,\widetilde{e}^{\,a}(x)\,.
\end{align}
Here, $K_{\lambda/2}$ is a heat kernel labeled by the positive parameter $\lambda$.
The degree-odd map $\eta$ that satisfies \eqref{hodge-kodaire eta} for these $\pi$ and $\iota$ is defined by
\begin{align}
    \eta=(1-K_{\lambda})\,h\,.
\end{align}
Let us see the example of the stub in scalar field theory. In such a case, we define
\begin{equation}
    K_{\lambda}=e^{-\lambda(-\partial^2+m^2)}\,,
\end{equation}
by using the heat kernel of a scalar field. Then, the basis vectors of $\mathcal{H}$ is projected as follows:
\begin{equation}
    \pi\, c(x)=\iota\, c(x)=e^{-\frac{\lambda}{2}(-\partial^2+m^2)}\,c(x)\,,\quad \pi\, d(x)=\iota\, d(x)=e^{-\frac{\lambda}{2}(-\partial^2+m^2)}\,d(x)\,,
\end{equation}
where we used the basis vectors defined in \S\ref{scalar}. We will see that this projection serves as the regularization by analyzing the contact terms in the Ward-Takahashi identities. In addition, note that the limit $\lambda\rightarrow 0$ corresponds to the original theory without stubs.  

Unfortunately, these $\pi,\iota$ and $\eta$ do not satisfy the conditions~\eqref{piiota1} and~\eqref{anihilation} except $\eta^2=0$. 
In this case, the operator $\vb{M}'$ does not become coderivations.
According to~\cite{Schnabl:2023dbv}, the solution is simple: we simply pretend that these conditions hold.
More precise treatment is done in Appendix~A of~\cite{Erbin:2023hcs}.
Alternatively, we can ensure that the conditions are met by defining the product appropriately with respect to $K_{\lambda/2}$ and $K_\lambda$.
If we define the products as 
\begin{align}
    K_{\lambda/2}\cdot K_{\lambda/2}=K_{\lambda/2}\cdot K_\lambda=K_\lambda\cdot K_{\lambda/2}=K_\lambda\cdot K_\lambda=K_\lambda\,,
\end{align}
then the operator $\pi\iota$ can be treated as a projection.

We define the symprectic form $\bra{\omega'}$ on $\mathcal{H}'$ by
\begin{align}
    \bra{\omega'}=\bra{\omega}(\iota\otimes\iota)\,.\label{omegastub}
\end{align}
By using the operators $\vb{M}'$ and the symplectic form $\bra{\omega'}$, the stubbed action $S'$ on $\mathcal{H}'$ is defined as
\begin{align}
   S'&=-\sum_{n=0}^\infty \frac{1}{n+1}\bra{\,\omega'}(\,\mathbb{I}\otimes\pi_1\, \vb{M}'_n\,)\,\bm{\pi}\,\frac{1}{1-\Phi}\,.
\end{align}
This action is quasi-isomorphic to the original action and reproduces the original action with the limit $\lambda\rightarrow0$.
Starting from this stubbed action, we can construct the counterparts of the correlation function $\vb{f\,1}$ in the stubbed theory $\vb{f}'\,\vb{1}$ as 
\begin{align}
    \vb{f}'\,\vb{1}=\frac{1}{\vb{I}+\vb{h}'\,(\,\vb{m}'+i\hbar\, \vb{U}'\,)}\,\vb{1}\,,
\end{align}
where $\vb{m}'$ and $\vb{h}'$ are defined by
\begin{equation}
    \begin{split}
    \vb{m}'&=\vb{M}'-{\bm{\pi}}\,\vb{Q}\,{\bm{\iota}}\,,\\
    \vb{h}'&=\bm{\pi}\,\vb{h}\,\bm{\iota}\,.
    \end{split}
\end{equation}
Here, the operators $\vb{h}'$ is defined to satisfy
\begin{align}
    &{\bm{\pi}}\,\vb{Q}\,{\bm{\iota}}\, \vb{h}'+ \vb{h}'\,{\bm{\pi}}\,\vb{Q}\,{\bm{\iota}}=\vb{I}\,.
\end{align}
As discussed in \S~\ref{correlation function}, the algebraic Schwinger-Dyson equation holds for $\vb{f}'$:
\begin{align}
    (\,\vb{M}'+i\hbar \vb{U}'\,)\,\vb{f}'\,\vb{1}=0\,.
\end{align}
Let us consider the transformation $\vb{A}'$ defined by
\begin{align}
    \vb{A}'=\bm{\pi}\,\vb{A}\,\bm{\iota}\,,
\end{align}
as counterpart of the symmetry transformation $\vb{A}$.
This $\vb{A}'$ is not a symmetry transformation because it does not commute with $\vb{M}'$ for $\lambda>0$.
However, for $\lambda\rightarrow0$, it commutes with $\vb{M}'$ and becomes a symmetry transformation.
By repeating the arguments of \S~\ref{sec:noether} using $\vb{f}'\,\vb{1}$ and $\vb{A}'$, and finally setting $\lambda\rightarrow0$, we obtain the regularized anomalous term:
    \begin{equation}
    \begin{split}
     &-i\hbar\,\lim_{\lambda\rightarrow0}\,\sum_a\,\bra{\,\omega'}\big(\,\pi_1\,\vb{A}'\,\bm{\pi}\, \vb{e}_a(x)\,{\bm{\iota}}\otimes \pi\,\widetilde{e}^{\,a}(x)\,\big)\, \vb{f}'\,\vb{1}\\
     &= -i\hbar\,\lim_{\lambda\rightarrow0}\,\sum_a\,\bra{\,\omega}\,\big(\,\iota\,\pi\,\pi_1\,\vb{A}\,{\bm{\iota\,\pi}} \,\vb{e}_a(x)\,{\bm{\iota}}\otimes \iota\,\pi\,\widetilde{e}^{\,a}(x)\,\big)\, \vb{f}'\,\vb{1}\,.
     \end{split}
\end{equation}
Since the  projection $\iota\,\pi$ satisfies
\begin{equation}
    \begin{split}
    \bra{\omega}(\,\iota\,\pi\otimes\mathbb{I}\,)&=\bra{\omega}(\,\mathbb{I}\otimes\iota\,\pi\,)=\bra{\omega}(\,\iota\pi\otimes\iota\pi\,)\,,\\
    \iota\,\pi\, e_a(x)\otimes \widetilde{e}^{\,a}(x)&= e_a(x)\otimes \iota\,\pi\,\widetilde{e}^{\,a}(x)=\iota\,\pi\, e_a(x)\otimes \iota\,\pi\,\widetilde{e}^{\,a}(x)\,,
\end{split}
\end{equation}
the regularized anomalous term can be also expressed as
\begin{align}
     -i\hbar\,\lim_{\lambda\rightarrow0}\,\sum_a\bra{\,\omega}\big(\pi_1\,\vb{A}\,{\bm{\iota\,\pi}} \,\vb{e}_a(x){\bm{\iota}}\otimes \widetilde{e}^{\,a}(x)\big)\, \vb{f}'\,\vb{1}\,.
\end{align}
In particular, when the symmetry transformation $\vb{A}$ is  linear as in the examples in the previous section, it can be simplified as
\begin{align}
     -i\hbar\,\lim_{\lambda\rightarrow0}\sum_{a} \bra{\omega}\big(A\,K_{\lambda}\, e_{a}(x)\otimes \widetilde{e}^{\,a}(x)\big)\,.
\end{align}
Due to the smearing effect by the heat kernel over the delta function, this term is regularized.

Let us compute the anomaly for the example in the previous section.
We first consider the case of the $U(1)$ symmetry in complex scalar theories. The symmetry transformation $A$ is defined by~\eqref{def-A-cpx}, and the stub is chosen as
\begin{equation}
    \begin{split}
    \pi\, c(x)=\iota\, c(x)=e^{-\frac{\lambda}{2}(-\partial^2+m^2)}\,c(x)\,,&\quad \pi\, d(x)=\iota\, d(x)=e^{-\frac{\lambda}{2}(-\partial^2+m^2)}\,d(x)\,,\\
    \pi\, c^\dag(x)=\iota\, c^\dag(x)=e^{-\frac{\lambda}{2}(-\partial^2+m^2)}\,c^\dag(x)\,,&\quad \pi\, d^\dag(x)=\iota\, d^\dag(x)=e^{-\frac{\lambda}{2}(-\partial^2+m^2)}\,d^\dag(x)\,.
    \end{split}
\end{equation}
Then, the anomalous term can be evaluated as
\begin{equation}
    \begin{split}
    &-i\,\hbar\,\lim_{\lambda\rightarrow0}\sum_{a\in\{c,c^\dag\}} \bra{\omega}\big(A\,K_{\lambda}\,e_a(x)\otimes \widetilde{e}^{\,a}(x)\big)\\
     &=\hbar\,\alpha\,\lim_{\lambda\rightarrow0}\lim_{y\rightarrow x}\big[e^{-\lambda(-\partial_x^2+m^2)}\,\delta^d(x-y)-e^{-\lambda(-\partial_x^2+m^2)}\,\delta^d(x-y)\big]\\
     &=i\,\hbar\,\alpha\,\lim_{\lambda\rightarrow0}\qty[\,\int\frac{\dd[d]k}{(2\pi)^d}\,e^{-\lambda(k^2+m^2)}-\int\frac{\dd[d]k}{(2\pi)^d}\,e^{-\lambda(k^2+m^2)}\,]\\
     &=0\,,
     \end{split}
\end{equation}
where we carried out the Wick rotation in the third line.
Therefore, we  confirm that there is no anomaly in this case.

Let us next consider the case of the energy-momentum tensor in scalar theories. The symmetry transformation $A$ is defined by~\eqref{trsf em tensor} and the stub is the same as above.
Then, the anomalous term can be evaluated as
\begin{equation}
    \begin{split}
     &-i\,\hbar\lim_{\lambda\rightarrow0} \bra{\omega}\big(A\,K_{\lambda}\,c(x)\otimes d(x)\big)\\
     &={}-i\,\hbar\, a_\mu\, \lim_{\lambda\rightarrow0}\lim_{y\rightarrow x}e^{-\lambda(-\partial_x^2+m^2)}\,\partial_x^\mu\, \delta^d(x-y)\\
     &=i\,\hbar\, a_\mu\, \lim_{\lambda\rightarrow0}\,\int\frac{\dd[d]k}{(2\pi)^d}\,k^\mu\, e^{-\lambda(k^2+m^2)}\\
     &=0\,.
     \end{split}
\end{equation}
Again, we carried out the Wick rotation in the third line. Therefore, we  confirm that there is no anomaly in this case.

Finally, let us consider the axial $U(1)$ symmetry in vector-like gauge theories. The symmetry transformation is defined by~\eqref{chiral sym A}, and the stub can be chosen as
\begin{equation}
    \begin{split}
    \pi\,\theta_s(x)=\iota\, \theta_s(x)=(e^{-\lambda \not D^2(x)})_{st}\,\theta_t(x)\,,
    &\quad\pi\,\lambda_s(x)=\iota\,\lambda_s(x)=(e^{-\lambda \not D^2(x)})_{st}\,\lambda_t(x)\,,\\
    \pi\,\overline{\theta}_s(x)=\iota\, \overline{\theta}_s(x)=\overline{\theta}_t(x)\,(e^{-\lambda \overleftarrow{\not D}^2(x)})_{ts}\,,&\quad
    \pi\overline{\lambda}_s(x)=\iota\,\overline{\lambda}_s(x)=\overline{\lambda}_t(x)\,(e^{-\lambda \overleftarrow{\not D}^2(x)})_{ts}\,.
    \end{split}
\end{equation}
The anomalous term can be calculated as 
\begin{equation}
    \begin{split}
    &-i\,\hbar\,\lim_{\lambda\rightarrow 0}\big[\bra{\omega}\big(A\,K_\lambda\, \theta_s(x)\otimes \overline{\lambda}_s(x)\big)+\bra{\omega}\big(A\,K_\lambda\, \overline{\theta}_s(x)\otimes \lambda_s(x)\big)\big]\\
    &=\hbar\,\alpha\,\lim_{\lambda\rightarrow 0}\,\lim_{y\rightarrow x}\,\mathrm{tr}\,\big[\gamma_5\,e^{-\lambda \not D^2(x)}\,\delta^4(x-y)+\delta^{4}(x-y)\,e^{-\lambda \overleftarrow{\not D}^2(x)}\,\gamma_{5}\big]\\
    &=\hbar\,\alpha\,\lim_{\lambda\rightarrow 0}\,\int\,\frac{\dd[4]k}{(2\pi)^4}\,\mathrm{tr}\,\big[\gamma_5\,e^{-ikx}\,e^{-\lambda \not D^2(x)}\,e^{ikx}+\gamma_5\,e^{ikx}\,e^{-\lambda \overleftarrow{\not D}^2(x)}\,e^{-ikx}\big]\,.
    \end{split}
\end{equation}
As is well known, this quantity can be evaluated as follows.
Since
\begin{align}
    e^{-ikx}\,D_\mu\, e^{ikx}=D_\mu+i\,k_\mu\,,
\end{align}
we have 
\begin{align}
    e^{-ikx}\,e^{-\lambda \not D^2(x)}\,e^{ikx}=e^{-\lambda(k-iD)^2+\frac{i\lambda}{2}\gamma^\mu\gamma^\nu F_{\mu\nu}}
\end{align} 
where $F_{\mu\nu}$ denotes the field strength of background field $A_\mu$
\begin{equation}
    F_{\mu\nu}=i\,[D_\mu\,,D_\nu]\,.
\end{equation} 
Integration over $k$ yields a factor $\lambda^{-2}$ and the formulas
\begin{align}
    \mathrm{tr}\,\gamma_5=\mathrm{tr}\,\gamma_5\gamma^\mu\gamma^\nu=0\,,\quad
    \mathrm{tr}\,\gamma_5\gamma^\mu\gamma^\nu\gamma^\rho\gamma^\sigma=-4\,i\,\epsilon^{\mu\nu\rho\sigma}\,,
\end{align}
leads
\begin{equation}
    \begin{split}
        &-i\,\hbar\,\lim_{\lambda\rightarrow 0}\big[\bra{\omega}\big(A\,K_\lambda\, \theta_s(x)\otimes \overline{\lambda}_s(x)\big)+\bra{\omega}\big(A\,K_\lambda\, \overline{\theta}_s(x)\otimes \lambda_s(x)\big)\big]\\
        &=\hbar\,\alpha\,\lim_{\lambda\rightarrow 0}\,\int\,\frac{\dd[4]k}{(2\pi)^4}\,\mathrm{tr}\,\big[\gamma_5\,e^{-ikx}\,e^{-\lambda \not D^2(x)}\,e^{ikx}+\gamma_5\,e^{ikx}\,e^{-\lambda \overleftarrow{\not D}^2(x)}\,e^{-ikx}\big]\\
        &={}-\frac{i\,\hbar\,\alpha}{64\,\pi^{2}}\,\mathrm{tr}[\gamma_{5}\gamma^{\mu}\gamma^{\nu}\gamma^{\rho}\gamma^{\sigma}]F_{\mu\nu}F_{\rho\sigma}\\
        &={}-\frac{\hbar\,\alpha}{16\,\pi^2}\epsilon^{\mu\nu\rho\sigma}\,F_{\mu\nu}\,F_{\rho\sigma}\,,
    \end{split}
\end{equation}
where we carried out the Wick rotation in the third line.
This quantity is a well-known anomaly.\footnote{The ordinary derivation in quantum field theory is written in many textbooks. See, for example,~\S 77 of~\cite{srednicki2007quantum}. }

\section{Conclusions and discussion}\label{discussion}
\setcounter{equation}{0}
In this paper, we introduced the identity
\begin{align}
    (\,\vb{Q}+\vb{m}+i\hbar\,\vb{U}\,)\,\vb{f\,1}=0\,,
\end{align}
which we call the \textit{algebraic Schwinger-Dyson equation} and confirm that this is exactly the Schwinger-Dyson equation in terms of homotopy algebras. Then, we applied this identity to the formula for the expectation values of the derivative of the Noether current, and obtained the Ward-Takahashi identity in terms of homotopy algebras:
\begin{align}
     \expval{\,\partial_\mu\, j^\mu(x)\,}={}-i\,\hbar\,\sum_a\bra{\omega}(\pi_1\,\vb{A\,e}_a(x)\otimes \widetilde{e}^{\,a}(x))\,\vb{f\,1}\,.\label{single-WT-c}
\end{align}

In our formula, the terms associated with
\begin{equation}
    \sum_a\bra{\omega}(\pi_1\vb{A\,e}_a(x)\otimes \widetilde{e}^{\,a}(x))\,\vb{f\,1}
\end{equation}
contain divergence. As we discussed in \S\ref{general-WT} and \S\ref{sec:reg}, these terms are related to contact terms and anomalies, and we need appropriate regularizations. In our treatment, adding stubs makes the theory finite, and then we consider the original theory as the limit of the stub parameter $\lambda\rightarrow 0$. Moreover, these divergences are related to the renormalization of the composite operators. The process we make a replacement
\begin{equation}
    \frac{1}{1-\Phi}\,\longrightarrow\,\vb{f\,1}
\end{equation}
in \S\ref{sec:noether} corresponds to take the expectation values of fields. We formally dealt with this process in the general case. In the examples described in \S\ref{example}, the results are consistent with the results by ordinary prescriptions. We expect that our treatment is consistent with the ordinary prescriptions in quantum field theory in general. The treatment of these divergences originates from how to define the composite operator in terms of homotopy algebras, and it is important to consider this problem deeply.

Let us also comment on the extension to the gauge theory. In our example, we treated the background gauge field but the extension of the dynamical gauge field is straightforward. In this paper, the gauge is fixed at the beginning since we are considering correlation functions. This prescription provides a simple algebraic description, but the gauge symmetry is implicit. It is interesting and important to consider the direct relation to~\cite{Matsunaga,Chiaffrino:2021uyd}.

Let us mention future works. The motivation of this work is to apply this method to string field theory. Since the description of homotopy algebras is universal, we hope that this work contributes to finding and analyzing symmetries and anomalies in string field theory. Our work is based on the perturbative expansion, however, recently one of the authors are working on the nonperturbative description of correlation functions in homotopy algebras~\cite{Konosu-Okawa1}. Combined with these works, we hope that our work contributes to describing the nonperturbative anomaly from the homotopy algebraic point of view. 

We can also consider to extend this work to the higher form symmetry~\cite{Gaiotto:2014kfa}. In the context of higher form symmetry, the  understanding of the Ward-Takahashi identity plays an important role in describing the symmetry. Our formula enables us to deal with the Ward-Takahashi identity from the algebraic point of view. We expect that the algebraic approach for the Ward-Takahashi identity provides  effective tools to analyze the problem.

Lastly, it is interesting to develop our method to define quantum field theory from a mathematical point of view. It is important to relate our method to the formalism developed by Costello and Gwilliam using 
 factorization algebras~\cite{Costello:2016vjw,Costello:2021jvx}.

 We hope that our research will contribute to revealing the quantum aspect of quantum field theory and  string field theory.

\bigskip

\noindent
{\normalfont \bfseries \large Acknowledgments}

\medskip
We would like to thank Yuji Okawa for discussions and comments on the earlier version of the draft, and Hiroaki Matsunaga, Hayato Kanno, and Masashi Kawahira for discussions. 
J.~T.~Y. would like to thank Hiroshi Kunitomo and Taichi Tsukamoto for discussions.
The work of J.~T.~Y. is supported in part by JSPS KAKENHI Grant No. JP23KJ1311.

\appendix
\section{Proof: \eqref{cyclic}$\Leftrightarrow$\eqref{cyclic another}}\label{app: cyclic}
\setcounter{equation}{0}
For notational simplicity, in this section index $a$ and $x$ are written collectively as $A$.
First, we show \eqref{cyclic}$\Rightarrow$\eqref{cyclic another}.
Let us consider the quantity
\begin{equation}
     \sum_A\,\big[\,M_n(\,e_A\otimes\mathbb{I}^{\otimes n-1}\,)\,\otimes\,\tilde{e}_A\,+\,M_n(\,\widetilde{e}_A\otimes\mathbb{I}^{\otimes n-1}\,)\,\otimes\,e_A\,\big].
\end{equation}
By using the normalization~\eqref{nomalization of U}, this quantity can be rewritten as
\begin{align}
&\sum_A\,\big[\,M_n(\,e_A\otimes\mathbb{I}^{\otimes n-1}\,)\,\otimes\,\widetilde{e}^A\,+\,M_n(\,\widetilde{e}_A\otimes\mathbb{I}^{\otimes n-1}\,)\,\otimes\,e_A\,\big]\nonumber\\
    &=\sum_{A,B}\big[\,(-1)^{e_B}\widetilde{e}^B\,\otimes\,\omega\,\big(\,e_B,M_n(e_A\otimes\mathbb{I}^{\otimes n-1})\big)\otimes\,\widetilde{e}^A\nonumber\\
    &\qquad\quad+(-1)^{\widetilde{e}^B}e_B\,\otimes\,\omega\,\big(\,\widetilde{e}^B,M_n(e_A\otimes\mathbb{I}^{\otimes n-1})\big)\otimes\,\widetilde{e}^A\nonumber\\
    &\qquad\quad+(-1)^{e_B}\widetilde{e}^B\,\otimes\,\omega\,\big(\,e_B,M_n(\widetilde{e}_A\otimes\mathbb{I}^{\otimes n-1})\big)\otimes\,e_A\nonumber\\
    &\qquad\quad+(-1)^{\widetilde{e}_B}e_B\,\otimes\,\omega\,\big(\,\widetilde{e}^B,M_n(\widetilde{e}^A\otimes\mathbb{I}^{\otimes n-1})\big)\otimes\,e_A\,\big]\nonumber\\
    &=\sum_{A,B}\big[\,(-1)^{e_A e_B}\,\widetilde{e}_B\,\otimes\,\widetilde{e}^A\,\otimes\,\omega\,\big(\,e_B,M_n(e_A\otimes\mathbb{I}^{\otimes n-1})\big)\nonumber\\
    &\qquad\quad+(-1)^{e_A\widetilde{e}_B}\, e_B\,\otimes\,\widetilde{e}^A\,\otimes\,\omega\,\big(\,\widetilde{e}^B,M_n(e_A\otimes\mathbb{I}^{\otimes n-1})\big)\nonumber\\
    &\qquad\quad+\,(-1)^{\widetilde{e}_A e_B}\,\widetilde{e}_B\,\otimes\,e_A\,\otimes\,\omega\,\big(\,e_B,M_n(\widetilde{e}^A\otimes\mathbb{I}^{\otimes n-1})\big)\nonumber\\
    &\qquad\quad+(-1)^{\widetilde{e}^A\widetilde{e}_B}\, e_B\,\otimes\,e_A\,\otimes\,\omega\,\big(\,\widetilde{e}_B,M_n(\widetilde{e}^A\otimes\mathbb{I}^{\otimes n-1})\big)\,\big]\,.\label{proof of cyclic rewitten}
\end{align}
In the second equality, we used that $\omega(\dots)$ is an element of $\mathcal{H}^{\otimes 0}$ and can be commuted with any element of $\mathcal{H}$ with appropriate sign.
If we assume \eqref{cyclic}, $\omega$'s inputs can be replaced cyclically with appropriate sign and we find that the first and third terms in the left-hand side of \eqref{proof of cyclic rewitten} are 
\begin{align}
    &\sum_B-(-1)^{\widetilde{e}_B}\widetilde{e}^B\,\otimes\,\big[(-1)^{e_A}\widetilde{e}^A\,\otimes\, \omega\,\big(\,e_A,M_n(\mathbb{I}^{\otimes n-1}\otimes e_B)\big)\nonumber\\
    &\qquad\qquad\qquad\qquad+(-1)^{\widetilde{e}^A}e_A\,\otimes\, \omega\,\big(\,\widetilde{e}^A,M_n(\mathbb{I}^{\otimes n-1}\otimes e_B)\big)\big]\nonumber\\
    &=\sum_B-(-1)^{\widetilde{e}^B}\widetilde{e}^B\,\otimes\,M_n(\mathbb{I}^{\otimes n-1}\otimes e_B)\,,
\end{align}
where we used \eqref{nomalization of U} again.
In the same way, the second and fourth terms in the left-hand side of \eqref{proof of cyclic rewitten} are 
\begin{align}
    \sum_B-(-1)^{e_B}e_B\,\otimes\,M_n(\mathbb{I}^{\otimes n-1}\otimes \widetilde{e}^B)\ .
\end{align}
This completes the proof of \eqref{cyclic}$\Rightarrow$\eqref{cyclic another}.
On the other hand, the proof of \eqref{cyclic another}$\Rightarrow$\eqref{cyclic} can be obtained by following the above proof in reverse.

\section{The proof for the Schwinger-Dyson equations for scalar field theory}\label{old-SD}
\setcounter{equation}{0}
Following~\cite{Okawa:2022sjf,Konosu:2023pal,Konosu:2023rkm}, let us show that the formula
\begin{equation}
    \pi_{n}\,\vb{f}\,\vb{1}=\langle\,\Phi^{\otimes n}\,\rangle
\end{equation}
satisfies the Schwinger-Dyson equations.
In the path integral formalism, correlation functions are defined by
\begin{equation}
    \langle\,\varphi(x_{1})\varphi(x_{2})\ldots\varphi(x_{n})\,\rangle=\frac{1}{Z}\,\int\mathcal{D}\varphi\,\varphi(x_{1})\varphi(x_{2})\ldots\varphi(x_{n})e^{\frac{i}{\hbar}S}\,,
\end{equation}
where
\begin{equation}
    Z=\int\mathcal{D}\varphi\,e^{\frac{i}{\hbar}S}\,.
\end{equation}
Since
\begin{equation}
    \frac{1}{Z}\,\int\mathcal{D}\varphi\,\frac{\delta}{\delta\varphi(x_{n})}\qty[\varphi(x_{1})\varphi(x_{2})\ldots\varphi(x_{n-1})e^{\frac{i}{\hbar}S}]=0\,,
\end{equation}
the Schwinger-Dyson equations
\begin{equation}
    \sum_{i=1}^{n-1}\langle\,\varphi(x_{1})\ldots\varphi(x_{i-1})\delta^{d}(x_{i}-x_{n})\varphi(x_{i+1})\ldots\varphi(x_{n-1})\,\rangle+\frac{i}{\hbar}\langle\,\varphi(x_{1})\varphi(x_{2})\ldots\varphi(x_{n-1})\frac{\delta S}{\delta \varphi(x_{n})}\,\rangle=0\label{SD-scalar}
\end{equation}
are derived. 

Let us show~\eqref{SD-scalar} in terms of homotopy algebras. Since
\begin{equation}
    \pi_{n}\,{\vb f}^{-1}\,{\vb f}\,{\vb 1}=0
\end{equation}
for $n>0$, the relations
\begin{equation}
    \pi_{n}\,\vb{f}\,\vb{1}+\pi_{n}\,\vb{h}\,\vb{m}\,\vb{f}\,\vb{1}+i\hbar\,\pi_{n}\,\vb{h}\,\vb{U}\,\vb{f}\,\vb{1}=0
\end{equation}
hold.
We act the operator
\begin{equation}
    \bra{\,\omega_{n}(x_{1},\,\ldots,x_{n})}=(-1)^{n}\bra{\omega}(\,d(x_{1})\otimes\mathbb{I}\,)\cdots\bra{\omega}(\,d(x_{n})\otimes\mathbb{I}\,)\label{op-scalar}
\end{equation}
on the above relations:
\begin{equation}
        \bra{\,\omega_{n}(x_{1},\,\ldots,x_{n})}\,\pi_{n}\,\vb{f}\,\vb{1}
        +\bra{\,\omega_{n}(x_{1},\,\ldots,x_{n})}\,\pi_{n}\,\vb{h}\,\vb{m}\,\vb{f}\,\vb{1}
        +i\hbar\bra{\,\omega_{n}(x_{1},\,\ldots,x_{n})}\,\pi_{n}\,\vb{h}\,\vb{U}\,\vb{f}\,\vb{1}=0\,.\label{pre-SD-scalar}
\end{equation}
Since
\begin{equation}
    \pi_{n}\,{\vb f}\,{\vb 1}=\langle\,\Phi^{\otimes n}\,\rangle\,,
\end{equation}
the first term of the left-hand side is
\begin{equation}
     \bra{\,\omega_{n}(x_{1},\,\ldots,x_{n})}\,\pi_{n}\,\vb{f}\,\vb{1}=\langle\,\varphi(x_{1})\,\varphi(x_{2})\,\ldots\,\varphi(x_{n})\,\rangle\,.
\end{equation}
Let us next consider the second term of the left-hand side of~\eqref{pre-SD-scalar}.
\begin{equation}
    \pi_{n}\,\vb{h}\,\vb{m}\,\vb{f}\,\vb{1}=\sum_{k=0}^{\infty}\,\mathbb{I}^{\otimes(n-1)}\otimes hm_{k}(\Phi^{\otimes k})\,.
\end{equation}
Notice that
\begin{equation}
    m_{k}(\Phi^{\otimes k})=-\int \dd[d]x\, d(x)\,\frac{\delta S_{\mathrm{int}}}{\delta \varphi(x)}\,,
\end{equation}
where $S_{\mathrm{int}}$ is the interaction part of the action. Then, we obtain
\begin{equation}
    \pi_{n}\,\vb{h}\,\vb{m}\,\vb{f}\,\vb{1}=-\int\dd[d]x\,\dd[d]y\,\Phi^{\otimes(n-1)}\otimes \Delta(x-y)\,c(y)\frac{\delta S_{\mathrm{int}}}{\delta \varphi(x)}\,.
\end{equation}
Acting the operator~\eqref{op-scalar},
\begin{equation}
    \bra{\,\omega_{n}(x_{1},\,\ldots,x_{n})}\,\pi_{n}\,\vb{h}\,\vb{m}\,\vb{f}\,\vb{1}=-\int\,\dd[d]x\,\left\langle\,\varphi(x_{1})\,\ldots\,\varphi(x_{n-1})\Delta(x-x_{n})\frac{\delta S_{\mathrm{int}}}{\delta \varphi(x)}\,\right\rangle\,.
\end{equation}
Let us next consider the third term of the left-hand side of~\eqref{pre-SD-scalar}:
\begin{equation}
    \begin{split}
        &i\hbar\,\pi_{n}\,\vb{h}\,\vb{U}\,\vb{f}\,\vb{1}\\
        &=i\hbar\,\sum_{i=1}^{n-1}\int\dd[d]x\,\langle\,\Phi^{\otimes (i-1)}\otimes c(x)\otimes\Phi^{\otimes (n-i-1)}\otimes hd(x)\,\rangle\\
        &=i\hbar\,\sum_{i=1}^{n-1}\,\int\dd[d]x\dd[d]y\,\langle\,\Phi^{\otimes (i-1)}\otimes c(x)\otimes\Phi^{\otimes (n-i-1)}\otimes \Delta(x-y)\,c(y)\,\rangle\,.\\
    \end{split}
\end{equation}
Then, we obtain
\begin{equation}
    i\hbar\bra{\,\omega_{n}(x_{1},\,\ldots,x_{n})}\,\pi_{n}\,\vb{h}\,\vb{U}\,\vb{f}\,\vb{1}=i\hbar\,\sum_{i=1}^{n-1}\int\dd[d]x\,\Delta(x_{i}-x_{n})\langle\,\varphi(x_{1})\ldots\varphi(x_{i-1})\varphi(x_{i+1})\ldots\varphi(x_{n-1})\,\rangle\,.
\end{equation}
Therefore, the relations~\eqref{pre-SD-scalar} become
\begin{equation}
    \begin{split}
        &\langle\,\varphi(x_{1})\,\varphi(x_{2})\,\ldots\,\varphi(x_{n})\,\rangle\\
        &\qquad-\int\,\dd[d]x\,\left\langle\,\varphi(x_{1})\,\ldots\,\varphi(x_{n-1})\Delta(x-x_{n})\frac{\delta S_{\mathrm{int}}}{\delta \varphi(x)}\,\right\rangle\\
        &\qquad +i\hbar\,\sum_{i=1}^{n-1}\int\dd[d]x\,\Delta(x_{i}-x_{n})\langle\,\varphi(x_{1})\ldots\varphi(x_{i-1})\varphi(x_{i+1})\ldots\varphi(x_{n-1})\,\rangle=0\,.
    \end{split}
\end{equation}
Acting the operator $(-\partial_{x_{n}}^{2}+m^2)$, we obtain
\begin{equation}
    -\left\langle\,\varphi(x_{1})\,\ldots\,\varphi(x_{n-1})\frac{\delta S}{\delta \varphi(x_{n})}\,\right\rangle+i\hbar\sum_{i=1}^{n-1}\delta(x_{i}-x_{n})\,\langle\,\varphi(x_{1})\ldots\varphi(x_{i-1})\varphi(x_{i+1})\ldots\varphi(x_{n-1})\,\rangle=0\,.
\end{equation}
This is exactly the Schwinger-Dyson equations~\eqref{SD-scalar}.


\small
\begin{thebibliography}{99}
\bibitem{Stasheff:I}
  J.~D.~Stasheff,
  ``Homotopy associativity of $H$-spaces. I,''
  Trans. of the Amer. Math. Soc. {\bf 108}, 275 (1963).

\bibitem{Stasheff:II}
  J.~D.~Stasheff,   ``Homotopy associativity of $H$-spaces. II,''
  Trans. of the Amer. Math. Soc. {\bf 108}, 293 (1963).

\bibitem{Getzler-Jones}
  E.~Getzler and J.~D.~S.~Jones,
  ``$A_\infty$-algebras and the cyclic bar complex,''
  Illinois~J.~Math {\bf 34}, 256 (1990).

\bibitem{Markl}
  M.~Markl,
  ``A cohomology theory for $A (m)$-algebras and applications,''
  J. Pure Appl. Algebra {\bf 83}, 141 (1992).


\bibitem{Penkava:1994mu}
  M.~Penkava and A.~S.~Schwarz,
  ``$A_\infty$ algebras and the cohomology of moduli spaces,'' Trans. Amer. Math. Soc. {\bf 169}, 91 (1995)
  [hep-th/9408064].
  

\bibitem{Gaberdiel:1997ia}
  M.~R.~Gaberdiel and B.~Zwiebach,
  ``Tensor constructions of open string theories. 1: Foundations,''
  Nucl. Phys. {\bf B505}, 569 (1997) [hep-th/9705038].
  

\bibitem{Zwiebach:1992ie}
B.~Zwiebach,
``Closed string field theory: Quantum action and the Batalin-Vilkovisky master equation,''
Nucl. Phys. B \textbf{390}, 33-152 (1993)
[arXiv:hep-th/9206084 [hep-th]].

\bibitem{Markl:1997bj}
M.~Markl,
``Loop homotopy algebras in closed string field theory,''
Commun. Math. Phys. \textbf{221}, 367-384 (2001)
[arXiv:hep-th/9711045 [hep-th]].

\bibitem{Maccaferri:2022yzy}
C.~Maccaferri and J.~Vo\v{s}mera,
``The classical cosmological constant of open-closed string field theory,''
JHEP \textbf{10}, 173 (2022)
[arXiv:2208.00410 [hep-th]].

\bibitem{Maccaferri:2023gcg}
C.~Maccaferri, A.~Ruffino and J.~Vo\v{s}mera,
``The nilpotent structure of open-closed string field theory,''
JHEP \textbf{08}, 145 (2023)
[arXiv:2305.02843 [hep-th]].

\bibitem{Kajiura:2004xu}
H.~Kajiura and J.~Stasheff,
``Homotopy algebras inspired by classical open-closed string field theory,''
Commun. Math. Phys. \textbf{263}, 553-581 (2006)
[arXiv:math/0410291 [math.QA]].

\bibitem{Kajiura:2005sn}
H.~Kajiura and J.~Stasheff,
``Open-closed homotopy algebra in mathematical physics,''
J. Math. Phys. \textbf{47}, 023506 (2006)
[arXiv:hep-th/0510118 [hep-th]].

\bibitem{Erler:2013xta}
T.~Erler, S.~Konopka and I.~Sachs,
``Resolving Witten`s superstring field theory,''
JHEP \textbf{04}, 150 (2014)
[arXiv:1312.2948 [hep-th]].

\bibitem{Erler:2016ybs}
T.~Erler, Y.~Okawa and T.~Takezaki,
``Complete Action for Open Superstring Field Theory with Cyclic $A_\infty$ Structure,''
JHEP \textbf{08}, 012 (2016)
[arXiv:1602.02582 [hep-th]].

\bibitem{Kunitomo:2019glq}
H.~Kunitomo and T.~Sugimoto,
``Heterotic string field theory with cyclic $L_\infty$ structure,''
PTEP \textbf{2019}, no.6, 063B02 (2019)
[erratum: PTEP \textbf{2020}, no.1, 019201 (2020)]
[arXiv:1902.02991 [hep-th]].

\bibitem{Kunitomo:2022qqp}
H.~Kunitomo,
``Open-closed homotopy algebra in superstring field theory,''
PTEP \textbf{2022}, no.9, 093B07 (2022)
[arXiv:2204.01249 [hep-th]].

\bibitem{Kajiura:2003ax}
H.~Kajiura,
``Noncommutative homotopy algebras associated with open strings,''
Rev. Math. Phys. \textbf{19}, 1-99 (2007)
[arXiv:math/0306332 [math.QA]].

\bibitem{Sen:2016qap}
A.~Sen,
``Wilsonian Effective Action of Superstring Theory,''
JHEP \textbf{01}, 108 (2017)
[arXiv:1609.00459 [hep-th]].

\bibitem{Erbin:2020eyc}
H.~Erbin, C.~Maccaferri, M.~Schnabl and J.~Vo\v{s}mera,
``Classical algebraic structures in string theory effective actions,''
JHEP \textbf{11}, 123 (2020)
[arXiv:2006.16270 [hep-th]].

\bibitem{Koyama:2020qfb}
D.~Koyama, Y.~Okawa and N.~Suzuki,
``Gauge-invariant operators of open bosonic string field theory in the low-energy limit,''
[arXiv:2006.16710 [hep-th]].

\bibitem{Arvanitakis:2020rrk}
A.~S.~Arvanitakis, O.~Hohm, C.~Hull and V.~Lekeu,
``Homotopy Transfer and Effective Field Theory I: Tree-level,''
Fortsch. Phys. \textbf{70}, no.2-3, 2200003 (2022)
[arXiv:2007.07942 [hep-th]].

\bibitem{Arvanitakis:2021ecw}
A.~S.~Arvanitakis, O.~Hohm, C.~Hull and V.~Lekeu,
``Homotopy Transfer and Effective Field Theory II: Strings and Double Field Theory,''
Fortsch. Phys. \textbf{70}, no.2-3, 2200004 (2022)
[arXiv:2106.08343 [hep-th]].

\bibitem{Bonezzi:2023xhn}
R.~Bonezzi, C.~Chiaffrino, F.~Diaz-Jaramillo and O.~Hohm,
``Tree-level Scattering Amplitudes via Homotopy Transfer,''
[arXiv:2312.09306 [hep-th]].

\bibitem{Konopka:2015tta}
S.~Konopka,
``The S-Matrix of superstring field theory,''
JHEP \textbf{11}, 187 (2015)
[arXiv:1507.08250 [hep-th]].

\bibitem{Kunitomo:2020xrl}
H.~Kunitomo,
``Tree-level S-matrix of superstring field theory with homotopy algebra structure,''
JHEP \textbf{03}, 193 (2021)
[arXiv:2011.11975 [hep-th]].

\bibitem{Erler:2020beb}
T.~Erler and H.~Matsunaga,
``Mapping between Witten and lightcone string field theories,''
JHEP \textbf{11}, 208 (2021)
[arXiv:2012.09521 [hep-th]].

\bibitem{Hohm:2017pnh}
O.~Hohm and B.~Zwiebach,
``$L_{\infty}$ Algebras and Field Theory,''
Fortsch. Phys. \textbf{65}, no.3-4, 1700014 (2017)
[arXiv:1701.08824 [hep-th]].

\bibitem{Jurco:2018sby}
B.~Jur\v{c}o, L.~Raspollini, C.~S\"amann and M.~Wolf,
``$L_\infty$-Algebras of Classical Field Theories and the Batalin-Vilkovisky Formalism,''
Fortsch. Phys. \textbf{67}, no.7, 1900025 (2019)
[arXiv:1809.09899 [hep-th]].

\bibitem{Nutzi:2018vkl}
A.~N\"utzi and M.~Reiterer,
``Amplitudes in YM and GR as a Minimal Model and Recursive Characterization,''
Commun. Math. Phys. \textbf{392}, no.2, 427-482 (2022)
[arXiv:1812.06454 [math-ph]].

\bibitem{Arvanitakis:2019ald}
A.~S.~Arvanitakis,
``The $L_\infty$-algebra of the S-matrix,''
JHEP \textbf{07}, 115 (2019)
[arXiv:1903.05643 [hep-th]].

\bibitem{Macrelli:2019afx}
T.~Macrelli, C.~S\"amann and M.~Wolf,
``Scattering amplitude recursion relations in Batalin-Vilkovisky\textendash{}quantizable theories,''
Phys. Rev. D \textbf{100}, no.4, 045017 (2019)
[arXiv:1903.05713 [hep-th]].

\bibitem{Jurco:2019yfd}
B.~Jur\v{c}o, T.~Macrelli, C.~S\"amann and M.~Wolf,
``Loop Amplitudes and Quantum Homotopy Algebras,''
JHEP \textbf{07}, 003 (2020)
[arXiv:1912.06695 [hep-th]].

\bibitem{Saemann:2020oyz}
C.~Saemann and E.~Sfinarolakis,
``Symmetry Factors of Feynman Diagrams and the Homological Perturbation Lemma,''
JHEP \textbf{12}, 088 (2020)
[arXiv:2009.12616 [hep-th]].

\bibitem{Okawa:2022sjf}
Y.~Okawa,
``Correlation functions of scalar field theories from homotopy algebras,''
JHEP \textbf{05}, 040 (2024)
[arXiv:2203.05366 [hep-th]].

\bibitem{Konosu:2023pal}
K.~Konosu and Y.~Okawa,
``Correlation functions involving Dirac fields from homotopy algebras I: the free theory,''
[arXiv:2305.11634 [hep-th]].

\bibitem{Konosu:2023rkm}
K.~Konosu,
``Correlation functions involving Dirac fields from homotopy algebras II: the interacting theory,''
[arXiv:2305.13103 [hep-th]].

\bibitem{Batalin:1981jr}
I.~A.~Batalin and G.~A.~Vilkovisky,
``Gauge Algebra and Quantization,''
Phys. Lett. B \textbf{102}, 27-31 (1981).

\bibitem{Batalin:1983ggl}
I.~A.~Batalin and G.~A.~Vilkovisky,
``Quantization of Gauge Theories with Linearly Dependent Generators,''
Phys. Rev. D \textbf{28}, 2567-2582 (1983)
[erratum: Phys. Rev. D \textbf{30}, 508 (1984)].

\bibitem{Schwarz:1992nx}
A.~S.~Schwarz,
``Geometry of Batalin-Vilkovisky quantization,''
Commun. Math. Phys. \textbf{155}, 249-260 (1993)
[arXiv:hep-th/9205088 [hep-th]].

\bibitem{Gwilliam:2012jg}
O.~Gwilliam and T.~Johnson-Freyd,
``How to derive Feynman diagrams for finite-dimensional integrals directly from the BV formalism,'' Topology and quantum theory in interaction, 175-185, Contemp. Math., 718, Amer. Math. Soc., Providence, RI, 2018 [arXiv:1202.1554 [math-ph]].

\bibitem{Chiaffrino:2021pob}
C.~Chiaffrino, O.~Hohm and A.~F.~Pinto,
``Homological quantum mechanics,''
JHEP \textbf{02}, 137 (2024)
[arXiv:2112.11495 [hep-th]].

\bibitem{Masuda:2020tfa}
T.~Masuda and H.~Matsunaga,
``Perturbative path-integral of string fields and the $A_{\infty }$ structure of the BV master equation,''
PTEP \textbf{2022}, no.11, 113B04 (2022)
[arXiv:2003.05021 [hep-th]].

\bibitem{Doubek:2017naz}
M.~Doubek, B.~Jur\v{c}o and J.~Pulmann,
``Quantum $L_\infty$ Algebras and the Homological Perturbation Lemma,''
Commun. Math. Phys. \textbf{367}, no.1, 215-240 (2019)
[arXiv:1712.02696 [math-ph]].

\bibitem{Erler:2016rxg}
T.~Erler,
``Supersymmetry in Open Superstring Field Theory,''
JHEP \textbf{05}, 113 (2017)
[arXiv:1610.03251 [hep-th]].

\bibitem{Matsunaga}
H.~Matsunaga, 
``Homotopy algebra \& symmetry generators in QFT,''
talk in the workshop of Homotopy Algebra of Quantum Field Theory and Its Application, Yukawa Institute for Theoretical Physics, Kyoto University, Japan, March 2021.


\bibitem{Schnabl:2023dbv}
M.~Schnabl and G.~Stettinger,
``Open string field theory with stubs,''
JHEP \textbf{07}, 032 (2023)
[arXiv:2301.13182 [hep-th]].

\bibitem{Erbin:2023hcs}
H.~Erbin and A.~H.~F\i{}rat,
``Open string stub as an auxiliary string field,''
SciPost Phys. \textbf{17}, 044 (2024)
[arXiv:2308.08587 [hep-th]].

\bibitem{Schnabl:2024fdx}
M.~Schnabl and G.~Stettinger,
``More on stubs in open string field theory,''
[arXiv:2402.00308 [hep-th]].

\bibitem{Maccaferri:2024puc}
C.~Maccaferri, R.~Poletti, A.~Ruffino and B.~Valsesia,
``Adding stubs to quantum string field theories,''
[arXiv:2403.10471 [hep-th]].

\bibitem{Chiaffrino:2021uyd}
C.~Chiaffrino and I.~Sachs,
``QFT with stubs,''
JHEP \textbf{06}, 120 (2022)
[arXiv:2108.04312 [hep-th]].


\bibitem{srednicki2007quantum}
M.~Srednicki,
``Quantum Field Theory,''
Cambridge University Press (2007).

\bibitem{Erler:2015uba}
T.~Erler,
``Relating Berkovits and A$_\infty$ superstring field theories; small Hilbert space perspective,''
JHEP \textbf{10}, 157 (2015)
[arXiv:1505.02069 [hep-th]].

\bibitem{crainic2004perturbation}
M.~Crainic,
``On the perturbation lemma, and deformations,''
[arXiv:math/0403266 [math.AT]].

\bibitem{Konosu-Okawa1}
K.~Konosu and Y.~Okawa,
``Nonperturbative correlation functions from homotopy algebras,''
[arXiv:2405.10935 [hep-th]].


\bibitem{Gaiotto:2014kfa}
D.~Gaiotto, A.~Kapustin, N.~Seiberg and B.~Willett,
``Generalized Global Symmetries,''
JHEP \textbf{02}, 172 (2015)
[arXiv:1412.5148 [hep-th]].

\bibitem{Costello:2016vjw}
K.~Costello and O.~Gwilliam,
``Factorization Algebras in Quantum Field Theory: Volume 1,''
Cambridge University Press (2016).


\bibitem{Costello:2021jvx}
K.~Costello and O.~Gwilliam,
``Factorization Algebras in Quantum Field Theory: Volume 2,''
Cambridge University Press (2021).

\bibitem{Kajiura:2001ng}
H.~Kajiura,
``Homotopy algebra morphism and geometry of classical string field theory,''
Nucl. Phys. B \textbf{630}, 361-432 (2002)
[arXiv:hep-th/0112228 [hep-th]].

\end{thebibliography}
\end{document}